\documentclass{SciPost}

\hypersetup{
    colorlinks,
    linkcolor={red!50!black},
    citecolor={blue!50!black},
    urlcolor={blue!80!black}
}

\usepackage[bitstream-charter]{mathdesign}
\DeclareSymbolFont{usualmathcal}{OMS}{cmsy}{m}{n}
\DeclareSymbolFontAlphabet{\mathcal}{usualmathcal}

\fancypagestyle{SPstyle}{
\fancyhf{}
\lhead{\colorbox{scipostblue}{\bf \color{white} ~SciPost Physics }}
\rhead{{\bf \color{scipostdeepblue} ~Submission }}

\fancyfoot[C]{\textbf{\thepage}}
}

\usepackage[normalem]{ulem}
\usepackage{cancel}
\usepackage[utf8]{inputenc}

\usepackage{amsmath,amssymb,amsfonts,stmaryrd}
\usepackage{physics}
\usepackage{dsfont}
\usepackage{graphicx}
\usepackage{xcolor}
\usepackage{graphicx}
\usepackage{cancel}
\usepackage{color}
\usepackage{tikz}
\usetikzlibrary{matrix}
\usepackage{relsize}
\usepackage{multirow}
\usepackage[all]{xy}
\hypersetup{colorlinks=true,citecolor=blue,linkcolor=blue, urlcolor=blue, breaklinks=true}
\usepackage[scr=boondox]{mathalpha}

\usepackage{comment}
\usepackage[compat=1.1.0]{tikz-feynman}

\usepackage{bm} 
\usepackage{environ}
\usepackage{url}

\usepackage{mathtools}

\newcommand{\hsig}{\widehat{\sigma}}
\newcommand{\htau}{\widehat{\tau}} 

\newcommand{\ssi}{\sigma}   
\newcommand{\st}{\tau}      
\newcommand{\hssi}{\widehat{\sigma}} 
\newcommand{\hst}{\widehat{\tau}}  

\newcommand{\ssip}{\sigma'}          
\newcommand{\stp}{\tau'}          
\newcommand{\ssipp}{\sigma''}         
\newcommand{\stpp}{\tau''} 

\newcommand{\hssip}{\widehat{\sigma}'}

\newcommand{\htt}{\widehat{t}}
\newcommand{\mt}{\mathbf{t}}
\newcommand{\hmt}{\widehat{\mathbf{t}}}
\newcommand{\half}{\frac{1}{2}}
\newcommand{\mN}{\mathcal{N}}
\newcommand{\KT}{\mathcal{N}_{\text{KT}}}
\newcommand{\NKW}{\mathcal{N}_{\text{KW}}}
\newcommand{\NKWbar}{\bar{\mathcal{N}}_{\text{KW}}}
\newcommand{\NKWtilde}{\tilde{\mathcal{N}}_{\text{KW}}}

\newcommand{\iprime}{i'}
\newcommand{\jprime}{j'}

\newcommand{\hu}{\widehat{u}}
\newcommand{\hW}{\widehat{W}}
\newcommand{\hA}{\widehat{A}}
\newcommand{\ha}{\widehat{a}}

\newcommand{\beq}{\begin{equation}}
\newcommand{\eeq}{\end{equation}}
\newcommand{\row}{\text{row}}
\newcommand{\col}{\text{col}}

\newcommand{\anc}[1]{{\footnotesize{\color{magenta}  (AN:)#1  }}}

\begin{document}

\pagestyle{SPstyle}

\begin{center}{\Large \textbf{\color{scipostdeepblue}{
Web of non-invertible dualities for (2+1) dimensional models with subsystem symmetries\\
}}}\end{center}

\begin{center}\textbf{
Avijit Maity\textsuperscript{1$\star$},
Vikram Tripath\textsuperscript{1$\S$} and
Andriy H. Nevidomskyy\textsuperscript{2,3,4$\dagger$}
}\end{center}

\begin{center}
{\bf 1} Department of Theoretical Physics, Tata Institute of Fundamental Research, Homi Bhabha Road, Colaba, Mumbai 400005, India
\\
{\bf 2} Division of Condensed Matter Physics and Materials Science, Brookhaven National Laboratory, Upton, NY 11973-5000, USA
\\
{\bf 3} Department of Physics and Astronomy, Rice University, Houston, Texas 77005, USA
\\
{\bf 4} Rice Center for Quantum Materials and Advanced Materials Institute, Rice University, Houston, Texas 77005, USA
\\[\baselineskip]
$\star$ \href{mailto:avijit.maity@tifr.res.in}{\small avijit.maity@tifr.res.in}\,,\quad
$\S$ \href{mailto:vtripathi@theory.tifr.res.in}{\small vtripathi@theory.tifr.res.in}\,,\quad
$\dagger$ \href{mailto:anevidoms@bnl.gov}{\small anevidoms@bnl.gov}
\end{center}

\section*{\color{scipostdeepblue}{Abstract}}
\textbf{\boldmath{%
We extend non-invertible duality concepts familiar from one-dimensional systems to two spatial dimensions by constructing a web of non-invertible dualities for lattice models with subsystem symmetries. Focusing on $\mathbb{Z}_2\times\mathbb{Z}_2$ subsystem symmetry on the square lattice, we construct two complementary non-invertible dualities: a duality that maps spontaneous subsystem symmetry-broken (SSSB) phases to the trivial phase (often referred to as the Kramers-Wannier (KW) duality in 1+1D models), and a generalized subsystem Kennedy-Tasaki (KT) transformation that maps SSSB phases to subsystem symmetry-protected topological (SSPT) phases while leaving the trivial phase invariant. Crucially, these dualities are boundary-sensitive. On open lattices, both the subsystem KW and KT transformations can be implemented as unitary, invertible operators. In particular, the subsystem KT map not only identifies the bulk Hamiltonians of the dual phases but also carries the spontaneous ground-state degeneracy of the SSSB phase directly onto the protected boundary degeneracy characteristic of the SSPT phase. In contrast, on closed manifolds, the subsystem KW/KT maps become intrinsically non-unitary and non-invertible when restricted to the original Hilbert space. We establish this non-invertibility from three complementary perspectives -- ground state degeneracy matching (applied to two copies of the Xu-Moore/Ising-plaquette model), analysis of the symmetry-twist sector mapping, and the fusion algebra of the duality operator. We further show that enlarging the Hilbert space to include twisted sectors allows the formulation of the subsystem KW duality as a projective unitary which preserves quantum transition probabilities, consistent with the recent formulations of the generalized Wigner theorem for non-invertible symmetries. We also show that the KT map faithfully transmits the algebraic content of bulk and edge invariants diagnosing strong SSPT order: although strictly local SSPT repair operators map to highly nonlocal objects in the dual SSSB phase, the essential commutation algebra and the bulk-edge correspondence remain intact. We conclude with field-theoretic consistency checks and discuss implications for the classification and detection of subsystem-protected phases. Our construction provides a concrete lattice realization of non-invertible subsystem dualities, highlighting the central role of symmetry-twist sectors in characterizing generalized symmetries and exotic phases of quantum matter.
}}

\vspace{\baselineskip}

\vspace{10pt}
\noindent\rule{\textwidth}{1pt}
\tableofcontents
\noindent\rule{\textwidth}{1pt}
\vspace{10pt}


\section{Introduction}
\paragraph*{Gapped phases of matter and their classification:}The classification of quantum phases of matter and their transitions is a central challenge in theoretical physics. 
Landau’s paradigm, based on the concept of spontaneous symmetry breaking (SSB), provides a powerful framework through the lens of the symmetries of the Hamiltonian and the associated local order parameters. 
Yet, it has become clear that this approach is limited, as it cannot fully capture exotic phases with robust topological features. Gapped systems without symmetry breaking can host intrinsic topological order \cite{Wen_VacuumDeg_PRB1989,Wen_FQHDeg_PRB1990,wen1990topological}, which are characterized by patterns of long-range entanglement \cite{Kitaev_TEE_PRL2006, Levin_TEE_PRL2006}. 
Examples of such long-range entangled (LRE) phases include fractional quantum Hall states \cite{Laughlin_QHE_PRB1983} and quantum spin liquids \cite{Savary_QSLReview_2016}. On the other hand, gapped systems can remain nontrivial even without symmetry breaking or intrinsic topological order, provided they are protected by global symmetries. 
These phases are referred to as symmetry-protected topological (SPT) phases of matter, which are short-range entangled (SRE) states \cite{Chen_SPTcohom_PRB2013,Chen_SPT_science,Gu_SPTtensor_PRB2009,Pollmann_SPT_2010PhRvB}. 
While different LRE phases cannot be connected by local unitary transformations, SPT phases can be continuously deformed into a trivial product state through a local unitary transformation, provided the process is allowed to break the protecting global symmetry \cite{Chen_LU_2010PhRvB,Verstraete_Renormalization_2005PhRvL,Vidal_Entanglement_PRL2007}. 
Examples of SPTs are the spin-$1$ Haldane gap phase \cite{Haldane_spin1_PRL1983}, the Affleck--Kennedy--Lieb--Tasaki (AKLT) model \cite{AKLT_PRL1987}, and topological insulators \cite{Hasan_TI_2010RvMP}. Their nontrivial nature is manifested in gapless boundary modes that are protected by global symmetry, as seen in the archetypical example of the AKLT chain, which hosts fractionalized spin edge modes \cite{kennedy1990exact}. Unlike conventional phases with long-range order detectable by local correlations, the AKLT state is distinguished by a nonlocal string order parameter \cite{Nijs_stringorder_PRB1989}. These properties are also the defining characteristics of the spin-$1$ Haldane gap phase, which can be viewed as a smooth deformation of the  AKLT state.

\begin{figure}[tb]
    \centering
    \includegraphics[width=0.50\linewidth]{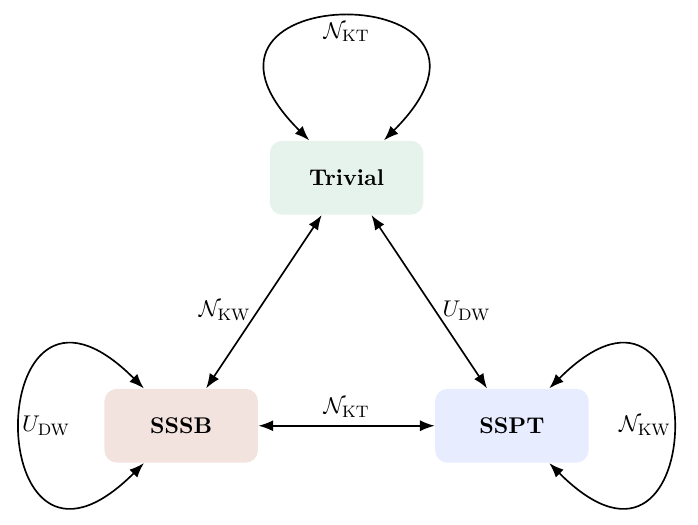}
    \caption{Schematic illustration of the Kramers--Wannier (KW), domain-wall decoration (DW) and Kennedy--Tasaki (KT) dualities in a model with subsystem symmetries such as the one studied in this work.  The dualities interrelate the phase with spontaneously broken subsystem symmetry (SSSB), the subsystem symmetry-protected topological (SSPT)  and the trivial phase.} 
    \label{fig:Dualities_N_KT}
\end{figure}

\paragraph*{Quantum dualities:}
Duality is a powerful concept in theoretical physics, where two seemingly different formulations can describe the same physical theory. The most well-known example is the so-called Kramers-Wannier (KW) duality transformation\footnote{Strictly speaking, the 1941 work by Kramers and Wannier~\cite{Kramers_duality_1941} demonstrated the duality at the level of the partition function of a finite-size classical 2D Ising model. Its extension to the self-duality of the quantum TFI chain model is more recent and can be demonstrated at the level of the bond algebra automorphism~\cite{Cobanera2010,Cobanera2011}, without referring to the classical Kramers--Wannier duality. 
Hence, some authors prefer not to use the KW acronym when referring to the latter TFI duality.} that relates the paramagnetic and ferromagnetic phases of the transverse-field Ising (TFI) chain \cite{Kramers_duality_1941, Kogut_gauge_RMP1979}. 
Similar to the KW duality transformation, Kennedy and Tasaki introduced in 1992 a nonlocal unitary transformation that maps spin-1 Hamiltonians realizing SPT on an open chain to the new Hamiltonians with global $\mathbb{Z}_2 \times \mathbb{Z}_2$ symmetry \cite{KT_1992hidden_Springer,KT_1992hidden_PRB}. 
Soon thereafter, Oshikawa revisited these results and provided a simple, compact expression valid for any integer spin \cite{oshikawa1992hidden}.
As a result, spontaneous breaking of this ``hidden'' $\mathbb{Z}_2 \times \mathbb{Z}_2$ symmetry in the KT-dual model is mapped onto the nonzero expectation value of the long-range string order parameter in the original system. 
At the same time, the ground-state degeneracy of the dual SSB system accounts for the degeneracy associated with the gapless edge modes of the original model. In summary, the Kennedy-Tasaki (KT) transformation reveals a duality between an SPT phase and a SSB phase in $\mathbb{Z}_2 \times \mathbb{Z}_2$ symmetric Hamiltonians. Fig.~\ref{fig:Dualities_N_KT} provides a schematic summary of the dualities

\paragraph*{Generalized symmetries:}
In recent years, the traditional notion of symmetry in physics has been significantly broadened through what is now referred to as the \textit{generalized symmetries} framework \cite{Gaiotto_Generalized_2015JHEP, McGreevy_Generalized_2023ARCMP}. While conventional symmetries are invertible and neatly organized by a group-theoretical structure, it is now recognized that symmetries can also be fundamentally non-invertible, with their algebraic properties captured more naturally by fusion categories \cite{Frohlich_KWdefect_PRL2004, Frohlich_defectRCFT_2007, Shao_NonInvertible_2023arXiv, SchaferNameki_NonInvertible_2024PhR, Bhardwaj_Categorical_PRL2024}. 
For example, there has been a recent resurgence of interest in the aforementioned KW duality of the transverse-field Ising chain. The duality relates the paramagnetic and ferromagnetic phases, which differ in their ground-state degeneracy. This mismatch implies that the duality cannot be realized by an ordinary unitary mapping. Instead, it is implemented by a non-unitary operator that follows the non-invertible fusion rules of the so-called ``Ising-category'' fusion rule \cite{Frohlich_KWdefect_PRL2004, Frohlich_defectRCFT_2007, Bhardwaj_gauging_2018JHEP, Aasen2016topological, Aasen_defect_duality_2020, Hsieh_Fermionic_PRL2021, Fukusumi_defectIsing_PRB2021, Fukusumi_Fermionization_SciPost2021, Lootens_Dualities_PRX2023, Li_2023_KT, Moradi_holography_SciPost2023, Seiberg_2024_majorana, Moradi_fractionalization_SciPost2025}. In recent work, the KW transformation has been reinterpreted as a process of gauging non-anomalous global symmetries \cite{Chang_defect_2019JHEP, Bhardwaj_gauging_2018JHEP, Aasen2016topological, Aasen_defect_duality_2020, Thorngren_fusion1_2019, Thorngren_fusion2_2021}. This realization has sparked an intensive wave of research, particularly focused on developing the mathematical foundations of non-invertible symmetries and understanding their role in quantum field theories \cite{Bhardwaj_non-invertible_scipost2023, Bhardwaj_sandwich_scipost2023, Roumpedakis_Gauging_2023CMaPh, Bhardwaj_GeneralizedCharges_2023arXiv, Bhardwaj_Gapped_SciPost2025, Lu-strange_2025arXiv, Hsin_exoticHall_SciPost2024, Bhardwaj_Hasse_2024arXiv, Bartsch_representation_SciPost2024}. More recently, the focus has begun to shift toward condensed-matter settings, where such exotic symmetries emerge concretely in finely tuned lattice models at fixed points, opening up new avenues for their physical realization \cite{Lootens_Dualities_PRX2023, Li_2023_KT, Cao_sub_KW_2023, Lootens_Dualities_PRXQuantum2024, Bhardwaj_LatticeModels_2024arXiv, Pace_T-duality_SciPostPhys2025, Choi_higher-form_SciPostPhys2025, Gorantla_Tensor_2024arXiv, Seiberg_LSM_SciPostPhys2024, Seifnashri_Cluster_PRL2024, Chatterjee_spinchains_SciPost2024, Aswin_KT_sub_PRB2024, Aksoy_wb_2024, Cao_Global_2025arXiv, Pace_modulated_SciPostPhys2025, Ebisu_modulated_PRB2025, Kim_modulated_SPT_SciPost_2025, Mana_Higher-order_arXiv2025, Chung_Qudit_2025arXiv}.

The KT transformation originally introduced in Refs.~\cite{KT_1992hidden_Springer,KT_1992hidden_PRB} was defined for spin-$1$ systems with open boundaries, and its extension to a ring geometry had remained unresolved until recently. In an insightful recent work \cite{Li_2023_KT}, the authors constructed the KT transformation on a ring through a combination of gauging and stacking SPT phases, treating both spin-$1$ and spin-$1/2$ chains and showing their equivalence. Remarkably, on a ring, the KT transformation was shown to be realized by non-unitary operators satisfying a non-invertible fusion rule, whereas on an interval with appropriate boundary conditions, the transformation becomes unitary. This establishes the KT transformation as a non-invertible duality transformation. 

\paragraph*{Subsystem symmetries:}
In recent years, a new type of global symmetry known as subsystem symmetry has emerged as playing an important role in the study of exotic lattice systems. Unlike ordinary global symmetries, where the symmetry generator acts on the entire spatial manifold, a subsystem symmetry acts only within a chosen subregion. Choosing different subregions typically produces independent conserved quantities, so on a lattice, the total number of conserved charges increases only subextensively with system size \cite{Xu_Moore_prl_2004,Xu_Moore_nuclB_2005}. This idea has gained renewed attention with the discovery of fracton topological order \cite{Chamon_Glassiness_PRB2005, Haah_stabilizer_PRB2011, BRAVYI2011839, Yoshida_fractal_PRB2013, Vijay_topologicalquantum_PRB2015, Vijay_Fracton_PRB2016}, a striking new type of topological order marked by subextensive degeneracy of ground states and quasiparticle excitations whose motion is severely restricted. 
Subsystem symmetries can also protect nontrivial phases, known as subsystem symmetry-protected topological (SSPT) phases \cite{You_SSPT_2018, Devakul_SSPT_classification_2018}. 
One of the earliest and most striking examples thereof is the square-lattice cluster model \cite{Raussendorf_MBQC_2002}.
In the presence of line-like subsystem symmetries, this model hosts an SSPT phase that can serve as a universal resource for measurement-based quantum computation \cite{Else_MBQC_2012NJPh, Raussendorf_UniversalQC_PRL2019}. This discovery not only connected subsystem symmetries to quantum information theory but also inspired further work showing that similar computational universality extends to other SSPT phases \cite{Devakul_UniversalQC_PRA2018, Stephen_2019Quantum}. In developing a classification framework for SSPT phases, Refs.~\cite{Devakul_UniversalQC_PRA2018, Stephen_2019Quantum} introduced the distinction between ``weak'' and ``strong'' SSPTs. The weak SSPT refers to phases that can be viewed as stacks or arrays of independent one-dimensional SPT chains, while strong SSPTs -- exemplified by the square-lattice cluster state -- cannot be smoothly connected to the limit of decoupled 1d SPTs without closing the bulk gap.

In this work, we extend the Kennedy-Tasaki transformation originally formulated in (1+1)D, to the setting of two spatial dimensions with $\mathbb{Z}_2 \times \mathbb{Z}_2$ SSPT phase.   We focus our attention on the example of the so-called cluster state Hamiltonian on a square lattice, which realizes a strong SSPT phase with symmetry-protected gapless edge modes. Upon applying the KT transformation, this model becomes dual to two copies of the Ising plaquette model, which harbours subextensive ground state degeneracy thanks to the $\mathbb{Z}_2 \times \mathbb{Z}_2$ \textit{subsystem} spontaneous symmetry breaking (SSSB). We thus establish the duality between SSPT and SSSB phases, which, when taken together with the generalized KW transformation, connects these to the trivial phase -- as shown schematically in
 Fig.~\ref{fig:Dualities_N_KT}. 

In the process of formulating the generalized KT transformation, we obtain several interesting results, notably demonstrating that the subsystem KT duality is non-invertible when formulated on a system with closed (periodic or twisted) boundary conditions; but, by contrast, this duality can be implemented via an invertible, unitary operator when open boundary conditions are imposed. This dichotomy, common also to the KW duality, which we also generalized to the same (2+1)D models with subsystem symmetry, underscores the importance of the boundary conditions when analyzing the generalized symmetries, as foreshadowed by the early bond-algebraic formulation of dualities in Refs.~\cite{Cobanera2010,Cobanera2011} and more recently clarified in the formulation of the generalized Wigner theorem~\cite{Wigner-theorem-Ortiz2025} encompassing non-invertible symmetries. We formulate the noninvertible fusion rules for the duality defect operators for the case of both the generalized KW and KT transformations in these models. 
Last but not least, we touch upon the classification of SSPTs~\cite{Devakul_SSPT_classification_2018} and analyze how the cohomological invariants can be viewed through the lens of the generalized KT transformation.

\vspace{-2mm}
\subsection{Summary of Main Results}

In view of the length of this paper, we shall summarize our main results here, referring the reader to the corresponding sections of the main text for details. We have attempted to make the paper as self-contained as possible, introducing the concepts of the KW and KT duality transformations first for the (1+1)D case, before extending the formalism to (2+1)D models with subsystem symmetries.

\vspace{-3mm}
\subsubsection{(Self)duality transformation for (2+1)D systems with subsystem symmetries} 

We begin by formulating in Section~\ref{sec:subsym_KW} the duality transformation for (2+1)D systems with one-dimensional subsystem symmetries, such as the symmetries acting on the rows and columns of a lattice. The resulting duality maps (subsystem) spontaneous symmetry broken phase onto a trivial phase of the model. By analogy with the one-dimensional TFI chain, this duality is often referred to as the (generalized) Kramers-Wannier transformation -- a nomenclature we adopt, albeit not historically accurate (as mentioned above, Kramers and Wannier formulated their duality for a \textit{classical} two-dimensional system~\cite{Kramers_duality_1941}). We apply the KW transformation to the concrete example of the Xu -- Moore model, also known as the Ising plaquette model in a transverse field \cite{Xu_Moore_prl_2004,Xu_Moore_nuclB_2005}. At the self-dual point, the transformation could be regarded as a symmetry of the model. Crucially, the properties of this symmetry depend on the boundary conditions -- we show that in a system with open boundaries, this is a conventional symmetry implemented by a unitary operator; by contrast, the symmetry becomes non-invertible, non-unitary if closed (periodic or twisted) boundary conditions are imposed. We conclude Sec.~\ref{sec:subsym_KW} by formulating the fusion rules for the subsystem defect operators of this noninvertible symmetry.

\vspace{-3mm}
\subsubsection{Generalized Kennedy-Tasaki transformation in (2+1)D}

Building on the subsystem KW duality developed above, we introduce in Secs.~\ref{sec:KT_closed} and \ref{sec:KT_SSPT_strong} a generalized KT transformation applicable to $(2+1)$-dimensional systems with subsystem symmetries, extending the construction of Ref.~\cite{Li_2023_KT} to two spatial dimensions. 
This generalized transformation establishes a one-to-one correspondence between $\mathbb{Z}_2 \times \mathbb{Z}_2$ SSPT phases and $\mathbb{Z}_2 \times \mathbb{Z}_2$ subsystem symmetry-broken (SSSB) phases. A schematic overview of these dualities is provided in Fig.~\ref{fig:Dualities_N_KT}.
We distinguish two cases: (i) the weak SSPT phase, which is mapped by the KT transformation to the collection of decoupled Ising chains, discussed in Sec.~\ref{sec:KT_closed} and (ii) strong SSPT, exemplified by the 2D cluster Hamiltonian, which is mapped by the KT transformation onto the Ising plaquette model, shown in Sec.~\ref{sec:KT_SSPT_strong}.

In Sec.~\ref{sec:SSPT_2d}, we introduce the two-dimensional cluster Hamiltonian with closed boundary conditions realizing a strong SSPT phase, and then analyze the open-boundary case, where the global subsystem symmetry generators fractionalize into local boundary operators obeying a nontrivial anticommutation algebra.

In Appendix~\ref{app:KT_QFT}, we establish from a field-theoretic perspective that the combined operation of gauging a subsystem symmetry and stacking an SSPT phase transforms an SSSB phase into an SSPT phase. Next, we present a careful formulation of the subsystem KT transformation on the square lattice for both open boundary and periodic/antiperiodic boundary conditions.
In the latter case of closed boundary conditions, our generalized subsystem KT transformation becomes non-unitary and non-invertible -- the fact that we establish through three complementary perspectives. First, in Sec.~\ref{sec:closed_KT_example}, we implement the subsystem KT transformation on two copies of the Xu--Moore model and show explicitly that it maps a $\mathbb{Z}_2 \times \mathbb{Z}_2$ SSSB phase to the square-lattice cluster Hamiltonian realizing a strong SSPT phase, while leaving the trivial paramagnet invariant. Here, we provide a ground-state perspective: when acting within the original Hilbert space, the subsystem KT transformation is intrinsically non-invertible. However, by extending the Hilbert space to include twisted sectors, the transformation regains unitarity and becomes fully invertible, in agreement with the statement of the generalized Wigner theorem that has been recently formulated to encompass the notion of non-invertible symmetries~\cite{Wigner-theorem-Ortiz2025}. 
Second, in Sec.~\ref{sec:symtwist_KT_sub}, we explicitly analyze how symmetry-twist sectors are mapped under the transformation, which already provides clear evidence that it does not act as a unitary operator when closed (periodoc or twisted) boundary conditions are imposed. 
Finally, in Sec.~\ref{sec:fusion_KT_sub}, we examine the fusion algebra between the subsystem KT duality operator and the subsystem symmetry generators. 
This fusion structure directly reveals the non-invertibility and non-unitarity of the transformation. 

We further examine in Sec.~\ref{sec:KT_open} the subsystem KT transformation on an open square lattice and demonstrate, through explicit calculation, that in this geometry the transformation acts as a \textit{unitary}, invertible operator. To illustrate this concretely, we analyze a $\mathbb{Z}_2 \times \mathbb{Z}_2$  SSSB phase of the Ising plaquette model and show that the subsystem KT transformation maps onto the cluster Hamiltonian on the square lattice, which realizes the strong SSPT phase. Moreover, within the ground-state manifold, we find that the subsystem symmetry generators act only on localized boundary modes of the open lattice. Remarkably, these boundary degrees of freedom form operators that satisfy the Pauli algebra, highlighting the effective edge structure generated by the transformation.

\subsubsection{$\mathbb{Z}_2 \times \mathbb{Z}_2$  SSPT, its bulk and edge invariants, and KT transformation}

The subsystem KT transformation we introduced in Sec.~\ref{sec:KT_SSPT_strong} and \ref{sec:KT_open}  connects not only the bulk Hamiltonians of the SSSB and SSPT phases but also their boundary degrees of freedom, revealing the duality between symmetry breaking and symmetry protection which we discuss in detail in Sec.~\ref{sec:gapped_SSPT_open}. We explicitly compute the edge symmetry operators of the strong SSPT phase and show that they realize a projective representation of the subsystem symmetry group. Under the KT transformation, these localized edge operators are mapped to decorated, extended subsystem strings traversing entire rows or columns in the dual SSSB phase, thereby making the hidden symmetry strings of the SSPT explicit. 

Sec.~\ref{sec:invariants} then formulates the bulk and edge invariants that diagnose strong SSPT phases \cite{You_SSPT_2018, Devakul_SSPT_classification_2018} and remain invariant under any \textit{linearly symmetric local unitary} (LSLU) evolution. The bulk invariant measures the phase accumulated when a half-space symmetry operator is commuted past a local corner ``repair'' operator introduced by truncating the subsystem symmetry to a finite region \cite{Devakul_SSPT_classification_2018}. The boundary analogue of the bulk invariant characterizes the projective algebra of the edge symmetry generators. The two quantities are connected by the bulk-edge correspondence: the bulk phase equals the commutation phase between short edge segments immediately to the left and to the right of a vertical cut. 

For the strong $\mathbb{Z}_2 \times \mathbb{Z}_2$ SSPT (2D cluster state) model, this correspondence yields a nontrivial value of the bulk invariant. In contrast, for a weak SSPT constructed from decoupled one-dimensional cluster states, all the bulk invariant values are trivial, as expected. On the boundary, the same information is encoded in a ``link diagram'' representation introduced in Ref.~\cite{Devakul_SSPT_classification_2018}, where nontrivial commutation phases arise only between neighboring boundary generators of different types. The parity of such links intersected by any cut reproduces the corresponding bulk invariant.

In Sec.~\ref{sec:bulk-edge} we show that although the subsystem KT transformation maps the entire SSPT phase to its dual SSSB phase, the quantities obtained by applying this transformation to the bulk diagnostic operators do not constitute intrinsic invariants of the SSPT phase itself. In fact, the subsystem KT map sends the strictly local repair operators of the SSPT description to highly extended, nonlocal objects in the dual picture. Nonetheless, we show that the essential commutation structure that distinguishes the strong phase is preserved under the transformation. In this way, the subsystem KT map faithfully transmits the algebraic content associated with subsystem-protected order, even though the transformed bulk/edge invariants no longer admit a simple local interpretation.

\subsection{Relation to Prior Work}

Prior work on subsystem dualities in the Ising plaquette model -- most notably Ref.~\cite{Cao_sub_KW_2023} -- has focused almost exclusively on \textit{closed} square lattices with periodic or antiperiodic boundary conditions. Ref.~\cite{Okuda_Anomaly_2024} developed a framework based on foliated cluster states to construct statistical models associated with general Calderbank-Shor-Steane (CSS) codes, and derived generalized Kramers-Wannier-Wegner-type \cite{Wegner_Duality_1971} dualities in the presence of twist-defect insertions. In particular, by taking the 2D quantum plaquette Ising model as the underlying CSS code, they related it to the  3D classical anisotropic plaquette Ising model and derived generalized dualities in which the dual theory involves a sum over twisted sectors. They also studied measurement-assisted implementations of Kramers–Wannier duality operators for general CSS codes and showed that the resulting fusion rules are non-invertible, extending earlier results for the 2D quantum plaquette Ising model discussed in Ref.~\cite{Cao_sub_KW_2023}. The structure of the generalized (Kramers--Wannier-like) subsystem duality on an open square lattice, however, has not been addressed, despite its importance for formulating boundary-sensitive subsystem dualities such as the generalized  Kennedy--Tasaki transformation.

In this work, we fill this gap by deriving the subsystem KW duality operator explicitly on the open square lattice and demonstrating, through direct calculation, that it defines a unitary and fully invertible mapping in this geometry. To place this in context, we revisit the Xu-Moore model, where the subsystem (self-)duality is realized. By analyzing its ground-state structure, we show that the duality transformation becomes intrinsically non-invertible when restricted to the original Hilbert space on closed manifolds, and that unitarity can be restored only after enlarging the Hilbert space to include twisted sectors, in agreement with recent observations \cite{Li_2023_KT}. This ground-state-based perspective serves as a conceptual and technical precursor to the analysis of the generalized subsystem KT transformation, where analogous issues of non-unitarity and non-invertibility again arise on closed manifolds.

Earlier works related to subsystem Kennedy-Tasaki transformations arose in the study of measurement-based quantum computation. In particular, Doherty and Bartlett \cite{Doherty_PRL_2009} mapped the 1D cluster Hamiltonian in a transverse field on an open chain to decoupled transverse-field Ising chains, and also related the 2D cluster Hamiltonian (realizing 2D strong $\mathbb{Z}_2 \times \mathbb{Z}_2$ SSPT phase) to copies of the plaquette Ising model (realizing 2D SSSB phase). Subsequently, You \textit{et al.} \cite{You_SSPT_2018} generalized this construction to three dimensions by mapping the 3D cluster Hamiltonian (realizing 3D strong $\mathbb{Z}_2 \times \mathbb{Z}_2$ SSPT) to two copies of the cubic Ising model. Although these works did not explicitly identify these mappings as subsystem Kennedy-Tasaki transformations, they already contained the essential structure of such dualities. More recently, a subsystem KT transformation for the strong SSPT phase under \emph{periodic} boundary conditions was explicitly formulated in Ref.~\cite{Aswin_KT_sub_PRB2024}, where its relation to SSPT and SSSB phases was emphasized.
The corresponding subsystem KW duality operator was realized there through a sequential circuit construction combined with a projection onto symmetry subspaces. Importantly, the KW transformation in that work acts within the same lattice and Hilbert space, mapping the theory to another formulation defined on an identical set of spins.

In our formulation, we have taken a different route, conceptually speaking, the one that emphasizes precise treatment of symmetry actions, twist sectors/boundary conditions -- the essential elements for defining the subsystem KT transformation as a genuine duality. We construct the explicit form of the subsystem KT operator for both the weak and strong SSPT cases, under both open and closed (periodic or antiperiodic) geometries. Closed geometries play a crucial role here, as twisting the boundary conditions serves as a sharp diagnostic of the underlying SPT order. The subsystem KW formalism used in our construction draws inspiration from  Ref.~\cite{Cao_sub_KW_2023}, which maps a model defined on the original lattice to a dual theory defined on the \textit{dual} lattice, rather than acting within the same lattice as in Ref.~\cite{Aswin_KT_sub_PRB2024} (see Sec.~\ref{sec:KW_symmetry} for more details).

Crucially, we demonstrate that while the generalized KT transformation is unitary and invertible on open lattices, it becomes intrinsically non-unitary and non-invertible on closed manifolds. The resulting fusion algebra differs from that of Ref.~\cite{Aswin_KT_sub_PRB2024} by additional twist-dependent terms that explicitly encode the symmetry-twist interplay central to the non-invertible character of the transformation. This unified framework establishes a complete correspondence between $\mathbb{Z}_2\times\mathbb{Z}_2$ SSSB and SSPT phases under the subsystem KT transformation, encompassing all boundary conditions.

Looking beyond the exclusive focus of Ref.~\cite{Aswin_KT_sub_PRB2024} on the strong SSPT phase, our work provides a unified framework that connects both weak and strong SSPTs through their common bulk-edge characterization. The quantities obtained by applying this map to the bulk diagnostic operators of SSPT  no longer have the meaning of the corresponding invariants in the SSSB phase. Indeed, the subsystem KT transformation sends the strictly local repair operators that define the SSPT bulk invariant to highly extended, nonlocal objects in the dual SSSB picture.
Nevertheless, we show that the \textit{essential commutation algebra} distinguishing the strong SSPT phase is preserved exactly under the duality. In this sense, the subsystem KT map faithfully transmits the algebraic content of subsystem-protected order, even though the transformed bulk and edge invariants no longer admit a simple local interpretation.

\section{Subsystem Kramers-Wannier duality transformation in $(2 + 1)$D spin systems}
\label{sec:subsym_KW}

In this section, we begin with a brief review of the subsystem $\mathbb{Z}_2$ symmetry in $(2+1)$D on a square lattice, together with the associated subsystem KW duality transformation \cite{Cao_sub_KW_2023}, as a prelude to the subsystem KT transformation to be discussed in Secs.~\ref{sec:KT_closed} and \ref{sec:KT_open}. We will later reformulate the subsystem KT transformation in terms of the KW transformation.

\subsection{Subsystem $\mathbb{Z}_2$ symmetry on a square lattice}
\label{sec:subsym_square}
\begin{figure}[tb]
    \centering
    \includegraphics[width=0.75\linewidth]{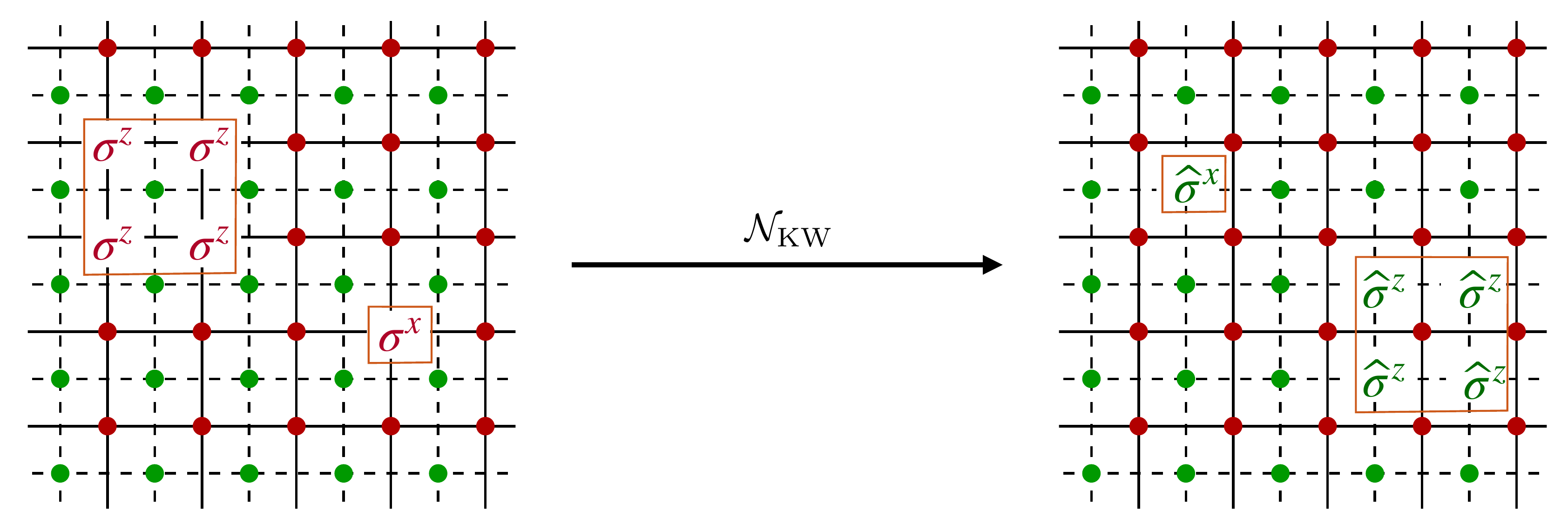}
    \caption{Depiction of the Xu--Moore model on the square lattice, with the red sites hosting the original Ising spins $\sigma_{i,j}$, while the green sites correspond to the dual degrees of freedom $\hssi_{i-\half,j-\half}$ at the centers of the plaquettes. The mapping of the plaquette operator and the transverse field term through the KW transformation is illustrated, following 
    Eq.~\eqref{eq:XM-dual-variables}.}
    \label{fig:KW_duality}
\end{figure}

\paragraph{Original lattice:}
We consider a closed $L_x \times L_y$ square lattice (i.e. a torus), with sites labeled by coordinates $(i,j)$, where $i = 1, \dots, L_x$ and $j = 1, \dots, L_y$, subject to the periodic identifications $i \sim i + L_x$ and $j \sim j + L_y$. Each site hosts a spin-$1/2$ degree of freedom, with a two-dimensional local Hilbert space spanned by the basis states $\ket{\ssi_{i,j}}$, where $\sigma_{i,j} = 0, 1$. We indicate the vertices of the original lattice in red, as shown in Fig.~\ref{fig:KW_duality}. The Pauli operators act on the state in the standard way,
\begin{align}
\label{eq:sigma_pauli}
    \sigma^z_{i,j} \ket{\ssi_{i,j}} = (-1)^{\ssi_{i,j}} \ket{\ssi_{i,j}},  \hspace{1cm} \sigma^x_{i,j} \ket{\ssi_{i,j}} = \ket{1-\ssi_{i,j}}.
\end{align}

As a concrete example of a model with subsystem symmetries, we consider the Xu-Moore model, also referred to as the transverse-field plaquette Ising (TFPI) model
\cite{Xu_Moore_prl_2004,Xu_Moore_nuclB_2005}, defined as follows:
\begin{align}
\label{eq:XM_closed}
    H^h_{\text{XM}} = - \sum_{i=1}^{L_x} \sum_{j=1}^{L_y} \left( \sigma^z_{i-1,j-1} \sigma^z_{i,j-1} \sigma^z_{i-1,j} \sigma^z_{i,j} + h \sigma^x_{i,j}\right),
\end{align}
where the first term encodes the interaction on plaquettes of the square lattice and the second term the effect of the transverse field. This model was first proposed in Refs.~\cite{Xu_Moore_prl_2004,Xu_Moore_nuclB_2005} as an effective description of ordering in arrays of Josephson-coupled $p \pm i p$ superconducting grains. The Xu-Moore model admits an onsite (non-anomalous) global \textit{subsystem} $\mathbb{Z}_2$ \textit{symmetry}, whose generators are the $\mathbb{Z}_2$ operators acting along every row and column of the lattice,
\begin{align}
\label{eq:subsym_operator}
    U^x_j = \prod_{i = 1}^{L_x} \sigma^x_{i,j}, \hspace{1cm} U^y_i = \prod_{j = 1}^{L_y} \sigma^x_{i,j}.
\end{align}

We denote the eigenvalues of $U^x_j, U^y_i$ as $(-1)^{u^x_j}, (-1)^{u^y_i}$, respectively, where $u^x_j, u^y_i = 0,1$. The $L_x + L_y$ symmetry operators are subject to a single constraint and are therefore not linearly independent:
\begin{align}
\label{eq:subsym_operator_contraint}
    \prod_{i = 1}^{L_x} U^y_i \prod_{j = 1}^{L_y} U^x_j = 1.
\end{align}
As a result, the number of independent symmetry operators is reduced to $L_x + L_y - 1$, and the Hilbert space (for a fixed boundary condition) is correspondingly decomposed into $2^{L_x + L_y - 1}$ symmetry sectors.

In addition, one can introduce subsystem $\mathbb{Z}_2$ defects along the time direction, resulting in twisted boundary conditions for the spins as in Ref.~\cite{Cao_sub_duality_2022}
\begin{align}
\label{eq:twists1}
    \ket{\ssi_{i+L_x,j}} = \ket{\ssi_{i,j} + t^x_j}, \quad \ket{\ssi_{i,j+L_y}} = \ket{\ssi_{i,j} + t^y_i}, \quad \ket{\ssi_{i+L_x,j+L_y}} = \ket{\ssi_{i,j}+t^{xy} + t^x_j+ t^y_i},
\end{align}
where $t^x_j, t^y_i, t^{xy} \in \{0,1\}$ specify the boundary conditions along the $j$-th row, the $i$-th column, and at the corner, respectively, for all $i,j$. $t = +1$ corresponding to periodic boundary conditions, and $t = -1$ corresponding to anti-periodic (twisted) boundary conditions. The additional twist parameter $t^{xy}$ allows us to specify the boundary conditions for the twist variables $t^x_j, t^y_i$ as 
\begin{align}
\label{eq:twists2}
    t^x_{j+L_y} = t^x_j + t^{xy}, \quad t^y_{i+L_x} = t^y_i + t^{xy}.
\end{align}
Although there are $L_x + L_y + 1$ twist parameters, a more careful analysis reveals that, for a Hamiltonian with subsystem $\mathbb{Z}_2$ symmetry, only certain combinations of these parameters are physically relevant \cite{Cao_sub_duality_2022},
\begin{align}
\label{eq:twists3}
    \mt^x_{j-\frac{1}{2}} \equiv t^x_{j-1} + t^x_j, \quad  \mt^y_{i-\frac{1}{2}} \equiv t^y_{i-1} + t^y_i.
\end{align}
Imposing the constraint
\begin{align}
    \sum_{j=1}^{L_y} \mt^x_{j-\frac{1}{2}} =  \sum_{i=1}^{L_x} \mt^y_{i-\frac{1}{2}} = t^{xy},
\end{align}
reduces the number of independent twist variables $(\mt^x_{j-\frac{1}{2}}, \mt^y_{i-\frac{1}{2}})$ to $L_x + L_y - 1$. Consequently, each symmetry sector is further decomposed into $2^{L_x + L_y - 1}$ twist sectors.

In summary, as pointed out in Ref.~\cite{Cao_sub_duality_2022}, the Hilbert space admits a natural decomposition into $4^{L_x + L_y - 1}$ distinct symmetry-twist sectors. Each sector is uniquely specified by the set of labels $(u^x_j, u^y_i, \mt^x_{j-\frac{1}{2}}, \mt^y_{i-\frac{1}{2}})$, which encode the subsystem $\mathbb{Z}_2$ symmetry charges and the corresponding twist variables. The symmetry-twist labels are subject to global constraints
\begin{align}
\label{eq:symtwist_constraint}
    \prod_{i = 1}^{L_x} (-1)^{u^y_i} \prod_{j = 1}^{L_y} (-1)^{u^x_j} = 1, \quad \prod_{i = 1}^{L_x} (-1)^{\mt^x_{j-\frac{1}{2}}} \prod_{j = 1}^{L_y} (-1)^{\mt^y_{i-\frac{1}{2}}} = 1.
\end{align}

\paragraph{Dual lattice:}
We also define a dual lattice, with dual spins $\hssi_{i-\half,\, j-\half} \in \{0,1\}$ located at the centers of the plaquettes of the original lattice, depicted 
in green in Fig.~\ref{fig:KW_duality}. The states $\ket{\hssi_{i-\half, j-\half}}$ in the dual local Hilbert space are acted upon by the Pauli operators $\hsig^z_{i-\half,\, j-\half}$ and $\hsig^x_{i-\half,\, j-\half}$, in the same manner as in Eq.~\eqref{eq:sigma_pauli},
\begin{align}
\label{eq:sigma_pauli_dual}
    \hsig^z_{i-\half,\, j-\half} \ket{\hssi_{i-\half, j-\half}} = (-1)^{\hssi_{i-\half, j-\half}}  \ket{\hssi_{i-\half, j-\half}}, \quad \hsig^x_{i-\half,\, j-\half} \ket{\hssi_{i-\half, j-\half}} =  \ket{1 - \hssi_{i-\half, j-\half}}.
\end{align}

The Xu-Moore model \eqref{eq:XM_closed} exhibits an exact self-duality between the large-$h$ and small-$h$ phases, which can be seen as follows. On an infinite square lattice, we can define the dual spins in terms of the original spins by 

\begin{align}
   \sigma^x_{i,j} & = \hsig^z_{i-\half,j-\half} \hsig^z_{i+\half,j-\half} \hsig^z_{i-\half,j+\half}\hsig^z_{i+\half,j+\half}, \nonumber\\
   \sigma^z_{i,j} & = \prod_{\iprime \leq i, \jprime \leq j} \hsig^x_{\iprime-\half,\jprime-\half}. 
   \label{eq:XM-dual-variables}
\end{align}

With the above, we can now demonstrate that, on an infinite square lattice, the Xu-Moore Hamiltonian \eqref{eq:XM_closed} admits the following representation in terms of the dual spins:
\begin{align}
\label{eq:XM_dual_infinite}
    \widehat{H}^h_{\text{XM}} = - \sum_{i} \sum_{j} \left( \hsig^x_{i-\half,j-\half} + h \hsig^z_{i-\half,j-\half} \hsig^z_{i+\half,j-\half} \hsig^z_{i-\half,j+\half} \hsig^z_{i+\half,j+\half}\right),
\end{align}
which has the same form as the original Hamiltonian, except that the roles of the term multiplied by $h$ are interchanged. This is an example of the KW duality in two dimensions.

The dual Hamiltonian \eqref{eq:XM_dual_infinite} respects an on-site global $\mathbb{Z}_2$ subsystem  symmetry, whose generators on a closed square lattice take the form
\begin{align}
\label{eq:subsym_operator_dual}
    \widehat{U}^x_{j-\half} = \prod_{i=1}^{L_x} \hsig^x_{i-\half,\, j-\half}, \quad  \widehat{U}^y_{i-\half} = \prod_{j=1}^{L_y} \hsig^x_{i-\half,\, j-\half}.
\end{align}
Mirroring the discussion preceding Eq.~\eqref{eq:symtwist_constraint}, the dual Hilbert space can also be decomposed into $4^{L_x+L_y-1}$ symmetry-twist sectors labeled by $(\widehat{u}^x_{j-\half}, \widehat{u}^y_{i-\half}, \hmt^x_j, \hmt^y_i)$ with the global constraints
\begin{align}
\label{eq:symtwist_constraint_dual}
     \prod_{i = 1}^{L_x} (-1)^{\widehat{u}^y_{i-\half}} \prod_{j = 1}^{L_y} (-1)^{\widehat{u}^x_{j-\half}} = 1, \quad \prod_{i = 1}^{L_x} (-1)^{\hmt^y_i} \prod_{j = 1}^{L_y} (-1)^{\hmt^x_j} = 1.
\end{align}
Here $(-1)^{\widehat{u}^x_{j-\half}}$ and $ (-1)^{\widehat{u}^y_{i-\half}}$ denote the eigenvalues of $ \widehat{U}^x_{j-\half}, \widehat{U}^y_{i-\half}$, respectively, while $\hmt^x_j, \hmt^y_i$ represent the corresponding twist variables.

The spin variables of the original lattice and the dual spins on the dual lattice should not be viewed as coexisting, independent degrees of freedom. Rather, they are connected through the KW transformation, so that the configuration in one representation completely determines the configuration in the other.

\subsection{Subsystem KW  transformation on a closed square lattice}
\label{sec:KW_operator_closed}

We may regard the subsystem KW transformation as a duality that exchanges the plaquette interaction with the transverse-field term, thereby establishing the mapping between the ordered (SSSB) and trivial phase, as illustrated schematically in Fig.~\ref{fig:Dualities_N_KT}. In recent years, several works \cite{Cao_sub_KW_2023, Aswin_KT_sub_PRB2024} have constructed explicit forms of the duality operator $\NKW$, often referred to as the subsystem KW duality operator, which implements this mapping. More precisely, our goal is to construct an operator $\NKW$ satisfying the following property: for any state $\ket{\psi} \in \mathcal{H}$,
\begin{align}
\label{eq:Pauli_map_KW}
    & \NKW \sigma^z_{i-1,j-1} \sigma^z_{i,j-1} \sigma^z_{i-1,j} \sigma^z_{i,j} \ket{\psi} = \hsig^x_{i-\half,j-\half} \NKW \ket{\psi}, \quad \forall \ket{\psi} \in \mathcal{H}, \nonumber\\ & \NKW \sigma^x_{i,j} \ket{\psi} = \hsig^z_{i-\half,j-\half} \hsig^z_{i+\half,j-\half} \hsig^z_{i-\half,j+\half} \hsig^z_{i+\half,j+\half} \NKW \ket{\psi}, \quad \forall \ket{\psi} \in \mathcal{H}.
\end{align}
We now briefly review the subsystem KW transformation, following Ref.~\cite{Cao_sub_KW_2023}, where the operator $\NKW$ is defined explicitly on a closed square lattice. KW duality operators have been explored both in the context of mapping between two theories \cite{Aasen2016topological,Tantivasadakarn2023hierarchy,Li_2023_KT,Cao_sub_KW_2023,Tantivasadakarn2024SPT} and as self-dual maps within a single theory \cite{Seiberg_2024_majorana}. Our discussion adopts the specific construction of Ref.~\cite{Cao_sub_KW_2023}, in which the subsystem KW duality operator $\NKW$ maps the theory on the original lattice, with spins $\ssi_{i,j}$, to a dual theory defined on the dual lattice, with spins $\hssi_{i-\half,j-\half}$.

The subsystem KW duality operator is then specified by its action on a basis state of the original Hilbert space $\mathcal{H}$, given explicitly as \cite{Cao_sub_KW_2023} 
\begin{align}
\label{eq:N_sub}
    \NKW \ket{\{\ssi_{i,j}\}} & = \frac{1}{2^{(L_x+L_y)/2}} \sum_{\{\hssi_{i-\half,j-\half}\}} (-1)^{\mathcal{C}_{\text{bulk}}^{\sigma\hsig} + \mathcal{C}_{\text{bdy}}^{\sigma\hsig}} \ket{\{\hssi_{i-\half,j-\half}\}} 
\end{align}
where the sum runs over all dual-spin configurations of $\hsig^z$, and the phase factors $\mathcal{C}_{\text{bulk}}^{\sigma\hsig}$ and $\mathcal{C}_{\text{bdy}}^{\sigma\hsig}$ are defined by
\begin{align}
\label{eq:C_bulk_bdy}
    \mathcal{C}_{\text{bulk}}^{\sigma\hsig} & = \sum_{i=1}^{L_x} \sum_{j=1}^{L_y} \left(\ssi_{i-1,j-1} + \ssi_{i,j-1} + \ssi_{i-1,j} + \ssi_{i,j}\right) \hssi_{i-\half,j-\half}, \nonumber\\
    \mathcal{C}_{\text{bdy}}^{\sigma\hsig} & = \sum_{j=1}^{L_y} \htt^x_{j-\half} \left(\ssi_{L_x,j} + \ssi_{L_x,j-1}\right) + \sum_{i=1}^{L_x} \htt^y_{i-\half} \left(\ssi_{i,L_y} + \ssi_{i-1,L_y}\right) + \htt^{xy} \ssi_{L_x,L_y},
\end{align}
Alternatively, the action of $\NKW$ can be written in a slightly different form
\beq
\label{eq:N_sub2}
     \NKW \ket{\{\ssi_{i,j}\}} = \frac{1}{2^{(L_x+L_y)/2}} \sum_{\{\hssi_{i-\half,j-\half}\}} (-1)^{\widetilde{\mathcal{C}}^{\sigma\hsig}_{\text{bulk}} + \widetilde{\mathcal{C}}_{\text{bdy}}^{\sigma\hsig}} \ket{\{\hssi_{i-\half,j-\half}\}},
\eeq

with the phase factors $\widetilde{\mathcal{C}}_{\text{bulk}}^{\sigma\hsig}$ and $\widetilde{\mathcal{C}}_{\text{bdy}}^{\sigma\hsig}$  given by
\begin{align}
\label{eq:C_tilde_bulk_bdy}
    \widetilde{\mathcal{C}}_{\text{bulk}}^{\sigma\hsig} & = \sum_{i=1}^{L_x} \sum_{j=1}^{L_y} \ssi_{i,j} \left(\hssi_{i-\half,j-\half} + \hssi_{i+\half,j-\half} + \hssi_{i-\half,j+\half} + \hssi_{i+\half,j+\half}\right), \nonumber\\
    \widetilde{\mathcal{C}}_{\text{bdy}}^{\sigma\hsig} & = \sum_{j=1}^{L_y}  t^x_j \left(\hssi_{\half,j-\half} + \hssi_{\half,j+\half}\right) + \sum_{i=1}^{L_x} t^y_i \left(\hssi_{i-\half,\half} + \hssi_{i+\half,\half}\right) + t^{xy} \hssi_{\half,\half}.
\end{align}
It is straightforward to check that $\mathcal{C}_{\text{bulk}}^{\sigma\hsig} + \mathcal{C}_{\text{bdy}}^{\sigma\hsig} = \widetilde{\mathcal{C}}_{\text{bulk}}^{\sigma\hsig} + \widetilde{\mathcal{C}}_{\text{bdy}}^{\sigma\hsig}$, since one can obtain either form by appropriately relabelling the summation indices. This establishes the equivalence of the expressions in Eqs.~\eqref{eq:N_sub} and \eqref{eq:N_sub2}. We present both expressions here, as they will be useful for later applications in Sec.~\ref{sec:KT_sub}. The boundary terms in the exponents are fixed to faithfully preserve the structure of the symmetry-twist sectors under the mapping.

We next analyze the action of the subsystem KW duality operator $\NKW$ on Pauli operators, and convince ourselves that it indeed satisfies the Eq.~\eqref{eq:Pauli_map_KW}. By the definition of \eqref{eq:sigma_pauli_dual}, the operator $\hsig^x_{\iprime-\half,\jprime-\half}$ acts on any state in the dual Hilbert space as
\begin{align}
   \hsig^x_{\iprime-\half,\jprime-\half} \ket{\hssi_{i-\half,j-\half}} = \begin{cases} \ket{1-\hssi_{\iprime-\half,\jprime-\half}}, & i=\iprime, j= \jprime \\ \ket{\hssi_{i-\half,j-\half}}, & \text{otherwise}\end{cases}.
\end{align}
For convenience, we introduce the shorthand notation for the above statement as follows:
\begin{align}
    \hsig^x_{\iprime-\half,\jprime-\half} \ket{\{\hssi_{i-\half,j-\half}\}} = \ket{\{\hssi_{i-\half,j-\half}\}; \{1 - \hssi_{\iprime-\half,\jprime-\half}\}}.
\end{align}
To prove the Pauli operator mapping in \eqref{eq:Pauli_map_KW}, we start by noting the expression of $\NKW$ in Eq.~\eqref{eq:N_sub},
\begin{align}
    \hsig^x_{\iprime-\half,\jprime-\half} \NKW \ket{\{\ssi_{i,j}\}} = \frac{1}{2^{(L_x+L_y)/2}} \sum_{\{\hssi_{i-\half,j-\half}\}} (-1)^{\mathcal{C}_{\text{bulk}}^{\sigma\hsig} + \mathcal{C}_{\text{bdy}}^{\sigma\hsig}} \ket{\{\hssi_{i-\half,j-\half}\}; \{1 - \hssi_{\iprime-\half,\jprime-\half}\}}.
\end{align}
After performing the shift $\hssi_{\iprime-\half,\jprime-\half} \to 1- \hssi_{\iprime-\half,\jprime-\half}$, an additional phase factor arises from $\mathcal{C}_{\text{bulk}}^{\sigma\hsig} + \mathcal{C}_{\text{bdy}}^{\sigma\hsig}$ in Eq.~\eqref{eq:C_bulk_bdy}, and we ultimately obtain
\begin{align}
    \hsig^x_{\iprime-\half,\jprime-\half} & \NKW \ket{\{\ssi_{i,j}\}} \nonumber\\
    & = \frac{1}{2^{(L_x+L_y)/2}} \sum_{\{\hssi_{i-\half,j-\half}\}} (-1)^{\mathcal{C}_{\text{bulk}}^{\sigma\hsig} + \mathcal{C}_{\text{bdy}}^{\sigma\hsig}} (-1)^{\ssi_{\iprime-1,\jprime-1} + \ssi_{\iprime,\jprime-1} + \ssi_{\iprime-1,\jprime} + \ssi_{\iprime,\jprime}} \ket{\{\hssi_{i-\half,j-\half}\}} \nonumber\\
    & = \NKW \sigma^z_{\iprime-1,\jprime-1} \sigma^z_{\iprime,\jprime-1} \sigma^z_{\iprime-1,\jprime} \sigma^z_{\iprime,\jprime} \ket{\{\ssi_{i,j}\}},
    \label{eq:KW-action-on-operators}
\end{align}

which is exactly the first mapping in \eqref{eq:Pauli_map_KW}. The second mapping proceeds by an analogous argument.

The operator $\NKW^\dagger$ can likewise be defined, specifying its action on the basis states of the dual Hilbert space $\widehat{\mathcal{H}}$,
\begin{align}
\label{eq:N_dagger_sub}
    \NKW^\dagger \ket{\{\hssi_{i-\half,j-\half}\}} & = \frac{1}{2^{(L_x+L_y)/2}} \sum_{\{\ssi_{i,j}\}} (-1)^{\mathcal{C}_{\text{bulk}}^{\sigma\hsig} + \mathcal{C}_{\text{bdy}}^{\sigma\hsig}} \ket{\{\ssi_{i,j}\}} \nonumber\\
     & = \frac{1}{2^{(L_x+L_y)/2}} \sum_{\{\ssi_{i,j}\}} (-1)^{\widetilde{\mathcal{C}}_{\text{bulk}}^{\sigma\hsig} + \widetilde{\mathcal{C}}_{\text{bdy}}^{\sigma\hsig}} \ket{\{\ssi_{i,j}\}},
\end{align}
where the phases $\mathcal{C}_{\text{bulk}}^{\sigma\hsig}$, $\mathcal{C}_{\text{bdy}}^{\sigma\hsig}$ are defined in Eq.~\eqref{eq:C_bulk_bdy}, while $\widetilde{\mathcal{C}}_{\text{bulk}}^{\sigma\hsig}$, $\widetilde{\mathcal{C}}_{\text{bdy}}^{\sigma\hsig}$ are displayed in Eq.~\eqref{eq:C_tilde_bulk_bdy} above.

It is straightforward to verify that the duality operator $\NKW$ is Hermitian. Furthermore, the operator $\NKW$ defined in this manner does not depend on the specific form of the underlying Hamiltonian. Instead, the condition $[H, \NKW] \ket{\{\ssi_{i,j}\}} = 0$ can be employed to identify the class of Hamiltonians that are self-dual under the subsystem KW transformation. 

Specializing to the case of the Xu-Moore model in Eq.~\eqref{eq:XM_closed}, we note that the subsystem KW duality operator connects the pre- and post-transformation symmetry-twist sectors, as was shown in Ref.~\cite{Cao_sub_KW_2023} via
\begin{align}
\label{eq:mapping_symtwist_KW}
\widehat{u}^x_{j-\half} \stackrel{\text{\smaller KW}}{\longrightarrow} \mt^x_{j-\frac{1}{2}}, \quad \widehat{u}^y_{i-\half} \stackrel{\text{\smaller KW}}{\longrightarrow} \mt^y_{i-\frac{1}{2}}, \quad \hmt^x_j \stackrel{\text{\smaller KW}}{\longrightarrow} u^x_j, \quad \hmt^y_i \stackrel{\text{\smaller KW}} {\longrightarrow} u^y_i.
\end{align}

Now that we have established the mapping of the Pauli operators as per Eq.~\eqref{eq:KW-action-on-operators}, we turn to the fusion rules of the subsystem KW duality operator $\NKW$. Here,  we use the word ``fusion'' in a loose sense -- not as a fusion of topological defects, but rather as a product of operators acting on the Hilbert space. Following Ref.~\cite{Cao_sub_KW_2023}, the resulting expression is
\begin{align}
\label{eq:N_KW_fusion}
    \NKW^\dagger \times \NKW = \frac{1}{2} \prod_{i=1}^{L_x} \left( 1 + (-1)^{\hmt^y_i} U^y_i \right) \prod_{j=1}^{L_y} \left( 1 + (-1)^{\hmt^x_j} U^x_j \right) \equiv \hat{P}(\hmt^x,\hmt^y), \end{align}
where the right-hand side involves a projection onto a definite twist sector (labeled by ${\hmt^y_i}$ and ${\hmt^x_j} $) of the model, which we denote by the projector $\hat{P}$, making clear the non-unitary nature of the transformation.

\subsection{Does the action of subsystem KW duality constitute a symmetry?}
\label{sec:KW_symmetry}
    
Here, we highlight the distinction between the KW duality operator and the KW duality symmetry. The operator form, denoted by $\NKW$ in Eq.~\eqref{eq:N_sub}, acts as a nonlocal map between two distinct Hilbert spaces--one defined on sites and the other on plaquettes. To reformulate this duality as a genuine symmetry acting within a single Hilbert space, one can relabel the plaquette degrees of freedom as site variables, following the prescription of Ref.~\cite{Cao_sub_KW_2023},
\begin{align}
\label{eq:half-translation}
    \hssi_{i+\half,j+\half} \rightarrow \ssi'_{i,j}.
\end{align}
With this identification, the transformed operator --  denoted $\NKWbar$ -- acts internally within a single Hilbert space ${\cal H}$ and can therefore be regarded as a candidate for the duality symmetry of the lattice model. Its action on Pauli operators is given by \cite{Cao_sub_KW_2023}
\begin{align}
\label{eq:Pauli_map_KW_bar}
    & \NKWbar \sigma^z_{i,j} \sigma^z_{i+1,j} \sigma^z_{i,j+1} \sigma^z_{i+1,j+1} \ket{\psi} = \sigma^x_{i,j} \NKWbar \ket{\psi}, \quad \forall \ket{\psi} \in \mathcal{H}, \nonumber\\ & \NKWbar \sigma^x_{i,j} \ket{\psi} = \sigma^z_{i-1,j-1} \sigma^z_{i,j-1} \sigma^z_{i-1,j} \sigma^z_{i,j}  \NKWbar \ket{\psi}, \quad \forall \ket{\psi} \in \mathcal{H}.
\end{align}
Through these relations, one sees that $\NKWbar$ leaves the  Xu-Moore Hamiltonian at the self-dual point $h=1$ invariant, and the operator $\NKWbar$ commutes with the Hamiltonian. 
A key advantage of this single Hilbert space formulation is that one can now meaningfully define the self-fusion of this operator, which was shown in Ref.~\cite{Cao_sub_KW_2023} to take the form
\begin{align}
\label{eq:N_KW_bar_fusion}
    \NKWbar^\dagger \times \NKWbar = \frac{1}{2} \prod_{i=1}^{L_x} \left( 1 + (-1)^{\mt^y_i} U^y_i \right) \prod_{j=1}^{L_y} \left( 1 + (-1)^{\mt^x_j} U^x_j \right) T_{xy} \equiv \hat{P}(\mt^x,\mt^y) T_{xy},
\end{align}
which closely mirrors the composition $\NKW^{\dagger} \times \NKW$ in Eq.~\eqref{eq:N_KW_fusion}, but includes an additional factor $T_{xy}$ implementing a translation along the lattice diagonal, which can be traced to two half-translations in Eq.~\eqref{eq:half-translation}:
\begin{align}
\label{eq:T_xy}
   T_{xy} \ket{\{\ssi_{i,j}\}} = \ket{\{\ssi'_{i,j}=\ssi_{i+1,j+1}\}}.
\end{align}
This reformulation aligns with recent theoretical developments in non-invertible KW symmetries~\cite{Seiberg_2024_majorana, Seiberg_LSM_SciPostPhys2024, Aswin_KT_sub_PRB2024}, clarifying how the translation operator appears on the right-hand side of the composition $\NKWbar^\dagger \times \NKWbar$. 

In light of the above, can we declare $\NKWbar$ the symmetry of the Xu-Moore model? At first glance, the answer would appear to be affirmative, as it commutes with the Hamiltonian in Eq.~\eqref{eq:XM_closed}, but there is an important subtlety that is often overlooked -- in order for an operator to be a faithful symmetry, it must, in addition, also preserve the transition probabilities between all the states in the Hilbert space. This requirement appears in the statement of Wigner's theorem~\cite{Wigner_book}, ensuring that the symmetry action cannot be detected by any measurement. Consider now two arbitrary states $|\alpha \rangle$, $|\beta \rangle$ in the Hilbert space ${\cal H}$, which are transformed into $|\tilde{\alpha}\rangle = \NKWbar |\alpha\rangle$ and $|\tilde{\beta}\rangle = \NKWbar |\beta\rangle$, respectively.
By Wigner's argument, we require that the transition probability remain unchanged: $|\langle \tilde{\beta}| \tilde{\alpha}\rangle|^2 = |\langle \beta|\alpha \rangle |^2$. Consider the left-hand side:
\beq
|\langle \tilde{\beta}| \tilde{\alpha}\rangle|^2 = |\langle \beta | \NKWbar^\dagger \times \NKWbar | \alpha \rangle|^2 = |\langle \beta | \hat{P}(\mt^x,\mt^y) T_{xy} | \alpha \rangle|^2 \neq |\langle \beta|\alpha \rangle |^2
\eeq
for general $|\alpha\rangle$ and $|\beta\rangle$. Indeed, it suffices to choose $|\alpha\rangle=|\beta\rangle$ to be in the kernel of the projector operator (such that $\hat{P}(\mt^x,\mt^y) |\alpha\rangle = 0$) to see that the transition probability vanishes, violating Wigner's condition.  

Is there a way to ensure that the subsystem KW transformation acts as a faithful symmetry on the Xu-Moore model at the self-dual point $h=1$? The answer is provided by the generalization of Wigner's theorem to non-invertible symmetry which has been proven recently~\cite{Wigner-theorem-Ortiz2025}, stating that in order to preserve transition probabilities, the symmetry operator must be a composition of a unitary (or antiunitary) operator $\mathcal{U}$ and a projector operator $\tilde{P}$ defined on an \textit{enlarged} Hilbert space $\tilde{H} = \mathcal{H}\oplus \mathcal{H}_\perp$, such that this projector acts trivially on the original Hilbert space: $\tilde{P}|\alpha\rangle = |\alpha\rangle, \forall |\alpha\rangle \in \mathcal{H}$. Indeed, for a 1+1D transverse-field Ising model, one can explicitly construct a faithful symmetry operator of the desired form
\beq
\mathcal{N}_\text{TFI} = \mathcal{U}_\text{TFI} \tilde{P},
\eeq
by enlarging the Hilbert space by one qubit, $\tilde{H}= \mathcal{H}\oplus \mathcal{H}_\eta$, where the auxiliary Hilbert space $\mathcal{H}_\eta$ is said to contain a so-called $\eta$-defect of the $\mathbb{Z}_2$ symmetry -- for details see Refs.~\cite{Seiberg_LSM_SciPostPhys2024} and \cite{Wigner-theorem-Ortiz2025}. The analogous construction for the 2+1D Xu-Moore model with subsystem symmetries is a subject of a separate work~\cite{XM_Giridhar2025} by one of the present authors (A.H.N.).
In short, a \textit{bona fide} symmetry operator $\NKWtilde$, satisfying the generalized Wigner's theorem~\cite{Wigner-theorem-Ortiz2025} can be constructed, with the fusion rule of the form 
\begin{align}
\label{eq:N_KW_bar_proper}
    \NKWtilde^\dagger \times \NKWtilde =  \tilde{P}_{+} T_{xy},
\end{align}
where $\tilde{P}_{+}$ projects onto the definite eigenvalue $Z=+1$ of the $(L_x+L_y-1)$ ancillae qubits spanning the auxiliary Hilbert space (see Ref.~\cite{XM_Giridhar2025} for details):
\beq
\tilde{P}_{+} = \prod_{i=1}^{L_x} \left(\frac{1+Z^\text{col}_i}{2} \right)\prod_{j=2}^{L_y}  \left(\frac{1+Z^\text{row}_j}{2}\right).
\eeq
While this expression resembles the r.h.s. of Eq.~\eqref{eq:N_KW_bar_fusion}, the crucial difference is that the projector $\tilde{P}_{+}$ leaves the physical Hilbert space intact, instead performing a non-invertible projection onto the enlarged Hilbert space $\mathcal{H}_\perp$. There is a one-to-one correspondence between the $(L_x+L_y-1)$ ancillae that span the enlarged Hilbert space and 
the subsystem symmetries defined by the operators $\{U_j^x,U_i^y\}$ in Eq.~\eqref{eq:subsym_operator}, in fact the generalized KW transformation maps them onto one another, thus exchanging the ordered and disordered phases of the Xu -- Moore model. One can think of the ancillae as labeling the twist sectors $\{t_i^y,t_j^x\}$ of the model -- indeed, as seen from Eq.~\eqref{eq:mapping_symtwist_KW}, the duality transformation exchanges the subsystem symmetry sectors with the twist sectors of the dual model.

\subsection{KW duality as a gauging transformation}

In recent developments, the KW transformation has been understood in terms of gauging global symmetries \cite{Chang_defect_2019JHEP,Bhardwaj_gauging_2018JHEP,Aasen2016topological,Aasen_defect_duality_2020,Thorngren_fusion1_2019,Thorngren_fusion2_2021}. For a theory with a non-anomalous on-site 
symmetry, gauging promotes the global transformation into a local redundancy by introducing site-dependent symmetry actions. In the Euclidean spacetime formulation, this is implemented by coupling to background gauge variables placed on temporal links and spatial links of the original lattice. The gauge degrees of freedom thus live on the vertices of the dual lattice. The procedure involves two steps: (i) attaching to the partition function a phase whose exponent is the cup product of the gauge fields in the original and dual theories, and (ii) summing over all gauge-field configurations of the original theory to promote its background field to a dynamical one. The resulting gauged theory 
carries a new dual global symmetry, and its Hilbert space naturally decomposes into symmetry and twist sectors. Viewed in this light, the KW transformation precisely corresponds to this gauging step and is identified with the so-called $S$-transformation~\cite{Li_2023_KT}.

In our setting, the symmetry in question is not a conventional global one but a subsystem $\mathbb{Z}_2$ symmetry. The gauging procedure is entirely analogous: in Euclidean spacetime we introduce temporal link variables $A^z$ and spatial plaquette fields $A^{xy}$, as in the Xu-Moore model, and again proceed in two steps -- first attaching to the partition function a cup-product phase between the original and dual gauge fields, and then summing over all gauge configurations of the original theory to promote its background field to a dynamical one. A detailed discussion of this procedure for subsystem $\mathbb{Z}_2$ symmetries is presented in Appendix~\ref{app:gauging_KW} following Ref.~\cite{Cao_sub_duality_2022}. A more detailed exposition for the Xu -- Moore model, incorporating also the so-called ``half-gauging'' procedure to derive the duality defects, can be found in Ref.~\cite{XM_Giridhar2025}.

The analogous gauging procedure can also be applied to the dual theory Eq.~\eqref{eq:XM_dual_infinite} with $\widehat{\sigma}$ matter degrees of freedom, by introducing the $\widehat{A}$ gauge fields, detailed in Appendix~\ref{app:gauging_KW}. In this way, the  partition function of the original theory, written in terms of the variables $\{\sigma_{i,j}, A\}$, can be mapped to that of the dual theory expressed in terms of $\{\widehat{\sigma}_{i-\half, j-\half}, \widehat{A}\}$. Under this duality, the twist sectors of the original theory are exchanged with the subsystem symmetry sectors of the dual theory, and vice versa, as summarized in Eq.~\eqref{eq:mapping_symtwist_KW}. In Appendix~\ref{sec:sector_mapping_from_gauging}, we rederive the symmetry-twist sector mapping in Eq.~\eqref{eq:mapping_symtwist_KW}, which follows naturally from analyzing the partition function and from the gauging procedure.

Thus, the operator $\mathcal{N}_{\text{KW}}$ in Eq.~\eqref{eq:N_sub} is naturally understood as implementing the gauging of the subsystem $\mathbb Z_2$ symmetry. This perspective also explains why the exponent structures in Eqs.~\eqref{eq:N_sub}, \eqref{eq:C_bulk_bdy}, and \eqref{eq:C_tilde_bulk_bdy} closely resemble minimal coupling terms to background gauge fields.

\subsection{Subsystem KW transformation on an open square lattice}
\label{sec:KW_operator_open}
The subsystem KW transformation on an open square lattice has not been addressed in Ref.~\cite{Cao_sub_KW_2023}, where the focus was primarily on the closed lattice case (with periodic or antiperiodic boundary conditions).
Here, we focus on the open square lattice, since this setup is crucial for reformulating the subsystem KT transformation in this geometry in Sec.~\ref{sec:KT_closed}. While the operator $\NKW$ realizing the subsystem duality transformation is typically non-unitary and follows a non-invertible fusion rule, we will show that for certain open boundary conditions, $\NKW$ in fact acts as a unitary operator.

We consider an open $L_x \times L_y$ square lattice, where each site is labeled by coordinates $(i,j)$ with $i = 1, \dots, L_x$ and $j = 1, \dots, L_y$. The corresponding dual spins are located at the centers of the plaquettes, with coordinates $(i-\half,j-\half)$, as depicted in Fig.~\ref{fig:KW_duality}. Let us begin with the expression in Eq.~\eqref{eq:N_sub} and adapt it so that only the terms fully contained within the system are retained in the exponent, corresponding to a free boundary condition. Specifically,
\begin{align}
\label{eq:N_sub_open}
    \NKW^{\text{open}} \ket{\{\ssi_{i,j}\}}  = \frac{1}{2^{(L_x+L_y)/2}} \sum_{\{\hssi_{i-\half,j-\half}\}} (-1)^{\mathcal{C}_{\text{open}}^{\sigma\sigma'}} \ket{\{\hssi_{i-\half,j-\half}\}}
\end{align}
where
\begin{align}
\label{eq:C_open}
    \mathcal{C}_{\text{open}}^{\sigma\sigma'}  = & \sum_{i=2}^{L_x} \sum_{j=2}^{L_y} \ssi_{i-1,j-1} \hssi_{i-\half,j-\half} + \sum_{i=1}^{L_x} \sum_{j=2}^{L_y} \ssi_{i,j-1} \hssi_{i-\half,j-\half} \nonumber\\
    & + \sum_{i=2}^{L_x} \sum_{j=1}^{L_y} \ssi_{i-1,j} \hssi_{i-\half,j-\half} + \sum_{i=1}^{L_x} \sum_{j=1}^{L_y} \ssi_{i,j} \hssi_{i-\half,j-\half}.
\end{align}

The unitarity of $\NKW^{\text{open}}$ can be confirmed by explicitly evaluating $\bra{\{\ssi_{i,j}\}} \NKW^{\text{open}\, \dagger} \NKW^{\text{open}} \allowbreak \ket{\{\ssip_{i,j}\}}$. To illustrate this, we proceed with the calculation,
\begin{align}
\label{eq:NN_open}
    \bra{\{\ssi_{i,j}\}} \NKW^{\text{open}\,\dagger} \NKW^{\text{open}} \ket{\{\ssip_{i,j}\}} = \frac{1}{2^{(L_x+L_y)}} \sum_{\substack {\{\hssi_{i-\half,j-\half}\},\\
    \{\hssip_{i-\half,j-\half}\}}} \bra{\{\hssi_{i-\half, j-\half}\}} (-1)^{\mathcal{C}_{\text{open}}^{\mN^\dagger \mN}} \ket{\{\hssip_{i-\half, j-\half}\}},
\end{align}
where we have dropped the superscript indices on $C_\text{open}^{\sigma\sigma'}$ for the sake of brevity. The resulting phase factor
\begin{align}
\label{eq:C_NN_open}
    \mathcal{C}_{\text{open}}^{\mN^\dagger \mN} = & \sum_{i=2}^{L_x} \sum_{j=2}^{L_y} \ssi_{i-1,j-1} \hssi_{i-\half,j-\half} + \sum_{i=1}^{L_x} \sum_{j=2}^{L_y} \ssi_{i,j-1} \hssi_{i-\half,j-\half}  + \sum_{i=2}^{L_x} \sum_{j=1}^{L_y} \ssi_{i-1,j} \hssi_{i-\half,j-\half} \nonumber\\
     & + \sum_{i=1}^{L_x} \sum_{j=1}^{L_y} \ssi_{i,j} \hssi_{i-\half,j-\half} + \sum_{i=2}^{L_x} \sum_{j=2}^{L_y} \ssip_{i-1,j-1} \hssi_{i-\half,j-\half} + \sum_{i=1}^{L_x} \sum_{j=2}^{L_y} \ssip_{i,j-1} \hssi_{i-\half,j-\half} \nonumber\\
    & + \sum_{i=2}^{L_x} \sum_{j=1}^{L_y} \ssip_{i-1,j} \hssi_{i-\half,j-\half} + \sum_{i=1}^{L_x} \sum_{j=1}^{L_y} \ssip_{i,j} \hssi_{i-\half,j-\half}.
\end{align}

As $ \ket{\{\hssi_{i-\half, j-\half}\}}$ is an orthonormal basis, the inner product $\bra{\{\hssi_{i-\half, j-\half}\}} \ket{\{\hssip_{i-\half, j-\half}\}}$ equals 1 when $\hssip_{i-\half, j-\half} = \hssi_{i-\half, j-\half}$ for all $(i,j)$ and $0$ otherwise. Consequently, summing over all $\hssip_{i-\half, j-\half}$, we obtain from Eq.~\eqref{eq:NN_open}
\begin{align}
    \bra{\{\ssi_{i,j}\}} \NKW^{\text{open} \, \dagger} \NKW^{\text{open}} \ket{\{\ssip_{i,j}\}} = \frac{1}{2^{L_x+L_y}} \sum_{\substack {\{\hssi_{i-\half,j-\half}\}}}  (-1)^{\mathcal{C}_{\text{open}}^{\mN^\dagger \mN}},
\end{align}
where $\mathcal{C}_{\text{open}}^{\mN^\dagger \mN}$ in Eq.~\eqref{eq:C_NN_open} becomes
\begin{align}
    \mathcal{C}_{\text{open}}^{\mN^\dagger \mN} = & \sum_{i=2}^{L_x} \sum_{j=2}^{L_y} \left(\ssi_{i-1,j-1} + \ssip_{i-1,j-1} \right) \hssi_{i-\half,j-\half} + \sum_{i=1}^{L_x} \sum_{j=2}^{L_y} \left(\ssi_{i,j-1} + \ssip_{i,j-1} \right) \hssi_{i-\half,j-\half} \nonumber\\
     &  + \sum_{i=2}^{L_x} \sum_{j=1}^{L_y} \left(\ssi_{i-1,j} + \ssip_{i-1,j} \right) \hssi_{i-\half,j-\half}  + \sum_{i=1}^{L_x} \sum_{j=1}^{L_y} \left(\ssi_{i,j} + \ssip_{i,j} \right) \hssi_{i-\half,j-\half}.
\end{align}

Next, performing the sum over all $\hssi_{i-\half, j-\half}$ results in
\begin{align}
    \bra{\{\ssi_{i,j}\}} \NKW^{\text{open} \, \dagger} \NKW^{\text{open}} \ket{\{\ssip_{i,j}\}} = \frac{2^{L_x L_y}}{2^{L_x+L_y}}\prod_{i=1}^{L_x} \prod_{j=1}^{L_y}  \delta_{\ssi_{i,j},\ssip_{i,j}}.
\end{align}

If we rescale $\NKW^{\text{open}}$ properly, the prefactor $2^{L_x L_y-(L_x+L_y)}$ drops out and the expression reduces to the identity. This calculation confirms that
\beq
\label{eq:KW_fusion_open}
\NKW^{\text{open} \, \dagger} \NKW^{\text{open}} = I,
\eeq
thereby establishing that $\NKW^{\text{open}}$ acts as a unitary operator under the free boundary conditions. This behavior notably differs from that of the closed square lattice, where $\NKW$ is known to be non-unitary, non-invertible operator \cite{Li_defect_2024}, as we demonstrated above, see Eq.~\eqref{eq:N_KW_fusion}. 

If, instead of the duality operator $\NKW^{\text{open}}$, we consider the duality symmetry operator $\NKWbar^{\text{open}}$ equipped with half-translation (Eq.~\ref{eq:half-translation}), the fusion rule under open boundary conditions takes the form
\begin{align}
\NKWbar^{\text{open}\,\dagger} \NKWbar^{\text{open}} = T_{xy},
\end{align}
where $T_{xy}$ is the diagonal lattice translation defined in Eq.~\eqref{eq:T_xy}.

To better understand the transformation, it is helpful to explicitly derive the mapping between the Pauli operators. By performing a straightforward calculation, we find that
\begin{align}
\label{eq:Pauli_map_KW_open}
    \hsig^x_{i-\half, j-\half} \NKW^{\text{open}} \ket{\psi} & = \NKW^{\text{open}} \mathcal{O}_{i,j}^x \left( \left\{ \sigma^{x,z} \right\} \right) \ket{\psi}, \nonumber\\
    \hsig^z_{i-\half, j-\half} \NKW^{\text{open}} \ket{\psi} & = \NKW^{\text{open}} \mathcal{O}_{i,j}^z \left( \left\{ \sigma^{x,z} \right\} \right) \ket{\psi}.
\end{align} 
In other words, the dual lattice variables get mapped to 

\begin{align}
    \hsig^x_{i-\half, j-\half} \stackrel{\NKW^{\text{open}}}{\longrightarrow} \mathcal{O}_{i,j}^x \left( \left\{ \sigma^{x,z} \right\} \right), \quad \hsig^z_{i-\half, j-\half} \stackrel{\NKW^{\text{open}}}{\longrightarrow} \mathcal{O}_{i,j}^z \left( \left\{ \sigma^{x,z} \right\} \right)
\end{align}

where 
\begin{align}
    &  \mathcal{O}_{i,j}^x \left( \left\{ \sigma^{x,z} \right\} \right) = \begin{cases} \sigma^z_{i-1,j-1} \sigma^z_{i,j-1} \sigma^z_{i-1,j} \sigma^z_{i,j}, & i=2,\dots,L_x,\, j = 2,\dots,L_y\\\sigma^z_{1,j-1} \sigma^z_{1,j}, &i=1,\, j = 2,\dots,L_y \\\sigma^z_{i-1,1} \sigma^z_{i,1}, &i=2,\dots,L_x,\, j = 1 \\ \sigma^z_{1,1}, & i=1,\, j=1\end{cases}, \nonumber\\
    & \mathcal{O}_{i,j}^z \left( \left\{ \sigma^{x,z} \right\} \right) = \prod_{\iprime = i}^{L_x} \prod_{\jprime = j}^{L_y} \sigma^x_{\iprime, \jprime}.
\end{align}

For example, consider the spin located at the corner of the square lattice. According to the mapping in Eq.~\eqref{eq:Pauli_map_KW_open}, we have $\hsig^x_{\half,\half} \stackrel{\NKW^{\text{open}}}{\longrightarrow} \sigma^z_{1,1}$. Unlike in the bulk, where $\hsig^x$ maps to a product of four $\sigma^z$ operators, at the corner $\hsig^x_{\half,\half}$ transforms into a single spin operator $\sigma^z_{1,1}$ on the original lattice. This difference originates from the fact that 
$\mathcal{C}_{\text{open}}$ in Eq.~\eqref{eq:C_open} contains the term 
$\sigma^z_{1,1} \hsig^x_{\half,\half}$. We will make use of these maps in Sec.~\ref{sec:KT_open}.

\subsection{Application of the KW Transformation to the Xu-Moore Model}
\label{sec:Xu-Moore_model}
As a concrete example, we turn to the Xu-Moore (XM) model \cite{Xu_Moore_prl_2004,Xu_Moore_nuclB_2005} which exhibits the subsystem KW duality. In this setting, we focus on the ground state and present an argument illustrating why the KW transformation, when restricted to the original Hilbert space, is necessarily non-invertible. We further show that its unitarity can be recovered by enlarging the Hilbert space to incorporate the twisted sector, thereby allowing the transformation to act in a fully invertible manner \cite{Li_2023_KT}. Establishing the invertibility or unitarity through an argument focused on the ground state will prove useful in Sec.~\ref{sec:closed_KT_example}, where we address the invertibility and unitarity of the subsystem KT transformation.

\subsubsection{On a closed square lattice: A non-invertible transformation}
\label{sec:Xu-Moore_closed}
We begin by considering the Hamiltonian of the Xu-Moore model in Eq.~\eqref{eq:XM_closed}, defined on a closed $L_x \times L_y$ square lattice. As a next step, we apply the subsystem KW transformation to the Xu-Moore Hamiltonian. Using the mapping between Pauli operators provided in Eq.~\eqref{eq:Pauli_map_KW}, each term in the original Hamiltonian in \eqref{eq:XM_closed} is rewritten in terms of the dual variables, yielding the KW-dual Hamiltonian of the Xu-Moore model,
\begin{align}
\label{eq:XM_dual_closed}
    \widehat{H}^h_{\text{XM}} = - \sum_{i=1}^{L_x} \sum_{j=1}^{L_y} \left( \hsig^x_{i-\half,j-\half} + h \hsig^z_{i-\half,j-\half} \hsig^z_{i+\half,j-\half} \hsig^z_{i-\half,j+\half} \hsig^z_{i+\half,j+\half}\right).
\end{align}

By performing a simple relabeling of variables, in which the spins originally defined on the plaquettes are shifted to the sites, we find that
\begin{align}
    H^h_{\text{XM}} = h \widehat{H}^{1/h}_{\text{XM}}.
\end{align}
Hence, at the critical point $h=1$, the Xu-Moore model exhibits a subsystem KW duality symmetry.

Let us now briefly discuss the phases of the Xu-Moore model \cite{Xu_Moore_prl_2004,Xu_Moore_nuclB_2005}. The quantum model in Eq.~\eqref{eq:XM_closed} enjoys subsystem symmetry as
\begin{align}
    \left[ H^h_{\text{XM}}, U^x_j \right] = 0, \quad  \left[ H^h_{\text{XM}}, U^y_i \right] = 0,
\end{align}
where $U^x_j$, $U^y_i$ are defined in Eq.~\eqref{eq:subsym_operator}. Due to the global constraint in Eq.~\eqref{eq:subsym_operator_contraint}, there are $L_x+L_y-1$ independent conserved quantities. Consequently, the theory is divided into $2^{L_x+L_y-1}$ symmetry sectors, each labeled by the eigenvalues of the operators in Eq.~\eqref{eq:subsym_operator}. For $h \ll 1$, these symmetries are spontaneously broken, resulting in $2^{L_x+L_y-1}$ degenerate ground states. In the ordered phase ($h \ll 1$), the lowest-energy excitation is generated by flipping a single spin. Such a flip creates four defective plaquettes, which can be viewed as $\mathbb{Z}_2$ vortices located around the flipped site. Remarkably, this composite excitation can fractionalize into four excitations that move apart to occupy the vertices of a rectangle, with all spins inside the rectangle flipped relative to the ground-state configuration. 

In the large-$h$ limit ($h \gg 1$), the system has a unique ground state that is invariant under the $L_x+L_y-1$ symmetry transformations. The lowest mobile excitation in this disordered phase corresponds to flipping a single bond (in the $\sigma^x$ basis). Such a flipped bond can propagate only in the direction perpendicular to its orientation. Numerical studies show that the phase transition occurring at the self-dual point $(h=1)$ is of first-order type \cite{orus2009first}.

We now turn to a physical perspective on the non-unitarity of the subsystem KW transformation, presenting an argument based on an analysis of the ground state. Under the subsystem KW transformation, the ordered phase $h \ll 1$ of the original model \eqref{eq:XM_closed} is mapped to the dual Xu-Moore model \eqref{eq:XM_dual_closed} expressed in terms of dual spins residing in the disordered phase, a regime in which the ground state is unique. This mapping, therefore, takes the $2^{L_x+L_y-1}$ degenerate ground states of the original ordered-phase model and collapses them into a single state in the dual description. Such a drastic reduction of the ground-state manifold makes it clear that the subsystem KW transformation is intrinsically non-unitary and non-invertible. A similar conclusion follows if we start from the disordered phase ($h \gg 1$) of the original Xu-Moore model \eqref{eq:XM_closed}, which has a unique ground state; under the subsystem KW transformation, it maps to the ordered phase of the dual Xu-Moore model \eqref{eq:XM_dual_closed}, featuring $2^{L_x+L_y-1}$ degenerate ground states.

We now explain how unitarity can be restored by enlarging the Hilbert space to include states from the twisted sector. In the disordered phase $h \gg 1$ of the original Xu-Moore model \eqref{eq:XM_closed}, the system does not distinguish between periodic and twisted boundary conditions--the ground-state energy remains essentially unchanged in either case. This insensitivity implies that introducing twisted boundary conditions does not cost energy in the thermodynamic limit. If we formally enlarge the Hilbert space to include all $2^{L_x+L_y-1}$ twisted-sector configurations (see Sec.~\ref{sec:subsym_square} for details), the ground-state manifold correspondingly expands, yielding a $2^{L_x+L_y-1}$-fold degeneracy. 

In contrast, in the ordered phase of the original Xu-Moore model \eqref{eq:XM_closed} ($h \ll 1$), imposing a twisted boundary condition is energetically costly, as it necessarily introduces vortices with finite energy. This extra energy shifts the twisted-sector ground state above that of the untwisted (periodic) sector, preventing them from being degenerate. Consequently, even if we formally enlarge the Hilbert space to include the twisted sectors, the low-energy manifold remains unchanged: the ground-state degeneracy stays at $2^{L_x+L_y-1}$, arising solely from the spontaneous breaking of the subsystem symmetry. This extended Hilbert space then provides a natural setting in which the subsystem KW transformation can act unitarily (and hence invertibly).

\subsubsection{On an open square lattice: An invertible transformation}
We now focus on the subsystem KW transformation of the Xu-Moore Hamiltonian on an open square lattice. While we have already established its unitarity in this setting in Sec.~\ref{sec:KW_operator_open}, we present here two additional physical arguments in support of this result. This discussion also serves as a prelude to the more intricate subsystem KT transformation on an open square lattice, which we will address in Sec.~\ref{sec:KT_open}.

First, we note that due to the mapping in Eq.~\eqref{eq:Pauli_map_KW_open}, the standard Xu-Moore model on an open square lattice is not exactly self-dual. More precisely, we find that
\begin{align}
\label{eq:XM_open_1}
    H^h_{\text{open XM}} = & - \sigma^z_{1,1}  - \sum_{i=2}^{L_x} \sigma^z_{i-1,1} \sigma^z_{i,1} - \sum_{j=2}^{L_y} \sigma^z_{1,j-1} \sigma^z_{1,j} - \sum_{i=2}^{L_x} \sum_{j=2}^{L_y} \sigma^z_{i-1,j-1} \sigma^z_{i,j-1} \sigma^z_{i-1,j} \sigma^z_{i,j} \nonumber \\
    & - h \sum_{i=1}^{L_x} \sum_{j=1}^{L_y}  \sigma^x_{i,j},
\end{align}
is dual to
\begin{align}
\label{eq:XM_dual_open_1}
    \widehat{H}^h_{\text{open XM}} = & - \sum_{i=1}^{L_x} \sum_{j=1}^{L_y} \hsig^x_{i-\half,j-\half} - h\sum_{i=1}^{L_x-1} \sum_{j=1}^{L_y-1}  \hsig^z_{i-\half,j-\half} \hsig^z_{i+\half,j-\half} \hsig^z_{i-\half,j+\half} \hsig^z_{i+\half,j+\half} \nonumber\\
    & - h\sum_{i=1}^{L_x-1}  \hsig^z_{i-\half,L_y-\half} \hsig^z_{i+\half,L_y-\half} - h \sum_{j=1}^{L_y-1} \hsig^z_{L_x-\half,j-\half} \hsig^z_{L_x-\half,j+\half} - h \hsig^z_{L_x-\half,L_y-\half}.
\end{align}

Note the presence of the first three terms of the original Hamiltonian in Eq.~\eqref{eq:XM_open_1} and the last three terms of the dual Hamiltonian in Eq.~\eqref{eq:XM_dual_open_1} at the boundary. 
These are precisely the terms whose transformation properties under the subsystem symmetry generators defined in Eqs.~\eqref{eq:subsym_operator} and \eqref{eq:subsym_operator_dual} we shall now examine in detail. The first term in Eq.~\eqref{eq:XM_open_1} explicitly breaks $U^x_1$ and $U^y_1$; the second term explicitly breaks $U^y_i$ for $i = 1,\dots,L_x$; and the third term explicitly breaks $U^x_j$ for $j = 1,\dots,L_y$. Similarly, the third term in Eq.~\eqref{eq:XM_dual_open_1} explicitly breaks $\widehat{U}^y_{i-\half}$ for $i = 1,\dots,L_x$; the fourth term explicitly breaks $\widehat{U}^x_{j-\half}$ for $j = 1,\dots,L_y$; and the fifth term explicitly breaks $\widehat{U}^y_{L_x-\half}$ and $\widehat{U}^x_{L_y-\half}$. Physically, this means that the corner longitudinal magnetic field $\sigma^z_{1,1}$ in Eq.~\eqref{eq:XM_open_1} and $\hsig^z_{L_x-\half,L_y-\half}$ in Eq.~ \eqref{eq:XM_dual_open_1}, acts as a boundary pinning field that explicitly selects a single symmetry-broken configuration, even deep in the ordered phase. The subsystem symmetry is therefore no longer free to generate a large degeneracy, and the ground state remains unique. This removal of degeneracy directly resolves the hindrance to the unitarity of the subsystem KW transformation discussed in Sec.~\ref{sec:KW_operator_closed}, since the mapping between the ground states becomes one-to-one.

We now present our second physical perspective on the problem. An alternative approach to restoring unitarity is to ensure that the subsystem symmetry of the original spins is strictly preserved from the outset. To achieve this, we can deliberately omit the first three terms of the original Hamiltonian in Eq.~\eqref{eq:XM_open_1}, as these terms explicitly break certain subsystem symmetry generators. Once these terms are removed, Eq.~\eqref{eq:XM_open_1} takes the simplified form
\begin{align}
\label{eq:XM_open_2}
    H^h_{\text{open XM}} = - \sum_{i=2}^{L_x} \sum_{j=2}^{L_y} \sigma^z_{i-1,j-1} \sigma^z_{i,j-1} \sigma^z_{i-1,j} \sigma^z_{i,j}  - h \sum_{i=1}^{L_x} \sum_{j=1}^{L_y}  \sigma^x_{i,j},
\end{align}
which in the dual picture corresponds to
\begin{align}
\label{eq:XM_dual_open_2}
    \widehat{H}^h_{\text{open XM}} = & - \sum_{i=2}^{L_x} \sum_{j=2}^{L_y} \hsig^x_{i-\half,j-\half} - h\sum_{i=1}^{L_x-1} \sum_{j=1}^{L_y-1}  \hsig^z_{i-\half,j-\half} \hsig^z_{i+\half,j-\half} \hsig^z_{i-\half,j+\half} \hsig^z_{i+\half,j+\half} \nonumber\\
    & - h\sum_{i=1}^{L_x-1}  \hsig^z_{i-\half,L_y-\half} \hsig^z_{i+\half,L_y-\half} - h \sum_{j=1}^{L_y-1} \hsig^z_{L_x-\half,j-\half} \hsig^z_{L_x-\half,j+\half} - h \hsig^z_{L_x-\half,L_y-\half}.
\end{align}
But now there are two special cases:
\begin{itemize}
    \item First, let us consider the disordered phase ($h\gg1$) of the original spins described by Eq.~\eqref{eq:XM_dual_open_2}, which possesses a unique ground state. Under the subsystem KW duality mapping, this phase corresponds to the ordered phase ($h\gg1$) of the dual Hamiltonian in \eqref{eq:XM_dual_open_2}. But for $h\gg1$, the third term in Eq.~\eqref{eq:XM_dual_open_2} explicitly breaks $\widehat{U}^y_{i-\half}$ for $i = 1,\dots,L_x$; the fourth term explicitly breaks $\widehat{U}^x_{j-\half}$ for $j = 1,\dots,L_y$; and the fifth term explicitly breaks $\widehat{U}^y_{L_x-\half}$ and $\widehat{U}^x_{L_y-\half}$. Consequently, we once again observe that the end spin becomes polarized, resulting in a unique ground state.

    \item We now turn our attention to the ordered phase ($h \ll 1$) of the original Hamiltonian in Eq.~\eqref{eq:XM_dual_open_2}, characterized by $2^{L_x + L_y - 1}$ degenerate ground states. In this $h \ll 1$ limit, the first term of the dual Hamiltonian in \eqref{eq:XM_dual_open_2} dominates, corresponding to the disordered phase of the dual spins. However, this dominant term lacks certain transverse field contributions, such as $\hsig^x_{\half,\half}$, $\hsig^x_{i-\half,\half}$ for $i = 2,\dots,L_x$, and $\hsig^x_{\half,j-\half}$ for $j = 2,\dots,L_y$. These $L_x + L_y - 1$ boundary spins lack a preferred orientation, which implies the presence of spin-$\tfrac{1}{2}$ ``edge modes''. As a result, the ground state manifold is $2^{L_x + L_y - 1}$-fold degenerate. The mapping \eqref{eq:Pauli_map_KW_open} directly shows that spontaneous breaking of the global subsystem $\mathbb{Z}_2$ symmetry in the original system manifests as an edge state in the dual system. Concretely,
    \begin{align}
    \label{eq:U_map_KW_open}
      \NKW^{\text{open}} U^x_j \NKW^{\text{open} \, \dagger} & = \begin{cases} \hsig^z_{\half,j-\half} \hsig^z_{\half,j+\half}, & j = 1,\dots,L_y-1\\\hsig^z_{\half,L_y-\half}, &j = L_y\end{cases}, \nonumber\\
     \NKW^{\text{open}} U^y_{i} \NKW^{\text{open} \, \dagger} & = \begin{cases} \hsig^z_{i-\half,\half} \hsig^z_{i+\half,\half}, & i = 1,\dots,L_x-1\\\hsig^z_{L_x-\half,\half}, &i = L_x\end{cases}.
\end{align}

\end{itemize}

This demonstrates that the transformation preserves the ground-state structure. In the disordered phase of the original theory, the unique ground state is mapped to the unique ground state of the dual theory. In the ordered phase, the entire degenerate ground-state manifold with $2^{L_x+L_y-1}$ states is mapped onto the corresponding manifold in the dual theory. Thus, the transformation consistently acts as a unitary operator, as further confirmed in Sec.~\ref{sec:KW_operator_open}.

\section{Subsystem Kennedy-Tasaki transformation: Mapping SSSB to weak SSPT}
\label{sec:KT_closed}
So far, we have discussed the generalized subsystem Kramers--Wannier transformation, which relates the $\mathbb{Z}_2$ subsystem SSB  phase to the trivial phase. We now turn to the generalized (subsystem) Kennedy--Tasaki transformation, a duality mapping between a spontaneous symmetry-broken phase and an invertible topological phase protected by symmetry (SPT), as illustrated schematically in Fig.~\ref{fig:Dualities_N_KT}. It is important to point out that a global $\mathbb{Z}_2$ symmetry is insufficiently rich to protect a bosonic topological phase in one spatial dimension -- this can be understood in terms of classification of bosonic SPTs, wherein different SPT phases correspond to (equivalence classes of) the projective representations of the global symmetry group $G$  with unitary coefficients~\cite{Chen_SPT_science,Chen_SPTcohom_PRB2013}. Distinct SPT phases in $d$ space dimensions are thus captured by the elements of the cohomology group with coefficients in $U(1)$: $\mathcal{H}^{d+1}(G,U(1))$~\cite{Chen_SPTcohom_PRB2013}, which for $G=\mathbb{Z}_2$ turns out to be trivial in 1 space dimension.

The simplest \textit{global} symmetry group that offers such protection turns out to be the dihedral group $D_2$, which is isomorphic to $\mathbb{Z}_2 \times \mathbb{Z}_2$. 
In 1+1D, the calculation of the cohomology yields $\mathcal{H}^2(\mathbb{Z}_2 \times \mathbb{Z}_2, U(1)) = \mathcal{H}^3(\mathbb{Z}_2 \times \mathbb{Z}_2, \mathbb{Z}) = \mathbb{Z}_2$, reflecting the well known fact that in addition to the trivial phase, there is a unique nontrivial SPT, realized in the Haldane phase of spin-$1$ chain. This, in fact, is the underpinning of the original Kennedy--Tasaki transformation, which preserves the global dihedral symmetry group $\mathbb{Z}_2 \times \mathbb{Z}_2$ of the spin chain, while interchanging the SPT and the spontaneous symmetry broken phase.

This section is organized as follows: we first review the $1$d KT transformation for the global $\mathbb{Z}_2 \times \mathbb{Z}_2$ symmetry in Sec.~\ref{sec:SPT_KT_1d}, which allows us to introduce the notion of the domain wall decoration operator $U_{DW}$ and consequently relate the KT transformation to the KW duality studied in the previous section. We then transition to two space dimensions and consider the weak SSPT protected by the \textit{subsystem}  $\mathbb{Z}_2 \times \mathbb{Z}_2$ symmetry in Sec.~\ref{sec:weak_SSPT}. We then proceed to analyze the non-invertible subsystem KT transformation, which maps a SSSB phase of decoupled chains to a weak SSPT phase in Sec.~\ref{sec:weak_SSPT}. Finally, in the subsequent Sec.~\ref{sec:KT_SSPT_strong} we shall presents the generalization to the strong SSPT case, where we study the properties of the subsystem KT transformation in detail.

\subsection{Brief review of $1$d $\mathbb{Z}_2 \times \mathbb{Z}_2$ SPT and KT transformation}
\label{sec:SPT_KT_1d}

\subsubsection{$1$d $\mathbb{Z}_2 \times \mathbb{Z}_2$ SPT}
\begin{figure}[tb]
    \centering
    \includegraphics[width=0.90\linewidth]{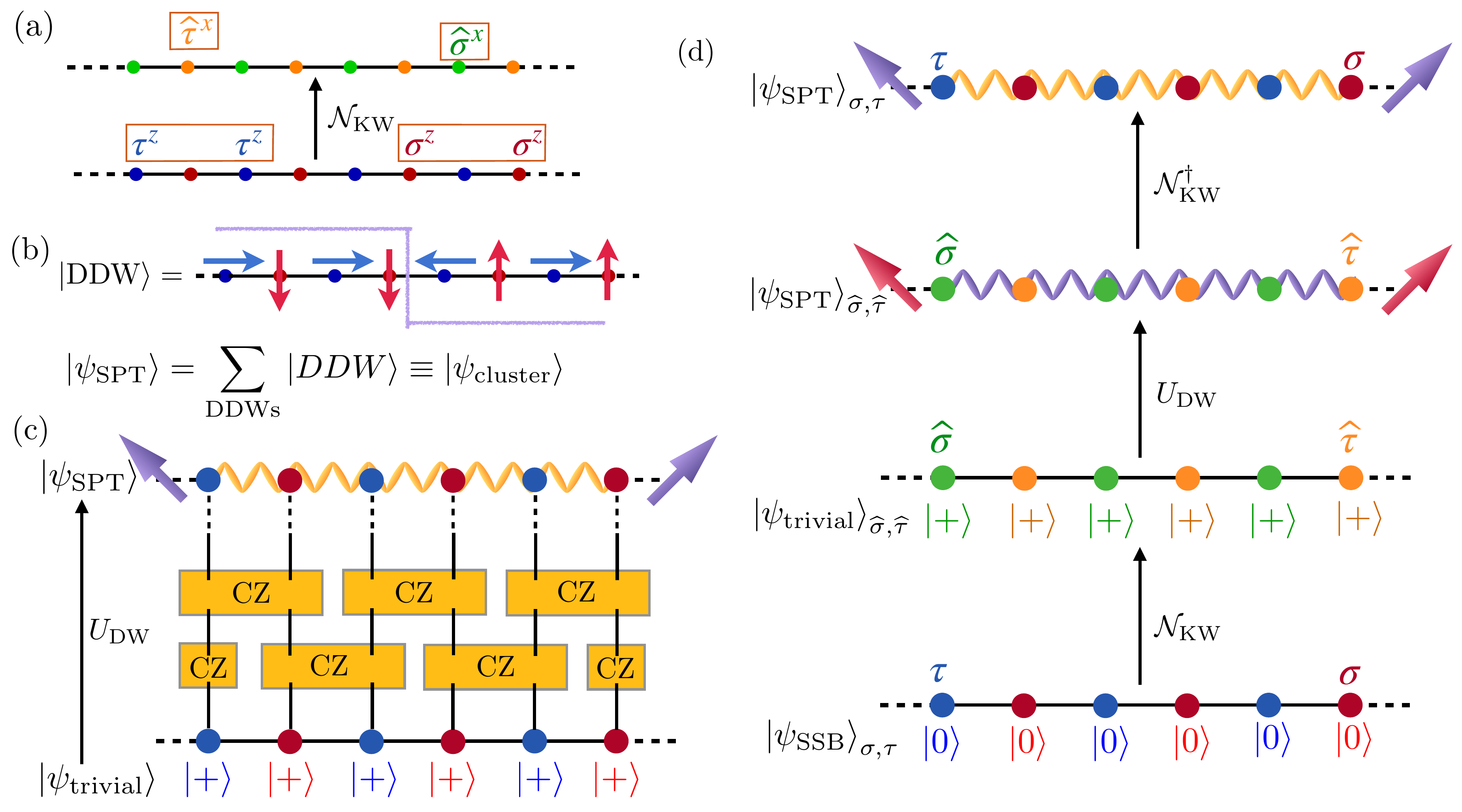}
    \caption{(\textbf{a}) The mapping of the Ising interaction terms through the $1$d KW transformation is illustrated; (\textbf{b}) Ground-state wave function showing the domain-wall decoration pattern; (\textbf{c}) Preparation of the $1$d cluster state $\ket{\Psi_{\mathrm{cluster}}}$ from the trivial state $\ket{\Psi_{\mathrm{trivial}}}$ using $U_{\text{DW}}$; (\textbf{d}) Illustration of the sequence of transformations that construct the 1D KT duality, which maps a $\mathbb{Z}_2 \times \mathbb{Z}_2$ symmetry-broken state to a $\mathbb{Z}_2 \times \mathbb{Z}_2$ SPT cluster state.}
    \label{fig:SPT_1d}
\end{figure}

We consider a spin chain with $L$ sites and $L$ links. Each site hosts a spin-$1/2$ state $\ket{\ssi_{j}}$, and each link hosts another spin-$1/2$ state 
$|\st_{j-\half} \rangle$,
so that each unit cell contains two spins, as depicted in Fig.~\ref{fig:SPT_1d}(a). The local Pauli operators act as follows (as before, we use the convention $\ssi_{j}=0,1$ and $\st_{j-\half}=0,1$):
\begin{align}
    \sigma^z_{j} \ket{\ssi_{j}} & = (-1)^{\ssi_{j}} \ket{\ssi_{j}},  \qquad \qquad \quad \sigma^x_{j} \ket{\ssi_{j}}  =  \ket{1-\ssi_{j}},\nonumber\\
    \tau^z_{j-\half} \ket{\st_{j-\half}}  & = (-1)^{\st_{j-\half}} \ket{\st_{j-\half}}, \; \tau^x_{j-\half} \ket{\st_{j-\half}} = \ket{1-\st_{j-\half}}. 
    \label{eq:Pauli-action}
\end{align}
The $\mathbb{Z}_2 \times \mathbb{Z}_2$ symmetry is generated by 
\begin{align}
\label{eq:sym_generator_1d}
    U_\sigma = \prod_{j=1}^L \sigma^x_j, \quad U_\tau = \prod_{j=1}^L \tau^x_{j-\half}.
\end{align}
The symmetry and twist sectors are labeled by $(u_\sigma, u_\tau, t_\sigma, t_\tau)$, where $u_\sigma, u_\tau$ are eigenvalues of the symmetry generators $U_\sigma, U_\tau$  and $t_\sigma, t_\tau$ specify the boundary conditions $\sigma_{j+L} = \sigma_j+t_\sigma $, $\tau_{j-\half+L} = \tau_{j-\half} + t_\tau$. Throughout the paper, we interchangeably use the notations $\ket{0} \equiv \ket{\uparrow}$, $\ket{1} \equiv \ket{\downarrow}$ for the eigenstates of the Pauli-$Z$ operator, and $\ket{+} \equiv \ket{\rightarrow}$, $\ket{-} \equiv \ket{\leftarrow}$ for the eigenstates of the Pauli-$X$ operator.

As a starting point, we consider the parent Hamiltonian of the $1$d \textit{cluster state}~\cite{cluster-state_Briegel2001,cluster_state_Raussendorf_Bravyi2005, Nielsen:2005syh, Son_cluster_1d_2012} on a ring:
\begin{align}
\label{eq:H_cluster_1d}
    H_{\mathrm{SPT}} = -\sum_{j=1}^L \sigma^z_{j-1} \tau^x_{j-\half} \sigma^z_{j} -\sum_{j=1}^L \tau^z_{j-\half} \sigma^x_{j} \tau^z_{j+\half}.
\end{align}
This Hamiltonian is exactly solvable and symmetric under a global $\mathbb{Z}_2^\sigma \times \mathbb{Z}_2^\tau$ symmetry generated by $U_\sigma$ and $U_\tau$ in Eq.~\eqref{eq:sym_generator_1d}. We represent the ground state of the $1$d cluster Hamiltonian in the basis of eigenstates of $\sigma_j^z \in \{\uparrow, \downarrow\}$ for site qubits and $\tau_{j+\frac{1}{2}}^x \in \{\rightarrow,\leftarrow\}$ for link qubits. The first term of the Hamiltonian \eqref{eq:H_cluster_1d} can be thought of as energetically enforcing the Gauss law constraint $\sigma_{j-1}^z \sigma_j^z = \tau_{j-\frac{1}{2}}^x$, which ties the configuration of the $\tau$ spins to the domain walls of the $\sigma$ spins. Specifically, when a domain wall occurs between neighboring sites (i.e., $\sigma_{j-1}^z \sigma_j^z = -1$), the corresponding $\tau^x$ link takes the value $-1$. This leads to the concept of \textit{decorated domain walls} (DDWs), where domain walls in the $\sigma$ configuration are ``decorated'' by excitations in the $\tau$ layer. A representative decorated configuration is illustrated in Fig.~\ref{fig:SPT_1d}b, where red arrows denote $\sigma^z$ values and blue arrows indicate the values of $\tau^x$. The first term in the cluster Hamiltonian ensures that all such DDW configurations satisfy local Gauss law on each link, and the second term (acting like the transverse field on $\sigma$ qubits) induces quantum fluctuations among the domain wall configurations. Consequently, the exact ground state is an equal-weight superposition of all valid decorated domain wall configurations. The resulting ground state is thus a linear superposition of all such DDW states:
\begin{align}
\label{eq:psi_DDW_1d}
    \ket{\Psi_{\mathrm{SPT}}} = \sum_{\text{DDWs}} \ket{\text{DDW}}.
\end{align}
This construction realizes a $\mathbb{Z}_2 \times \mathbb{Z}_2$ SPT phase~\cite{Son_cluster_1d_2012}, where each domain wall of one symmetry sector carries a charge under the other -- a structure we referred to above as domain-wall decoration. 

To contrast this with the trivial phase, we define the trivial Hamiltonian as 
\begin{align}
\label{eq:H_tri_1d}
    H_{\mathrm{trivial}} = -\sum_{j=1}^L \sigma^x_j - \sum_{j=1}^L \tau^x_{j-\half},
\end{align}
whose unique ground state is the trivial product state that can be written in the eigen-basis of $\sigma_j^z$ and $\tau_{j-\half}^z$
\begin{align}
\label{eq:psi_tri_1d}
    \ket{\psi_{\mathrm{trivial}}} = \bigotimes_j \ket{+}_{\sigma_j^x} \otimes \ket{+}_{\tau_{j-\half}^x} = \sum_{\{\ssi_{j}, \st_{j-\half}\}} \ket{\ssi_{j}, \st_{j-\half}},
\end{align}
where the sum runs over all $\sigma_j^z$ and $\tau_{j-\half}^z$ configurations. This state is fully unentangled and symmetric under the same $\mathbb{Z}_2^\sigma \times \mathbb{Z}_2^\tau$ symmetry.

The SPT/cluster ground state is related to the trivial one: $\ket{\Psi_{\mathrm{SPT}}} = U_{\mathrm{DW}} \ket{\Psi_{\mathrm{trivial}}}$  via a symmetric, finite-depth quantum circuit \cite{chen2014symmetry, Scaffidi_gaplessSPT_2017, Li_defect_2024}, illustrated in Fig.~\ref{fig:SPT_1d}c:
\begin{align}
    U_{\mathrm{DW}} & = \prod_{j=1}^{L} \mathrm{CZ}_{\sigma_{j-1}, \tau_{j-\half}} \cdot \mathrm{CZ}_{\sigma_{j}, \tau_{j-\half}} \nonumber\\
    & = \prod_{j=1}^{L} \exp \left[\frac{i \pi}{4} \left(1-\sigma^z_{j-1}\right) \left(1-\tau^z_{j-\half}\right)\right] \exp \left[\frac{i \pi}{4} \left(1-\sigma^z_{j}\right) \left(1-\tau^z_{j-\half}\right)\right],
    \label{eq:UDW-1d}
\end{align}
where $\mathrm{CZ}_{ab}$ is the Controlled-Z gate that introduces a $(-1)$ phase if both qubits $a$ and $b$ are in the $\ket{\downarrow}$ state. The unitary $U_{\mathrm{DW}}$ maps the trivial Hamiltonian $H_{\mathrm{trivial}}$ in \eqref{eq:H_tri_1d} to the cluster Hamiltonian $H_{\mathrm{SPT}}$ in \eqref{eq:H_cluster_1d}.

In the $(\sigma^z, \tau^z)$ basis, the SPT state takes the form
\begin{align}
    \ket{\psi_{\mathrm{SPT}}} = \frac{1}{\sqrt{2^L}} \sum_{\{\ssi_{j}, \st_{j-\half}\}} (-1)^{N_{\downarrow\downarrow}} \ket{\ssi_{j}, \st_{j-\half}},
\end{align}
where $N_{\downarrow\downarrow}$ counts the number of $(\downarrow, \downarrow)$ configurations across each pair of $\sigma$ and $\tau$ qubits connected by a CZ gate. The nontrivial sign structure introduced by $U_{\mathrm{DW}}$ ensures that when one changes basis on the $\tau$-qubits from $\tau^z$ to $\tau^x$, this wavefunction precisely enforces the constraint $\tau^x_{j-\half} = \sigma^z_{j-1} \sigma^z_{j}$, recovering the decorated domain-wall structure in Eq.~\eqref{eq:psi_DDW_1d}. Thus, $U_{\mathrm{DW}}$, known as the domain-wall decoration operator, attaches the charges of one $\mathbb{Z}_2$ symmetry to the domain walls of the other. Moreover, this construction provides an explicit realization of \textit{SPT stacking}: starting from two decoupled trivial symmetric phases ($\sigma$ and $\tau$) in Eq.~\eqref{eq:psi_tri_1d}, the circuit $U_{\mathrm{DW}}$ entangles them into a joint, nontrivial SPT phase protected by the combined $\mathbb{Z}_2\times \mathbb{Z}_2$ symmetry.

Both $H_{\text{SPT}}$ and $H_{\text{trivial}}$ describe gapped phases with short-range correlations and share the same bulk spectrum under periodic boundary conditions. On an open chain, however, $H_{\text{SPT}}$ supports spin-$\half$ edge modes, yielding a fourfold ground-state degeneracy. Within the ground-state subspace, the $\mathbb{Z}_2 \times \mathbb{Z}_2$ symmetry operators factorize as $U_{\sigma,\tau} \sim U^L_{\sigma,\tau} \otimes U^R_{\sigma,\tau}$ with \cite{Li_2023_KT}
\begin{align}
\label{eq:sym_edge_1d}
U^L_\sigma &= \tau^z_{\frac{1}{2}}, \quad  U^R_\sigma = \tau^z_{L-\frac{1}{2}} \sigma^x_L, \nonumber\\
U^L_\tau &= \tau^x_{\frac{1}{2}} \sigma^z_1, \quad U^R_\tau = \sigma^z_L.
\end{align}
At each boundary, the localized symmetry generators anticommute, $U^a_\sigma U^a_\tau = - U^a_\tau U^a_\sigma $ for $ ( a=L,R )$, realizing a projective representation of $\mathbb{Z}_2 \times \mathbb{Z}_2$.
The projective classes are therefore in one-to-one correspondence with elements of
$\mathcal{H}^2[\mathbb{Z}_2 \times \mathbb{Z}_2, U(1)] = \mathbb{Z}_2$, which classifies 1D bosonic SPT phases protected by $\mathbb{Z}_2 \times \mathbb{Z}_2$ symmetry.

\subsubsection{$1$d Kennedy--Tasaki transformation}

Thus far, our discussion has covered two mappings: the KW duality operator, which sends the $\mathbb{Z}_2 \times \mathbb{Z}_2$ phase to the trivial phase, and the DDW operator $U_{\mathrm{DW}}$, which transforms the trivial phase into the $\mathbb{Z}_2 \times \mathbb{Z}_2$ SPT phase.  Next, we discuss the $1$d Kennedy--Tasaki duality operator, which maps the $\mathbb{Z}_2 \times \mathbb{Z}_2$ SSB phase to the SPT phase protected by the same symmetry. Below, we largely follow the exposition in Ref.~\cite{Li_2023_KT}.

From the field-theory viewpoint, the KW transformation corresponds to the so-called $S$-operation, which gauges the $\mathbb{Z}_2 \times \mathbb{Z}_2$ symmetry, while $U_{\mathrm{DW}}$ realizes the so-called $T$-operation, corresponding to stacking the system with a gapped $\mathbb{Z}_2 \times \mathbb{Z}_2$ SPT phase (see Appendix~\ref{app:gauging_KW} and \ref{app:KT_QFT} for details). 
In a general field-theoretic framework, Ref.~\cite{Li_2023_KT} demonstrated that the combined $STS$ transformation implements the KT duality, mapping the $\mathbb{Z}_2 \times \mathbb{Z}_2$ SSB phase to the associated $\mathbb{Z}_2 \times \mathbb{Z}_2$ SPT phase.

In the lattice model, the  $STS$ transformation takes the form $\KT^{1\text{d}} = \NKW^\dagger U_{\text{DW}} \NKW$, which defines the KT duality operator. Its action on the Hilbert-space basis states is as follows \cite{Li_2023_KT}:
\beq
\label{eq:N_KT_1d}
  \KT^{1\text{d}} \ket{\{\ssi_{i}, \st_{i-\half}\}} = \frac{1}{2^{L-1}} \sum_{\{\ssip_i, \stp_{i-\half}\}} (-1)^{\mathcal{A}_\text{KT}(\{\sigma,\sigma';\tau,\tau'\})} \ket{\{\ssip_{i}, \stp_{i-\half}\}},
\eeq
where the configuration-dependent phase factor is given by
\beq
\mathcal{A}_\text{KT}^{1\text{d}}(\{\sigma,\sigma';\tau,\tau'\}) = \sum_{j=1}^L (\ssi_j+\ssip_j) (\st_{j-\half} + \st_{j+\half} + \stp_{j-\half} +\stp_{j+\half}) + (\st_{\half} + \stp_{\half}) (t_\sigma + t_\sigma').
\eeq
It is well defined for a closed boundary condition. To convince ourselves that this operator transforms the SSB into the SPT phase, consider the parent Hamiltonian of the $\mathbb{Z}_2\times\mathbb{Z}_2$ SSB state:
\begin{align}
\label{eq:H_SSB_1d}
    H_{\mathrm{SSB}} = -\sum_{j=1}^L \sigma^z_{j-1} \sigma^z_{j} -\sum_{j=1}^L \tau^z_{j-\half} \tau^z_{j+\half},
\end{align}

As shown in Ref.~\cite{Li_2023_KT}, acting with the $1$d  duality operator $\NKW$ maps $H_{\mathrm{SSB}}$ to the trivial dual Hamiltonian \mbox{$\widehat{H}_{\text{trivial}} = -\sum_{j=1}^L \hsig^z_{j-\half} -\sum_{j=1}^L \htau^x_{j}$} (see Fig.~\ref{fig:SPT_1d}a). Here, $\hsig$ and $\htau$ are the KW dual spins of $\sigma$ and $\tau$, residing on links and sites, respectively. Applying $U_{\text{DW}}$ maps the $\widehat{H}_{\text{trivial}}$ to the SPT Hamiltonian in the dual-spin basis, 
\begin{align}
    \widehat{H}_{\mathrm{SPT}} = -\sum_{j=1}^L \hsig^z_{j-\half} \htau^x_{j} \hsig^z_{j+\half} -\sum_{j=1}^L \htau^z_{j-1} \hsig^x_{j-\half} \htau^z_{j}.
\end{align} 
Finally, applying the operator $\mN_\text{KW}^\dagger$ once more returns the SPT Hamiltonian in terms of the original spin variables, namely 
$H_{\mathrm{SPT}}$ in our Eq.~\eqref{eq:H_cluster_1d}. Fig.~\ref{fig:SPT_1d}d illustrates the sequence of transformations that realize the $1$d KT duality, mapping a $\mathbb{Z}_2 \times \mathbb{Z}_2$ SSB state to the $\mathbb{Z}_2 \times \mathbb{Z}_2$ SPT cluster state.

Although $ \KT^{1\text{d}}$ is unitary on an open chain, it becomes non-unitary on a ring \cite{Li_2023_KT}. The easiest way to see this is by noting that $H_{\mathrm{SSB}}$ exhibits a four-fold degenerate ground state, while $H_{\mathrm{SPT}}$ has a unique ground state, meaning that the $ \KT^{1\text{d}}$ transformation must be non-invertible. Alternatively, note that the algebraic form $\KT^{1\text{d}} = \NKW^\dagger U_{\text{DW}} \NKW$, taken together with the fact that the $\NKW$ transformation is non-invertible on closed boundary conditions, leads to the same conclusion.

For a detailed discussion of the symmetry-twist sector mapping under $ \KT^{1\text{d}}$ and the associated fusion rules that further establish its non-invertibility, see Ref.~\cite{Li_2023_KT}. Here, we briefly summarize the essential structure of this mapping. Consider a state belonging to a symmetry-twist sector labeled by $[(u_\sigma, t_\sigma), (u_\tau, t_\tau)]$. Under the action of the KT transformation, this state is mapped to a new sector labeled by \cite{Li_2023_KT}
\begin{align}
\label{eq:symm-twist-KT}
    [(u_\sigma', t_\sigma'),(u_\tau', t_\tau')] = [(u_\sigma, t_\sigma+u_\tau), (u_\tau, t_\tau+u_\sigma)].
\end{align}

This structure will prove particularly useful when analyzing the corresponding symmetry-twist sector mapping for the weak subsystem KT transformation introduced in the next subsection.

\subsection{$2$d weak $\mathbb{Z}_2 \times \mathbb{Z}_2$ SSPT and weak subsystem KT transformation}
\label{sec:weak_SSPT}
Having reviewed the one-dimensional KT transformation and its role in mapping a $\mathbb{Z}_2\times\mathbb{Z}_2$ SSB phase to a $\mathbb{Z}_2\times\mathbb{Z}_2$ SPT phase, we now extend this framework to construct a weak SSPT phase in two dimensions.
Before proceeding, we briefly review the construction of weak SSPT phases following Refs.~\cite{You_SSPT_2018, Devakul_SSPT_classification_2018}. 
A two-dimensional weak SSPT phase can be constructed by stacking one-dimensional $\mathbb{Z}_2 \times \mathbb{Z}_2$ cluster chains in Eq.~\eqref{eq:H_cluster_1d} along both the horizontal and vertical directions of a square lattice. 
Each lattice site at coordinates $(i,j)$ therefore hosts two independent spin-$1/2$ degrees of freedom: $\sigma_j^{\text{row}}$ belonging to the horizontal chain that runs along row $j$, and $\sigma_i^{\text{col}}$ belonging to the vertical chain that runs along column $i$. Similarly, each horizontal (vertical) link carries $\tau^\text{row}$ ($\tau^\text{col}$) spin that forms part of the horizontal (vertical) cluster state. Because the two sets of chains act on disjoint spins, the total system represents a direct product of $N_{\text{chains}} = L_x + L_y$ decoupled one-dimensional SPTs.

The Hamiltonian of this weak SSPT is given by 
\begin{align}
\label{eq:H_SSPT_weak}
    H_{\text{SSPT}}^\text{weak}
= & -\sum_{i=1}^{L_x}\sum_{j=1}^{L_y}
\left[
\sigma^{z,\row}_{i-1,j}\tau^{x,\row}_{i-\half,j}\sigma^{z,\row}_{i,j} + \tau^{z,\row}_{i-\half,j}\sigma^{x,\row}_{i,j}\tau^{z,\row}_{i+\half,j} \right] \nonumber\\
& -\sum_{i=1}^{L_x}\sum_{j=1}^{L_y} \left[ \sigma^{z,\col}_{i,j-1}\tau^{x,\col}_{i,j-\half}\sigma^{z,\col}_{i,j} + \tau^{z,\col}_{i,j-\half}\sigma^{x,\col}_{i,j}\tau^{z,\col}_{i,j+\half}
\right].
\end{align}

The total symmetry group of this Hamiltonian is a product $(\mathbb{Z}_2\times\mathbb{Z}_2)^{N_{\text{chains}}}$, where each pair of $\mathbb{Z}_2$’s acts as the on-site symmetry for an individual $1$d cluster state:
\begin{align}
\label{eq:subsym_operator_weak}
    U^{\text{row}}_{\sigma,j} = \prod_{i = 1}^{L_x} \sigma^{x,\text{row}}_{i,j}, & \quad U^{\text{col}}_{\sigma,i} = \prod_{j = 1}^{L_y} \sigma^{x,\text{col}}_{i,j}, \nonumber\\
    U^{\text{row}}_{\tau,j} = \prod_{i = 1}^{L_x} \tau^{x,\text{row}}_{i-\half,j}, & \quad U^{\text{col}}_{\tau,i} = \prod_{j = 1}^{L_y} \tau^{x,\text{col}}_{i,j-\half}.
\end{align}
We regard the group generated by these unitary operators as the subsystem $\mathbb{Z}_2\times\mathbb{Z}_2$ symmetry group of the full (2+1)D model. Importantly, although every lattice site lies at the intersection of one row and one column subsystem, each spin is flipped by only one of them. In contrast, within the strong SSPT phase, each spin is flipped under both the horizontal and vertical subsystem symmetries. This feature -- the non-overlapping microscopic action of the subsystem symmetries -- is what distinguishes the weak SSPT from the strong SSPT introduced below in Section~\ref{sec:KT_SSPT_strong}.

The weak SSPT phase inherits its key features from the one-dimensional cluster chains that comprise it. For a system with open boundaries, each $1$d chain terminating at an edge contributes a twofold degeneracy associated with its SPT end modes. Consequently, a system of  size $L_x \times L_y$ exhibits a subextensive ground-state degeneracy that scales as $2^{L_x+L_y}$, 
protected by the subsystem symmetries: any local perturbation preserving all symmetries can only act within an individual chain, leaving the collective edge degeneracy intact.

The structure of the boundary symmetry action in the weak SSPT directly reflects the symmetry fractionalization pattern of the $1$d cluster state discussed earlier in Sec.~\ref{sec:SPT_KT_1d}. Each terminated chain at the boundary carries the same projective representation of $\mathbb{Z}_2\times\mathbb{Z}_2$, characterized by the nontrivial cohomology class in $\mathcal{H}^2[\mathbb{Z}_2\times\mathbb{Z}_2,U(1)] = \mathbb{Z}_2$, as established in Eq.~\eqref{eq:sym_edge_1d}. In the weak SSPT, every subsystem chain contributes an independent copy of this projective representation, and the total boundary Hilbert space forms a tensor product of these decoupled projective doublets. Since the subsystem symmetries associated with different rows and columns act on disjoint spins, all such projective representations commute, ensuring the robustness of the edge manifold.
This stacked realization of symmetry fractionalization embodies the essential nature of the weak SSPT: it is a direct product of lower-dimensional SPT chains, each carrying the same $\mathbb{Z}_2\times\mathbb{Z}_2$ projective class, yet lacking any intrinsic two-dimensional entanglement between subsystems.

Having discussed the weak SSPT, we now introduce the weak subsystem KT transformation as a product of $1$d KT transformations acting on every row and every column,
\begin{align}
\KT^{\text{weak}} = \left(\prod_{i=1}^{L_x}  \KT^{(\col,i)} \right) \left(\prod_{j=1}^{L_y}  \KT^{(\row,j)} \right).
\end{align}
where each $\KT^{(\row,j)}$ ( $\KT^{(\col,i)}$, respectively) acts only on the $1$d row (column) chain according to Eq.~\eqref{eq:N_KT_1d}. Importantly, this mapping does not couple the two directions: each row (column) transforms according to its own one-dimensional KT rule. Consequently, the subsystem KT transformation factorizes into a product of decoupled one-dimensional mappings acting along the two orthogonal sets of subsystem lines.

The transformation  ${\KT^{\text{weak}}}^\dagger$ maps the weak SSPT Hamiltonian in Eq.~\eqref{eq:H_SSPT_weak} to a system of decoupled Ising chains harboring the SSSB ground state:
\begin{align}
\label{eq:KT_SSSB}
H_{\text{SSSB}}^\text{weak}
= & -\sum_{i=1}^{L_x}\sum_{j=1}^{L_y}
\left[
\sigma^{z,\row}_{i-1,j}\sigma^{z,\row}_{i,j} + \tau^{z,\row}_{i-\half,j}\tau^{z,\row}_{i+\half,j} \right] \nonumber\\
& -\sum_{i=1}^{L_x}\sum_{j=1}^{L_y} \left[ \sigma^{z,\col}_{i,j-1}\sigma^{z,\col}_{i,j} + \tau^{z,\col}_{i,j-\half}\tau^{z,\col}_{i,j+\half}
\right].
\end{align}

This model enjoys the subsystem $\mathbb{Z}_2\times\mathbb{Z}_2$ symmetry whose generators are defined in Eq.~\eqref {eq:subsym_operator_weak}. Since the subsystem KT transformation acts independently along each one-dimensional subsystem line, the correspondence between symmetry and twist sectors follows the same rule as in the one-dimensional case [see Eq.~\eqref{eq:symm-twist-KT}], applied separately to each row and each column. The mapping thus preserves the separability of the subsystem directions: symmetry-twist data associated with rows and columns evolve independently under the transformation, reflecting its one-dimensional character. The overall transformation is non-invertible on a torus -- being a product of non-invertible $1$d KT maps -- but becomes unitary on an open rectangular geometry, where it preserves the spectrum of the system chain by chain.

At this point, it is useful to distinguish between the SSSB ground state realized in a system of \emph{decoupled Ising chains} and the SSSB phase appearing in the \emph{Xu--Moore} model (a.k.a. \emph{Ising plaquette} model in transverse field), discussed in Sec.~\ref{sec:subsym_square} for the $\mathbb{Z}_2$ case and in Sec.~\ref{sec:KT_sub} for the $\mathbb{Z}_2 \times \mathbb{Z}_2$ generalization. 
First, the subsystem symmetry generators of the decoupled Ising wires in Eq.~\eqref{eq:subsym_operator_weak} and those of the Ising plaquette model in Eq.~\eqref{eq:subsym_operator} are distinct, as previously discussed (see the discussion below Eq.~\eqref{eq:subsym_operator_weak}). Second, the symmetry operators in the Ising plaquette model are subject to the constraint in Eq.~\eqref{eq:subsym_operator_contraint}, whereas those in the decoupled Ising wires are unconstrained. Third, Ref.~\cite{Brandon_subsystem_Scipost2023} demonstrated that these two SSSB phases are genuinely different: by introducing and tuning inter-wire coupling terms, one can drive a phase transition from the decoupled Ising chain phase to two copies of the Ising plaquette model. 
To see this explicitly, consider adding inter-wire coupling terms to the decoupled Ising chain Hamiltonian [Eq.~\eqref{eq:KT_SSSB}]:
\begin{align}
    H_C = - \sum_{i=1}^{L_x}\sum_{j=1}^{L_y} K_\sigma \sigma^{x,\row}_{i,j} \sigma^{x,\col}_{i,j} - \sum_{i=1}^{L_x}\sum_{j=1}^{L_y} K_\tau \tau^{x,\row}_{i-\half,j} \tau^{x,\col}_{i,j-\half}.
\end{align}
Perturbation theory up to fourth order shows~\cite{Brandon_subsystem_Scipost2023} that the resulting effective Hamiltonian coincides exactly with two copies of the Ising plaquette model in Eq.~\eqref{eq:H_SSSB}. The effective Pauli operators in Eq.~\eqref{eq:H_SSSB} can be expressed in terms of the microscopic wire operators as follows \cite{Brandon_subsystem_Scipost2023}:
\begin{align}
    \sigma^z_{i,j} & = \sigma^{z,\row}_{i,j} \sigma^{z,\col}_{i,j}, \quad \sigma^x_{i,j} = \sigma^{x,\row}_{i,j} = \sigma^{z,\col}_{i,j}, \nonumber\\
    \tau^{z}_{i-\half,j-\half} & =\tau^{z,\row}_{i-\half,j} \tau^{z,\col}_{i,j-\half}, \quad \tau^{x}_{i-\half,j-\half}  =\tau^{x,\row}_{i-\half,j} = \tau^{x,\col}_{i,j-\half}.
\end{align}

Another clear distinction between the two SSSB phases lies in the structure of their excitations. In the decoupled Ising chains, excitations correspond to domain walls on individual wires, whereas in the Ising plaquette model, flipping a single spin creates four fracton-like defective plaquettes for each spin species, as discussed in Sec.~\ref{sec:Xu-Moore_closed}.

\section{$2$d strong $\mathbb{Z}_2 \times \mathbb{Z}_2$ SSPT}
\label{sec:SSPT_2d}
Having discussed the 1D cluster Hamiltonian in Sec.~\ref{sec:SPT_KT_1d} and the two-dimensional weak SSPT in Sec.~\ref{sec:weak_SSPT}, we now develop a systematic formulation of the \emph{strong} SSPT on the square lattice. This section is organized as follows. In Sec.~\ref{sec:2D_cluster}, we introduce the two-dimensional cluster Hamiltonian on a lattice with closed boundary conditions, which realizes a strong SSPT phase. In Sec.~\ref{sec:SSPT_edge}, we consider the open-boundary case and show how the global subsystem symmetry generators fractionalize into local boundary operators along the edge of the square lattice. We further demonstrate that these boundary operators satisfy a nontrivial anticommutation algebra.

\subsection{2D cluster Hamiltonian}
\label{sec:2D_cluster}

\begin{figure}[tb]
    \centering
    \includegraphics[width=0.75\linewidth]{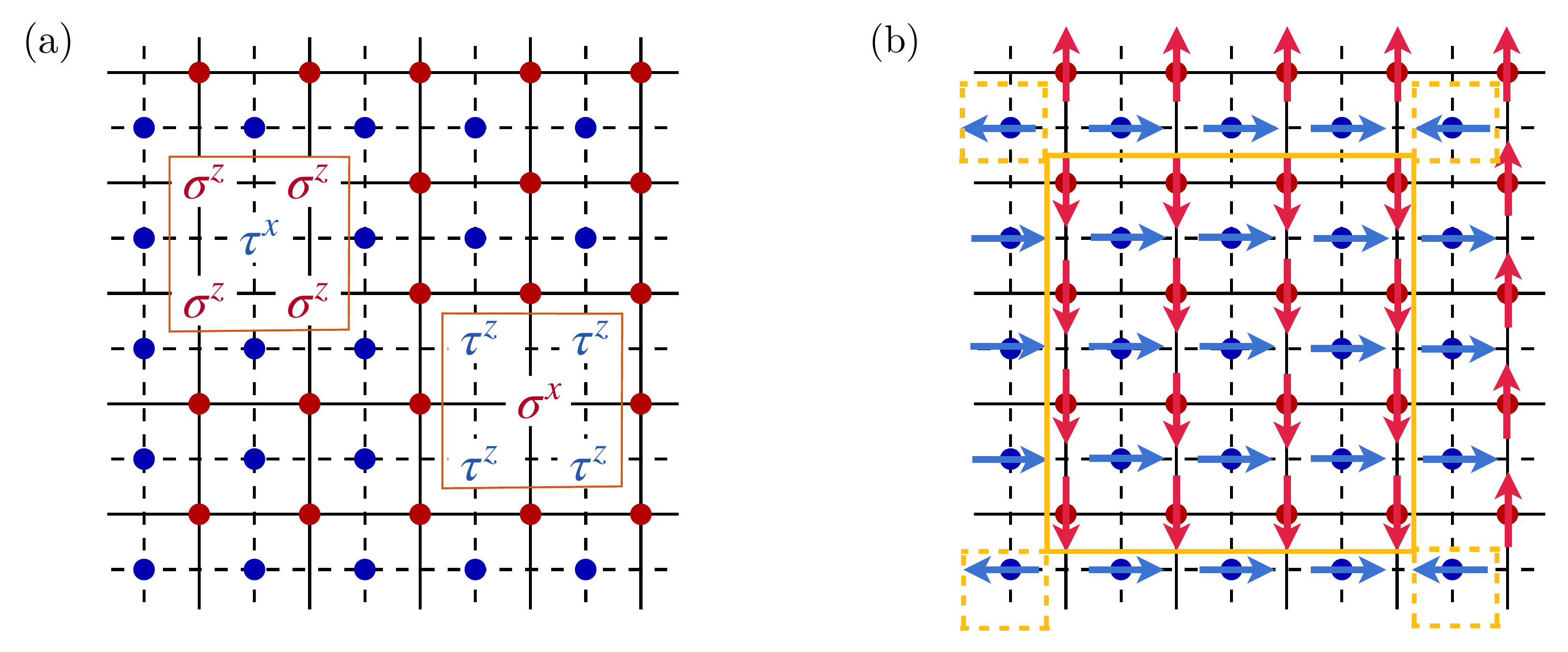}
    \caption{(\textbf{a}) The Pauli spin $\ssi_{i,j}$ resides on the red sites, while the Pauli spin $\tau_{i-\half,j-\half}$ is placed on the blue sites. It is important to emphasize that the blue $\st_{i-\half,j-\half}$ spins are part of the Hilbert space and are distinct from the dual space degrees of freedom $\hssi_{i-\half,j-\half}$ in the context of the subsystem KW transformation, shown in green in Fig.~\ref{fig:KW_duality}. 
    (\textbf{b}) A representative decorated domain wall configuration is illustrated. Solid yellow lines indicate domain walls between sites with opposite $\sigma^z$ values, while dashed yellow squares mark the corners where the domain wall is decorated $\tau^x=-1$ \cite{Zhou_Detecting_SSPT_PRB2022}.}
    \label{fig:SSPT_2d}
\end{figure}

We begin by reviewing the essential ingredients of the strong SSPT construction, following the framework introduced in Refs.~\cite{You_SSPT_2018, Devakul_SSPT_classification_2018}. We now turn to a closed square lattice of sizes $L_x \times L_y$ arranged on a torus, where each site is labeled by coordinates $(i,j)$ with $i = 1, \ldots, L_x$ and $j = 1, \ldots, L_y$. As in the previous section, at each site resides a spin-$1/2$ degree of freedom, described by a two-dimensional local Hilbert space with basis states $\ket{\ssi_{i,j} =  0,1}$. In addition, the center of every plaquette hosts a spin-$1/2$ degree of freedom, with its own two-dimensional local Hilbert space spanned by the basis states $|\st_{i-\half,j-\half}= 0,1\rangle$. We place the $\ssi_{i,j}$ spins on the red sites (vertices), and the $\st_{i-\half,j-\half}$ spins on the blue sites (plaquettes), as shown in Fig.~\ref{fig:SSPT_2d}a. Consequently, every unit cell contains two spin-$1/2$ degrees of freedom. This is in sharp contrast to the  Ising plaquette model discussed in Sec.~\ref{sec:subsym_KW}, where a unit cell carried only one spin-$1/2$, positioned either on a lattice site or at the center of a plaquette, but never on both. The local Hilbert space at each spin-$1/2$ degree of freedom is acted upon by the standard Pauli operators,
\begin{align}
\label{eq:sigma_pauli_tau}
    \sigma^z_{i,j} \ket{\ssi_{i,j}}  = (-1)^{\ssi_{i,j}} \ket{\ssi_{i,j}}, & \quad \sigma^x_{i,j} \ket{\ssi_{i,j}} = \ket{1-\ssi_{i,j}}, \nonumber\\
    \tau^z_{i-\half,j-\half} \ket{\st_{i-\half,j-\half}}  = (-1)^{\st_{i-\half,j-\half}} \ket{\st_{i-\half,j-\half}},  & \quad \tau^x_{i-\half,j-\half} \ket{\st_{i-\half,j-\half}} = \ket{1-\st_{i-\half,j-\half}}.
\end{align}

The Hamiltonian of the $2$d cluster state, whose ground state realizes a strong SSPT phase, is given by \cite{Raussendorf_MBQC_2002,cluster_state_Raussendorf_Bravyi2005, Devakul_SSPT_classification_2018}
\begin{align}
\label{eq:H_SSPT_2d}
    H_{\text{SSPT}}  = & - \sum_{i=1}^{L_x} \sum_{j=1}^{L_y} \sigma^z_{i-1,j-1} \sigma^z_{i,j-1} \sigma^z_{i-1,j} \sigma^z_{i,j} \tau^x_{i-\half,j-\half} \nonumber \\
    & -  \sum_{i=1}^{L_x} \sum_{j=1}^{L_y} \tau^z_{i-\half,j-\half} \tau^z_{i+\half,j-\half} \tau^z_{i-\half,j+\half} \tau^z_{i+\half,j+\half} \sigma^x_{i,j}.
\end{align}

As shown in Ref.~\cite{You_SSPT_2018}, the $2$d cluster model has the subsystem $\mathbb{Z}_2 \times \mathbb{Z}_2$ symmetry generated by the operators
\begin{align}
\label{eq:subsym_operator_sig_tau}
    U^x_{\sigma,j} = \prod_{i = 1}^{L_x} \sigma^x_{i,j}, & \quad U^y_{\sigma,i} = \prod_{j = 1}^{L_y} \sigma^x_{i,j}, \nonumber\\
    U^x_{\tau,j-\half} = \prod_{i = 1}^{L_x} \tau^x_{i-\half,j-\half}, & \quad U^y_{\tau,i-\half} = \prod_{j = 1}^{L_y} \tau^x_{i-\half,j-\half}.
\end{align}

In order to understand the ground state of the cluster Hamiltoninan, we build on the discussion of the $1$d $\mathbb{Z}_2 \times \mathbb{Z}_2$ SPT presented in Sec.~\ref{sec:SPT_KT_1d}. First, note that all terms in Eq.~\eqref{eq:H_SSPT_2d} commute. Hence, the ground state of the $2$d cluster model, $\ket{\psi_{\text{SSPT}}}$, is the common $+1$ eigenstate of all stabilizers. With periodic boundary conditions, the unique ground state can be explicitly constructed as follows. In Hamiltonian \eqref{eq:H_SSPT_2d}, the first term enforces the local Gauss law $\sigma^z_{i-1,j-1} \sigma^z_{i,j-1} \sigma^z_{i-1,j} \sigma^z_{i,j} =\tau^x_{i-\half,j-\half}$. Hence, the role of the first term of the Hamiltonian in \eqref{eq:H_SSPT_2d} is to ensure that the ground state realizes the decorated domain wall configuration, of which an example is depicted in Fig.~\ref{fig:SSPT_2d}b. The second term effectively flips the $\sigma^z$ spin, while simultaneously flipping the surrounding four $\tau^x$ spins. Thus, the ground state of $H_{\text{SSPT}}$ in \eqref{eq:H_SSPT_2d} can be viewed as a coherent superposition of all $\sigma^z$ domain-wall patterns such as depicted in Fig.~\ref{fig:SSPT_2d}b, with blue $\tau^x$ spins furnishing the corresponding decorations.

As in Sec.~\ref{sec:SPT_KT_1d}, one can define a domain wall decoration operator $U_{\text{DW}}$ that generates the SSPT state starting from the trivial state. Its explicit expression reads
\begin{align}
\label{eq:U_DW_SSPT}
    U_{\text{DW}}  =  & \prod_{i=1}^{L_x} \prod_{j=1}^{L_y} \exp \left[\frac{i \pi}{4} \left(1-\sigma^z_{i-1,j-1}\right) \left(1-\tau^z_{i-\half,j-\half}\right)\right] \exp \left[\frac{i \pi}{4} \left(1-\sigma^z_{i,j-1}\right) \left(1-\tau^z_{i-\half,j-\half}\right)\right] \nonumber \\
    & \times \exp \left[\frac{i \pi}{4} \left(1-\sigma^z_{i-1,j}\right) \left(1-\tau^z_{i-\half,j-\half}\right)\right] \exp \left[\frac{i \pi}{4} \left(1-\sigma^z_{i,j}\right) \left(1-\tau^z_{i-\half,j-\half}\right)\right].
\end{align}
This can be understood as a composition of four Controlled-Z operations between a given $\tau_{i-\half,j-\half}$ at the plaquette-center and each of the four corners ($\sigma$'s) surrounding it, in direct analogy to the 1-dimensional DW transformation in Eq.~\eqref{eq:UDW-1d}.

With periodic boundary conditions, the Hamiltonian $H_{\text{SSPT}}$ in \eqref{eq:H_SSPT_2d} admits a unique ground state. However, when defined on an open lattice, it supports spin-$\tfrac{1}{2}$ edge excitations that remain gapless, resulting in a sub-extensive ground-state degeneracy of $4^{L_x+L_y-1}$. A detailed discussion of the open geometry and its edge structure can be found in the next subsection.

\subsection{At the edge}
\label{sec:SSPT_edge}
On an open square lattice, the Hamiltonian $H_{\text{SSPT}}$ in Eq.~\eqref{eq:H_SSPT_2d} takes the explicit form
\begin{align}
\label{eq:H_SSPT_open}
    H_{\text{SSPT}}^{\text{open}}  = & - \sum_{i=2}^{L_x} \sum_{j=2}^{L_y} \sigma^z_{i-1,j-1} \sigma^z_{i,j-1} \sigma^z_{i-1,j} \sigma^z_{i,j} \tau^x_{i-\half,j-\half} \nonumber\\
    & -  \sum_{i=1}^{L_x-1} \sum_{j=1}^{L_y-1} \tau^z_{i-\half,j-\half} \tau^z_{i+\half,j-\half} \tau^z_{i-\half,j+\half} \tau^z_{i+\half,j+\half} \sigma^x_{i,j}.
\end{align}

This model is exactly solvable, since all the terms in Eq.~\eqref{eq:H_SSPT_open} mutually commute. The ground state is therefore the simultaneous eigenstate of these commuting operators -- so-called stabilizers of the cluster state -- with eigenvalue $+1$ (we use the $\sim$ symbol to denote operator relationships within the ground state manifold):
\begin{align}
\label{eq:GS_SSPT}
    \sigma^z_{i-1,j-1} \sigma^z_{i,j-1} \sigma^z_{i-1,j} \sigma^z_{i,j} \tau^x_{i-\half,j-\half} & \sim 1 \, \left(i=2,\dots,L_x,\, j=2,\dots,L_y \right), \nonumber\\
    \tau^z_{i-\half,j-\half} \tau^z_{i+\half,j-\half} \tau^z_{i-\half,j+\half} \tau^z_{i+\half,j+\half} \sigma^x_{i,j} & \sim 1 \, \left(i=1,\dots,L_x-1,\, j=1,\dots,L_y-1 \right).
\end{align}

\begin{figure}[tb]
    \centering
    \includegraphics[width=0.9\linewidth]{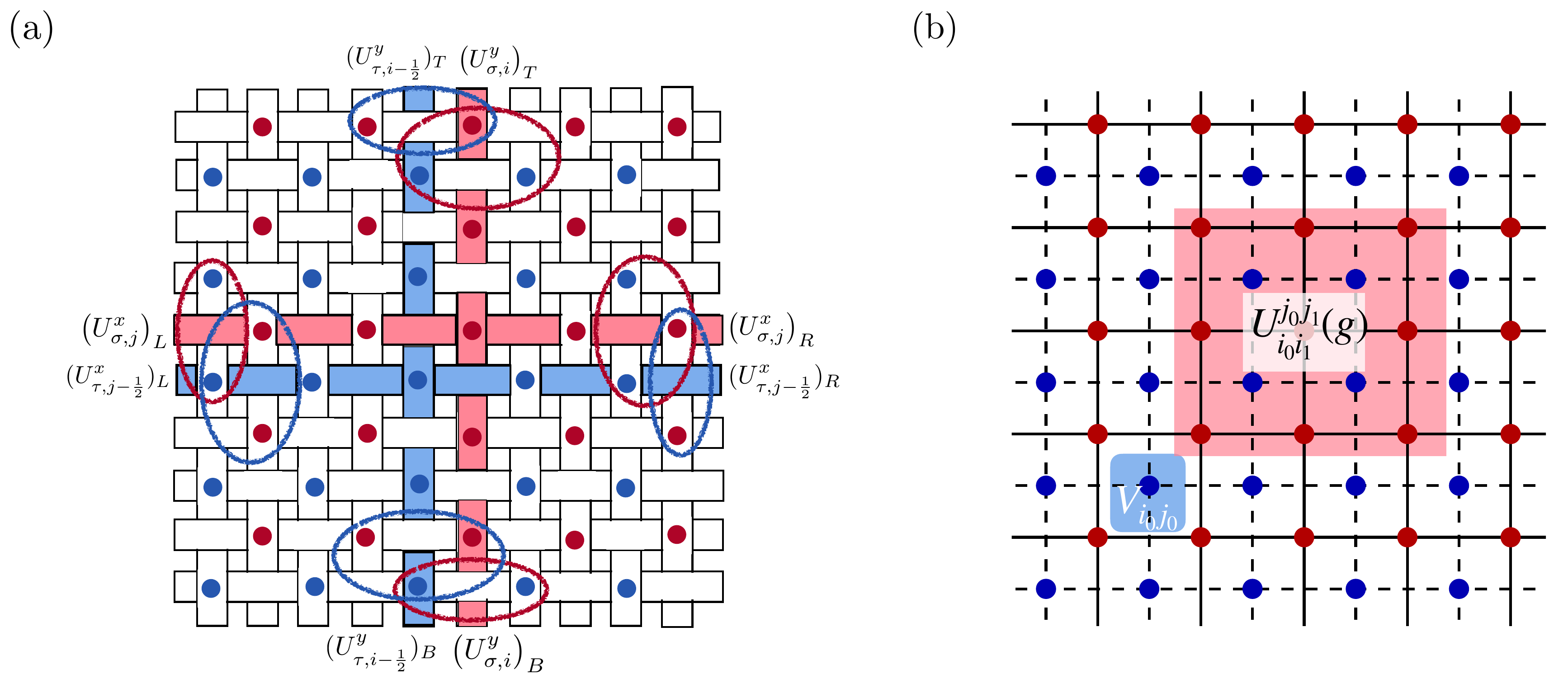}
    \caption{(\textbf{a}) Illustration of the fractionalization of subsystem symmetry generators: for the $\sigma$ spins, the symmetry acts along the $j$-th row (horizontal red strip) and the $i$-th column (vertical red strip); for the $\tau$ spins, it acts along the $(j-\tfrac{1}{2})$-th row (horizontal blue strip) and the $(i-\tfrac{1}{2})$-th column (vertical blue strip). The red ellipses indicate the corresponding edge operators for the $\sigma$ spins, while the blue ellipses indicate the edge operators for the $\tau$ spins, as defined in Eq.~\eqref{eq:edge_SSPT}. (\textbf{b}) Illustration of the truncated symmetry operator $U_{i_0i_1}^{j_0j_1}$ and local repair (corner) operator $V_{i_0 j_0}$.}
    \label{fig:Edge_SSPT}
\end{figure}

\paragraph*{Symmetry fractionalization in the SSPT:} 
Applying these relations, the subsystem $\mathbb{Z}_2 \times \mathbb{Z}_2$ symmetry generators, as defined in Eq.~\eqref{eq:subsym_operator_sig_tau}, can be equivalently represented \textit{within the ground-state subspace} as (see Fig.~\ref{fig:Edge_SSPT}a)
\begin{align}
\label{eq:symm_fractionalization_SSPT}
    U^x_{\sigma,j} & \sim \left(U^x_{\sigma,j}\right)_L \otimes \left(U^x_{\sigma,j}\right)_R, \nonumber\\
    U^y_{\sigma,i} & \sim \left(U^y_{\sigma,i}\right)_B \otimes \left(U^y_{\sigma,i}\right)_T, \nonumber\\ 
    U^x_{\tau,j-\half} & \sim \left(U^x_{\tau,j-\half}\right)_L \otimes \left(U^x_{\tau,j-\half}\right)_R, \\
    U^y_{\tau,i-\half} & \sim \left(U^y_{\tau,i-\half}\right)_B \otimes \left(U^y_{\tau,i-\half}\right)_T. \nonumber
\end{align}
Physically, this means that the bulk symmetry generators, when acting on the ground state subspace, \textit{fractionalize} into an action on the edges. Consider, for instance, the first equation above -- it states that the subsystem symmetry generator in the row $j$ 
acts as a product of two operators localized on the left and the right edge, respectively.
Concretely, the edge operators appearing on the right-hand side of Eq.~\eqref{eq:symm_fractionalization_SSPT} are defined as follows (see Fig.~\ref{fig:Edge_SSPT}a):
\begin{align}
\label{eq:edge_SSPT}
    \left(U^x_{\sigma,j}\right)_L & = \tau^z_{\half, j-\half} \tau^z_{\half, j+\half},  \, \left(U^x_{\sigma,j}\right)_R = \tau^z_{L_x-\half, j-\half} \tau^z_{L_x-\half, j+\half} \sigma^x_{L_x,j}, \, (j=1,\dots,L_y-1), \nonumber\\
    \left(U^y_{\sigma,i}\right)_B & = \tau^z_{i-\half,\half} \tau^z_{i+\half,\half},  \, \left(U^y_{\sigma,i}\right)_T = \tau^z_{i-\half,L_y-\half} \tau^z_{i+\half,L_y-\half} \sigma^x_{i,L_y}, \, (i=1,\dots,L_x-1),\nonumber\\
    \left(U^x_{\tau,j-\half}\right)_L & = \tau^x_{\half,j-\half} \sigma^z_{1,j-1} \sigma^z_{1,j},  \, \left(U^x_{\tau,j-\half}\right)_R = \sigma^z_{L_x,j-1} \sigma^z_{L_x,j}, \, (j=2,\dots,L_y), \nonumber\\
    \left(U^y_{\tau,i-\half}\right)_B & = \tau^x_{i-\half,\half} \sigma^z_{i-1,1} \sigma^z_{i,1},  \, \left(U^y_{\tau,i-\half}\right)_T = \sigma^z_{i-1,L_y} \sigma^z_{i,L_y}\, (i=1,\dots,L_x).
\end{align}
Here, $B, T, L, R$ indicate the bottom, top, left, and right corners of each subsystem symmetry generators, respectively. In the ground-state subspace, the symmetry generators therefore act only on localized modes along the boundary of the open square lattice. These boundary operators obey the Pauli algebra and at the same time commute with all the bulk stabilizers of the Hamiltonian $H_{\text{SSPT}}^{\text{open}}$. To illustrate, consider the left edge of the open system, where the structure of the boundary symmetry anticommutation algebra becomes most transparent,
\begin{align}
\label{eq:edge_anticommutation}
    \left(U^x_{\sigma,j}\right)_L (U^x_{\tau,j-\half})_L & = - (U^x_{\tau,j-\half})_L \left(U^x_{\sigma,j}\right)_L,\nonumber\\
    \left(U^x_{\sigma,j}\right)_L (U^x_{\tau,j+\half})_L & = - (U^x_{\tau,j+\half})_L \left(U^x_{\sigma,j}\right)_L.
\end{align}

This nearest-neighbor anticommutation arises from the local overlap of $\tau^z$ and $\tau^x$ operators on the shared boundary plaquettes as shown in Fig.~\ref{fig:Edge_SSPT}a. As a result, the $\sigma$- and $\tau$-type generators along the edge form a one-dimensional chain with a bipartite anticommutation pattern: each operator (say, $\sigma$) anticommutes only with its two adjacent operators of the opposite type (in this case, $\tau$) and otherwise commutes with the rest.

The same structure appears along the right, top, and bottom boundaries. Each edge thus supports an identical sequence of interlaced $\sigma$- and $\tau$-type operators linked by nearest-neighbor anticommutation. Counting the independent boundary modes, which amount to $2(L_x+L_y-1)$, shows that they generate a $4^{L_x+L_y-1}$-fold ground-state degeneracy, originating from the boundary of length $2(L_x+L_y-1)$. Consequently, the $4^{L_x+L_y-1}$-fold edge degeneracy on the boundary of this SSPT directly accounts for the $4^{L_x+L_y-1}$-fold degeneracy of the original SSSB phase.

\paragraph*{Projective representations of the subsystem symmetries:}
The algebra in Eq.~\eqref{eq:edge_anticommutation} shows that the boundary symmetry structure of the two-dimensional SSPT is intrinsically nontrivial:
the operator $\left(U^x_{\sigma,j}\right)_L$ not only anticommutes with $(U^x_{\tau,j-\half})_L$ from the row below, but also with $(U^x_{\tau,j+\half})_L$ from the row above.
Hence, each boundary symmetry operator simultaneously links to two neighboring subsystem symmetries, producing an extended pattern of mutual anticommutation along the entire edge -- sort of like the warps in a weave connect two neighbouring wefts.
Collectively, they form a non-factorizable projective representation of the subsystem symmetry group that extends continuously along the entire perimeter. In graphical terms, this boundary algebra can be viewed as a \textit{chain-like anticommutation graph} along each edge -- not a collection of isolated pairs, but an extended structure of alternating operators linked by their mutual anticommutation. It is this interlinked nature of the projective representation that holds the fabric of the strong SSPT together -- in juxtaposition to the weak SSPT studied in Sec.~\ref{sec:weak_SSPT} whose operators only anticommute in a given row or column, allowing the action of the symmetry operator to be confined to one thread.

This intertwined edge algebra marks a fundamental difference between the strong and weak SSPT phases.
In the weak SSPT, constructed by stacking decoupled one-dimensional cluster chains, each terminated chain contributes an independent pair of anticommuting edge operators: e.g. $U^x \sim U^x_L \otimes U^x_R$, thus realizing a tensor product of localized projective representations of $\mathbb{Z}_2\times\mathbb{Z}_2$ symmetry. In contrast, for the strong SSPT, the boundary symmetry generators are interlinked across neighboring sites and cannot be decomposed into independent $1$d doublets. The resulting edge representation is intrinsically collective and encodes a genuinely two-dimensional projective structure that cannot arise from any stack of $1$d SPTs.
This distinction between independently fractionalized $1$d edges in the weak SSPT and the weave-like, entangled anticommutation network of the strong SSPT provides a sharp diagnostic separating the two phases.

\section{Bulk and edge invariants}
\label{sec:invariants}
Above, we distinguished between weak and strong SSPT phases. Weak SSPTs, defined in Sec.~\ref{sec:KT_closed}, may be viewed as stacks of decoupled one-dimensional SPT chains protected by line-like subsystem symmetries, whereas strong SSPTs, such as the square-lattice cluster model in Eq.~\eqref{eq:H_SSPT_2d}, cannot be decomposed into such independent $1$d constituents, as is evident from the structure of the edge modes studied in the previous section~\ref{sec:SSPT_edge}.
While this distinction is physically intuitive, it had remained unclear how to formalize it in a general setting.
Ref.~\cite{Devakul_SSPT_classification_2018} introduced a precise criterion by defining an equivalence relation between two SSPT phases: two states are said to be LSLU equivalent if they can be connected through a finite-depth local unitary evolution that preserves all subsystem symmetries along each line. Under this equivalence, any weak SSPT -- being a simple stack of $1$d SPT chains -- can be smoothly deformed to a trivial product state, while a strong SSPT cannot. The resulting LSLU classification identifies distinct equivalence classes of strong SSPT phases with the non-trivial elements of the cohomology coset \cite{Devakul_SSPT_classification_2018}
\begin{align}
    \mathcal{C}[G_s] = \mathcal{H}^2[G_s^2,U(1)] / (\mathcal{H}^2[G_s,U(1)])^3,
\end{align}
where $G_s$ is the onsite symmetry group associated with the line-like subsystem symmetries and $\mathcal{H}^2[G_s,U(1)]$ denotes the second cohomology group of $G_s$ with coefficients in $U(1)$ (for a detailed review of the cohomology construct as it pertains to SPT classification, see Ref.~\cite{Chen_SPTcohom_PRB2013}). For our case, $G_s=\mathbb{Z}_2\times\mathbb{Z}_2$ and $\mathcal{C}(\mathbb{Z}_2\times\mathbb{Z}_2 )=\mathbb{Z}_2\times\mathbb{Z}_2\times\mathbb{Z}_2$ \cite{Devakul_SSPT_classification_2018}, meaning that there is a total of 8 equivalence classes of SSPTs (one of which is a weak SSPT that is LSLU equivalent to a trivial state). Each distinct class can be characterized by a set of \textit{bulk} and \textit{edge} invariants that remain unchanged under any LSLU transformation. The bulk invariant $\beta(g)$ introduced in Ref.~\cite{Devakul_SSPT_classification_2018} measures the commutation phase between a half-infinite subsystem symmetry and a truncated rectangular symmetry operator, while the corresponding edge invariant $\phi_{\mathrm{top}}$ encodes the projective algebra of the effective symmetry operators localized along the system’s boundary.

In the remainder of this section, we briefly review these invariants and explicitly compute $\beta(g)$ and $\phi_{\mathrm{top}}$ for both the weak and strong SSPT constructions following Ref.~\cite{Devakul_SSPT_classification_2018}. This section will be useful in Sec.~\ref{sec:bulk-edge}, where we show that the subsystem KT duality transformation preserves the values of these invariants.

\subsection{Bulk invariants}
\label{sec:bulk_invariants}
In Ref.~\cite{Devakul_SSPT_classification_2018}, the bulk invariant $\beta(g)$ was formulated on an infinite square lattice, where the truncated symmetry operations and the associated corner unitaries can be defined without boundary effects. Here, however, we focus on a finite $L_x\times L_y$ square lattice with open boundaries. Working on a finite lattice will prove essential when analyzing the effect of the subsystem KT transformation, which is a strictly unitary operation well-defined only in the presence of physical edges.

We begin by introducing a few notational conventions that will be used throughout this section. The on-site subsystem symmetry group is $G_s=\mathbb{Z}_2\times\mathbb{Z}_2 = \{1,g_\sigma,g_\tau,g_\sigma g_\tau\}$, where $g_\sigma$ and $g_\tau$ are the two generators of the group. Their local representations act as
\begin{align}
    u_{i,j} (g_\sigma) = \sigma_{i,j}^x, \quad  u_{i,j} (g_\tau) = \tau_{i-\half,j-\half}^x.
\end{align}
In this notation, the subsystem $\mathbb{Z}_2\times\mathbb{Z}_2$ symmetry generators introduced in Eq.~\eqref{eq:subsym_operator_sig_tau} take the compact form
\begin{align}
\label{eq:subsym_operator_general}
    U^x_{j} (g) = \prod_{i = 1}^{L_x} u_{i,j}(g), & \quad  U^y_{i} (g) = \prod_{j = 1}^{L_y} u_{i,j}(g).
\end{align}

Following Ref.~\cite{Devakul_SSPT_classification_2018}, the bulk invariant $\beta(g)$ can be defined using a truncated symmetry operator that applies the subsystem symmetry to a finite rectangular region of the lattice $[i_0,i_1] \times[j_0,j_1]$ (see Fig.~\ref{fig:Edge_SSPT}b):
\begin{align}
    U_{i_0i_1}^{j_0j_1}(g) = \prod_{i=i_0}^{i_1} \prod_{j=j_0}^{j_1} u_{i,j}(g).
\end{align}

Because the symmetry is applied only inside this region, the operator $U_{i_0i_1}^{j_0j_1} (g)$ acting on the ground state $\ket{\psi}$ creates local excitations confined to the four corners of the rectangle: bottom-left (BL), bottom-right (BR), top-right (TR), and top-left (TL).
These excitations can be removed by local unitaries $V_{i,j}(g)$ supported near each corner, which satisfy the following relation (see Fig.~\ref{fig:Edge_SSPT}b) \cite{Devakul_SSPT_classification_2018}
\begin{align}
    V_{ij}(g) \equiv V_{ij}^{BL}(g) = V_{ij}^{TL}(g^{-1}) =  V_{ij}^{TR}(g) = V_{ij}^{BR}(g^{-1}),
\end{align}
leading to the consistency condition \cite{Devakul_SSPT_classification_2018}
\begin{align}
    V_{i_0j_0}(g)
    V_{i_0j_1}(g^{-1})
    V_{i_1j_1}(g)
    V_{i_1j_0}(g^{-1})
    U^{j_0j_1}_{i_0i_1}(g) \ket{\psi} = \ket{\psi}.
\end{align}
The above equation means that the local unitaries $V_{ij}(g)$ transform projectively under $G_s$.

We now introduce the corresponding half-space symmetry operators, defined by extending the subsystem symmetry to the semi-infinite region on one side of a chosen coordinate, say to the right (indicated by the superscript $R$) of the $i$th column:
\begin{align}
    \mathcal{S}_i^R (g) = \prod_{i'=i}^{L_x} \prod_{j=1}^{L_y} u_{i',j}(g). 
\end{align}
Analogous half-space operators $\mathcal{S}_i^L (g)$, $\mathcal{S}_j^T (g)$, and $\mathcal{S}_j^B (g)$ can be defined for the left, top, and bottom regions, respectively \cite{Devakul_SSPT_classification_2018}; in what follows, we focus primarily on $\mathcal{S}_i^R (g)$.

The associated commutation phase between a half-space symmetry operator and a local corner operator defines the invariant
\begin{align}
\label{eq:beta_def}
    \beta_{ij}^{R}(g) &= {}_{\text{\smaller{SSPT}}}\bra{\psi} \mathcal{S}^{\dagger R}_{i}(g)V_{ij}^{\dagger}(g) \mathcal{S}^{R}_{i}(g) V_{ij}(g)\ket{\psi}_{\text{\smaller{SSPT}}}.
\end{align}
This  U$(1)$ phase is independent of the choice of corner or orientation and cannot be modified by the LSLU circuit. Hence, the set of $\beta_{ij}^{R}(g)$ invariants for the elements of the group $g\in G_s$ serves as a bulk topological invariant distinguishing inequivalent strong SSPT phases (the case of $\beta(g)=1\; \forall g\in G_s$ corresponds to the trivial, weak SSPT).

\paragraph{Evaluation for the $2$d cluster model:}
For the $2$d strong SSPT (cluster-state) Hamiltonian in Eq.~\eqref{eq:H_SSPT_2d}, a convenient choice of local repair operators is
\begin{align}
\label{eq:repair_operator}
    V_{ij}(g_\sigma) = \tau^z_{i-\half, j-\half}, \quad V_{ij}(g_\tau) = \sigma^z_{i-1,j-1}, \quad V_{ij}(g_\sigma g_\tau) = \tau^z_{i-\half, j-\half} \sigma^z_{i-1,j-1}.
\end{align}
To see this explicitly, consider the bottom-left (BL) corner located at $(i_0,j_0)$. Acting with the truncated symmetry operator $U_{i_0i_1}^{j_0j_1}(g)$ on the $2$d cluster Hamiltonian in Eq.~\eqref{eq:H_SSPT_2d} flips the sign of the local stabilizer term 
\begin{align}
    \sigma^z_{i_0-1,j_0-1} \sigma^z_{i_0,j_0-1} \sigma^z_{i_0-1,j_0} \sigma^z_{i_0,j_0} \tau^x_{i_0-\half,j_0-\half},
\end{align}
changing it to its negative. This local violation can be repaired by acting with $\tau^z_{i_0-\half,j_0-\half}$, which restores the eigenvalue of the affected stabilizer. Hence, the appropriate local repair operator at the BL corner is 
\begin{align}
     V_{ij}(g_\sigma) = \tau^z_{i-\half, j-\half}.
\end{align}
By applying the same reasoning to the other symmetry generators $g_\tau$ and $g_\sigma g_\tau$, we obtain the full set of repair operators listed in Eq.~\eqref{eq:repair_operator}.

Evaluating $\beta_{ij}^{R}(g)$ for the $2$d cluster model in Eq.~\eqref{eq:H_SSPT_2d} using the local repair operators defined in Eq.~\eqref{eq:repair_operator}, we obtain for the elements of $\mathbb{Z}_2\times\mathbb{Z}_2 = \{1,g_\sigma,g_\tau,g_\sigma g_\tau\}$:
\begin{align}
\label{eq:beta_SSPT}
    \beta_{ij}^{R}(1) = \beta_{ij}^{R} (g_\sigma) = \beta_{ij}^{R} (g_\tau)=1, \quad \beta_{ij}^{R} (g_\sigma g_\tau) = -1.
\end{align}
For example, the combined half-space symmetry operator takes the form
\begin{align}
\label{eq:half_space_sym}
\mathcal{S}_i^R(g_\sigma g_\tau)
= \prod_{i'=i}^{L_x} \prod_{j=1}^{L_y}
\sigma^x_{i',j}\tau^x_{i'-\half,j-\half}.
\end{align}
In the corresponding local repair operator $V_{ij}(g_\sigma g_\tau)$ in Eq.~\eqref{eq:repair_operator}, the factor $\sigma^z_{i-1,j-1}$ commutes with all the terms in $\mathcal{S}_i^R(g_\sigma g_\tau)$ in Eq.~\eqref{eq:half_space_sym}, whereas $\tau^z_{i-\half,j-\half}$ anticommutes with $\tau^x_{i-\half,j-\half}$ contained in the half-space symmetry.
This single anticommutation contributes an overall minus sign, giving
\begin{align}
\beta_{ij}^{R}(g_\sigma g_\tau) = -1.
\end{align}
The other $\beta_{ij}^{R}(g)$ values in Eq.~\eqref{eq:beta_SSPT} can be obtained analogously. Since $\beta_{ij}^{R} (g_\sigma g_\tau) \neq 1$, the cluster model realizes a non-trivial strong SSPT phase.

By contrast, for the weak SSPT Hamiltonian in Eq.~\eqref{eq:H_SSPT_weak}, one finds
\begin{align}
    \beta_{ij}^{R}(1) = \beta_{ij}^{R} (g_\sigma) = \beta_{ij}^{R} (g_\tau) = \beta_{ij}^{R} (g_\sigma g_\tau) =1,
\end{align}
confirming its trivial (weak) character under this invariant.

\subsection{Edge invariants}
\label{ssec:edge_invariants}

In the previous subsection, we introduced following Ref.~\cite{Devakul_SSPT_classification_2018} the bulk invariant $\beta(g)$, which encodes the non-trivial U$(1)$ phase accumulated when a half-space symmetry operator is commuted past a local corner repair operator. As discussed there, $\beta(g)$ captures an intrinsic bulk property that remains invariant under any linearly symmetric local unitary (LSLU) evolution. 
The same topological information can also be extracted from a purely boundary perspective through the \emph{edge invariant} $\phi_{\mathrm{top}}$, which characterizes the projective representation of subsystem symmetries along the (say, top) edge.

Although we have discussed the edge symmetry algebra along the left boundary in Sec.~\ref{sec:gapped_SSPT_open}, here we focus on the \emph{top} boundary, following Ref.~\cite{Devakul_SSPT_classification_2018}. 
Analogous to one-dimensional SPT chains, the top edge of the $2$d SSPT realizes a local projective representation of the extensive vertical symmetry group $G_y = (G_s)^{L_x}$, generated by the edge operators $U^{\mathrm{top}}_i(g)$ corresponding to the vertical subsystem symmetries $U^y_i(g), i=1,\ldots L_x$ defined in Eq.~\eqref{eq:subsym_operator_general}. 
Let $h^g_i \in G_y$ denote the element represented by $U^{\mathrm{top}}_i(g)$. 
Their algebra defines the \emph{edge commutation phase}
\begin{align}
    U^{\mathrm{top}}(h_i)\,U^{\mathrm{top}}(h_i') 
    = \phi_{\mathrm{top}}(h_i,h_i')\, U^{\mathrm{top}}(h_i')\,U^{\mathrm{top}}(h_i),
\end{align}
where $\phi_{\mathrm{top}}(h,h')=\pm1$ for $\mathbb{Z}_2$ symmetries. 
A value $\phi_{\mathrm{top}}(h,h')=-1$ indicates that the two symmetry generators anticommute when acting within the edge subspace. 
Because the underlying representation is local, $\phi_{\mathrm{top}}(h,h')=+1$ whenever $|i-i'|$ exceeds the correlation length. In a trivial weak-SSPT phase, this remains true for an arbitrary choice of $i$ and $i'$, regardless of the separation. A nontrivial value of $\phi_{\mathrm{top}}(h,h')$, on the other hand -- for suitably separated columns $\{i,i'\}$ -- would indicate a nontrivial SSPT edge invariant.

\paragraph{Relation between $\phi_{\mathrm{top}}$ and $\beta(g)$:}
The bulk invariant can be re-expressed entirely in terms of the edge projective data as \cite{Devakul_SSPT_classification_2018}
\begin{align}
    \beta(g)
    = \phi_{\mathrm{top}} \left(\tilde{h}^g_{\mathrm{left}},\, \tilde{h}^g_{\mathrm{right}}\right),
    \label{eq:beta_from_phi_top}
\end{align}
where $\tilde{h}^g_{\mathrm{left/right}}$ are short-width (size $\sim\xi$) products of edge symmetry generators immediately adjacent to a chosen vertical cut. 
This equality follows from identifying the truncated half-space symmetry in the bulk with its restriction to the top boundary and noticing that the corresponding corner repair operator acts on the same support as $\tilde{h}^g_{\mathrm{right}}$. 
The quantity $\phi_{\mathrm{top}}(\tilde{h}^g_{\mathrm{left}},\tilde{h}^g_{\mathrm{right}})$ therefore captures the same commutation phase as $\beta(g)$ in the bulk definition.

To make this relation more transparent, it is helpful to visualize the edge projective representation as a simple \emph{link diagram}.  
Place all edge generators $\{h_i^{g_\sigma},h_{i-\half}^{g_\tau}\}$ as vertices ordered along the edge.  
A link is drawn between two vertices whenever $\phi_{\mathrm{top}}(h,h')=-1$; i.e., whenever the corresponding edge operators anticommute.  
The global symmetry constraint -- that the full product of vertical and horizontal subsystem symmetries commute -- implies that each vertex connects to an \emph{even number} of vertices of each type.  

Now, consider introducing a vertical cut that divides the edge into two halves.  
The cross-commutator between the two edge blocks on either side of the cut is
\begin{align}
\label{eq:phi_top_def}
    \phi_{\mathrm{top}} \left(\tilde{h}^g_{\mathrm{left}}, \tilde{h}^g_{\mathrm{right}}\right)
    = \prod_{\substack{i\in{\rm left}\\ j\in{\rm right}}}
      \phi_{\mathrm{top}} \left(h_i^g,h_j^g\right)
    = (-1)^{N_{\text{cross}}(g)},
\end{align}
where $N_{\text{cross}}(g)$ counts the number of $\phi_{\mathrm{top}}=-1$ links crossing the cut. 
Thus, the bulk invariant $\beta(g)$ defined from the half-space symmetry is equivalently equal to the parity of links that cross any given cut:
\begin{align}
    \beta(g) = (-1)^{N_{\mathrm{cross}}(g)}.
\end{align}
Since LSLUs can only add or remove \emph{pairs} of nearby links, the parity $N_{\mathrm{cross}}(g) \bmod 2$ is cut-independent and invariant under all allowed local transformations. 
This provides a direct edge-based interpretation of the bulk topological invariant.

\paragraph{Evaluation for the $2$d cluster model:}
Let us now compute $\phi_{\mathrm{top}}$ explicitly for the $2$d cluster model.  
On the top boundary ($y=L_y$), the vertical subsystem symmetries act as
\begin{align}
\label{eq:edge_Vtop}
U^{\mathrm{top}}_i(g_\sigma) = \left(U^y_{\sigma,i}\right)_T, \quad 
U^{\mathrm{top}}_i(g_\tau) = \left(U^y_{\tau,{i-\half}}\right)_T, \quad
U^{\mathrm{top}}_i(g_\sigma g_\tau) = \left(U^y_{\sigma,i}\right)_T \left(U^y_{\tau,{i-\half}}\right)_T,
\end{align}
where $(U^y_{\sigma,i})_T$ and $(U^y_{\tau,{i-\half}})_T$ are the top-edge symmetry components given in Eq.~\eqref{eq:edge_SSPT}.
Since $(U^y_{\sigma,i})_T$ contains $\sigma^x_{i,L_y}$ and $(U^y_{\tau,{i\pm\half}})_T$ contain $\sigma^z_{i,L_y}$, they anticommute for the same column index, yielding
\begin{align}
\label{eq:phi_top_rel}
\phi_{\mathrm{top}} \left(h^{g_\sigma}_i, h^{g_\tau}_{i\pm\half}\right) = -1,
\end{align}
while all other pairs commute, $\phi_{\mathrm{top}}=+1$.  

In the link-diagram language, each $\sigma$-type vertex connects to its two neighboring $\tau$-type vertices by a single link. If we draw a vertical cut through the edge, exactly one of these $\sigma$-$\tau$ links crosses the cut.  
According to the above relation, this gives
\begin{align}
    \beta(1)=\beta(g_\sigma)=\beta(g_\tau)=+1,\quad
    \beta(g_\sigma g_\tau)=-1,
\end{align}
in perfect agreement with the bulk calculation of Sec.~\ref{sec:bulk_invariants}. Thus, the nontrivial projective commutation pattern of the edge encodes precisely the same topological invariant $\beta(g)$ that characterizes the strong SSPT phase in the bulk.


\section{Subsystem Kennedy-Tasaki transformation: Mapping strong SSPT to SSSB}
\label{sec:KT_SSPT_strong}

Having discussed the one-dimensional KT transformation in Sec.~\ref{sec:SPT_KT_1d} and the two-dimensional weak subsystem KT transformation in Sec.~\ref{sec:weak_SSPT}, which maps a system of decoupled chains with spontaneously broken $\mathbb{Z}_2 \times \mathbb{Z}_2$ symmetry to a weak $\mathbb{Z}_2 \times \mathbb{Z}_2$ SSPT phase, we now develop a systematic formulation of the \emph{strong} subsystem KT transformation on the square lattice.
In this section, we treat both periodic and antiperiodic (twisted) boundary conditions, and extend the construction to open boundaries in Sec.~\ref{sec:KT_open}.
The resulting transformation maps a two-dimensional $\mathbb{Z}_2 \times \mathbb{Z}_2$ SSSB phase to a strong $\mathbb{Z}_2 \times \mathbb{Z}_2$ SSPT phase.

Formulating the subsystem KT operator with twisted boundary conditions allows us to demonstrate its non-unitary and non-invertible character from three complementary viewpoints. First, in Sec.~\ref{sec:closed_KT_example}, following the discussion of the Xu-Moore model in Sec.~\ref{sec:Xu-Moore_model}, we show that the subsystem KT transformation is non-invertible when restricted to the original Hilbert space. Its unitarity is restored only after enlarging the Hilbert space to include the twisted (antiperiodic) sector \cite{Li_2023_KT}.
Second, in Sec.~\ref{sec:symtwist_KT_sub}, we analyze the mapping of symmetry-twist sectors, which already shows that the operator fails to act unitarily on closed geometries. Finally, in Sec.~\ref{sec:fusion_KT_sub}, we study the fusion algebra between the subsystem KT operator and the subsystem symmetry generators; the appearance of additional twist-dependent contributions both signals non-unitarity and identifies states that are annihilated by the operator, thereby establishing non-invertibility. 

\subsection{Non-invertible subsystem Kennedy-Tasaki transformation}
\label{sec:KT_sub}

Having established the domain wall decoration operator $U_{\text{DW}}$ above in Eq.~\eqref{eq:U_DW_SSPT}, and the subsystem KW transformation $\NKW$ in Sec.~\ref{sec:subsym_KW},
we now turn to the study of the strong subsystem KT transformation in two dimensions. First, we would like to make a brief remark -- in the field-theoretic language, the  operator $U_{\text{DW}}$ implements the so-called $T$-operation, while the KW transformation $\NKW$ corresponds to the $S$ operation of gauging the $\mathbb{Z}_2\times \mathbb{Z}_2$ symmetry. The Kennedy-Tasaki transformation can be viewed as a composition $STS$ of these operations~\cite{Li_2023_KT}. In Appendix~\ref{app:KT_QFT}, we extend the construction of Ref.~\cite{Li_2023_KT} to two dimensions with subsystem symmetry and demonstrate that the  $STS$ transformation indeed maps an SSSB phase to a strong SSPT phase. 
In this section, however, we shift to a lattice viewpoint and study the subsystem KT transformation directly in spin-$\half$ systems, where the $STS$ transformation is realized explicitly on a torus geometry via the composition $\KT = \NKW^\dagger U_{\text{DW}} \NKW$. Fig.~\ref{fig:Dualities_N_KT} gives a schematic summary. 

We start by analyzing a Hamiltonian consisting of two copies of Ising plaquette interactions, whose ground state realizes a $\mathbb{Z}_2 \times \mathbb{Z}_2$ SSSB phase, 
\begin{align}
\label{eq:H_SSSB}
    H_{\text{SSSB}}  = - \sum_{i=1}^{L_x} \sum_{j=1}^{L_y} \sigma^z_{i-1,j-1} \sigma^z_{i,j-1} \sigma^z_{i-1,j} \sigma^z_{i,j} -  \sum_{i=1}^{L_x} \sum_{j=1}^{L_y} \tau^z_{i-\half,j-\half} \tau^z_{i+\half,j-\half} \tau^z_{i-\half,j+\half} \tau^z_{i+\half,j+\half}.
\end{align}
In this section, our aim is to explicitly construct the lattice expression of $\KT$, ensuring that it maps $H_{\text{SSSB}}$ in \eqref{eq:H_SSSB} to the strong SSPT Hamiltonian $H_{\text{SSPT}}$ given in Eq.~\eqref{eq:H_SSPT_2d}. 
Before proceeding, let us emphasize that the operator $\KT$, as defined in this section, is not tied to the specific form of any particular Hamiltonian, such as $H_{\text{SSSB}}$ above. Rather, it is characterized by the condition $H' \KT \ket{\{\sigma_{i,j},\st_{i-\half,j-\half}\}} = \KT H \ket{\{\sigma_{i,j},\st_{i-\half,j-\half}\}}$, which serves to identify the broader class of subsystem $\mathbb{Z}_2 \times \mathbb{Z}_2$ symmetric Hamiltonians that are mapped, under the subsystem KT transformation, to Hamiltonians realizing the strong SSPT phase. Nevertheless, it is instructive to anchor the discussion to a specific form of the Hamiltonian such as in Eq.~\eqref{eq:H_SSSB}, thus making the derivation more concrete.

The eigenvalues of the subsystem $\mathbb{Z}_2 \times \mathbb{Z}_2$ symmetry operators in Eq.~\eqref{eq:subsym_operator_sig_tau} are given by $(-1)^{u^x_{\sigma,j}}$, $(-1)^{u^y_{\sigma,i}}$, $(-1)^{u^x_{\tau,j-\half}}$, $(-1)^{u^y_{\tau,i-\half}}$ respectively. We can also define the twist variables $\mt^x_{\sigma,j-\half}, \mt^y_{\sigma,i-\half}$ following Eqs.~(\ref{eq:twists1}--\ref{eq:twists3}) which denote the boundary conditions for spins $\sigma_{i,j}$. To these, we need to add the analogous twist variables (we denote those by $\mt^x_{\tau,j}, \mt^y_{\tau,i}$) for the $\st_{i-\half,j-\half}$ degrees of freedom. In summary, the symmetry and twist sectors of the model are labeled by the set of variables
\begin{align}
\label{eq:sector_sigtau}
    \left\{u^x_{\sigma,j}, u^y_{\sigma,i}, u^x_{\tau,j-\half}, u^y_{\tau,i-\half}, \mt^x_{\sigma,j-\half}, \mt^y_{\sigma,i-\half}, \mt^x_{\tau,j}, \mt^y_{\tau,i} \right \}.
\end{align}

Upon applying the generalized subsystem duality transformation $\NKW$ which we derived in Sec.~\ref{sec:subsym_KW}, the theory defined on the original lattice with spins $\ssi_{i,j}$ gets mapped to a dual theory living on the dual lattice, with spins $\hssi_{i-\half,j-\half}$. It is important to stress that the dual spins $\hssi_{i-\half,j-\half}$ thus obtained are distinct from the $\st_{i-\half,j-\half}$ spins in Eq.~\eqref{eq:H_SSSB}, even though they occupy the same spatial location (namely, the centers of each plaquette). Similarly, we denote the subsystem KW-duals of the spins $\st_{i-\half,j-\half}$ by $\hst_{i,j}$ which live on the vertices of the lattice, but are distinct from the spins $\sigma_{i,j}$. 

Likewise, the Hilbert space of the dual theory with spins $\hssi_{i-\half,j-\half}, \hst_{i,j}$ is similarly divided into symmetry-twist sectors labeled by
\begin{align}
    \left \{\hu^x_{\sigma,j-\half},\hu^y_{\sigma,i-\half}, \hu^x_{\tau,j}, \hu^y_{\tau,i}, \hmt^x_{\sigma,j}, \hmt^y_{\sigma,i}, \hmt^x_{\tau,j-\half}, \hmt^y_{\tau,i-\half} \right \}.
\end{align}

Following the definition of the subsystem KW transformation for a single $\mathbb{Z}_2$ symmetry in Eq.~\eqref{eq:N_sub}, we now extend the construction to the case of a subsystem $\mathbb{Z}_2 \times \mathbb{Z}_2$ symmetry. Applying this generalized transformation, we arrive at two equivalent expressions for the generalized KW operator
\begin{align}
\label{eq:N_Z2Z2}
    \NKW \ket{\{\ssi_{i,j}, \st_{i-\half,j-\half}\}} & = \frac{1}{2^{L_x+L_y}} \sum_{\{\hssi_{i-\half,j-\half}, \hst_{i,j}\}} (-1)^{\mathcal{A}_{\text{KW}} (\{\sigma,\hsig;\tau,\htau\})} \ket{\{\hssi_{i-\half,j-\half}, \hst_{i,j}\}} \nonumber\\
    & = \frac{1}{2^{L_x+L_y}} \sum_{\{\hssi_{i-\half,j-\half}, \hst_{i,j}\}} (-1)^{\widetilde{\mathcal{A} }_{\text{KW}} (\{\sigma,\hsig;\tau,\htau\})} \ket{\{\hssi_{i-\half,j-\half}, \hst_{i,j}\}}.
\end{align}
Here, the phase $\mathcal{A}_{\text{KW}}$ combines bulk and boundary contributions, $\mathcal{A}_{\text{KW}} = \mathcal{C}_{\text{bulk}}^{\sigma\hsig} + \mathcal{C}_{\text{bdy}}^{\sigma\hsig} + \mathcal{C}_{\text{bulk}}^{\tau\htau} + \mathcal{C}_{\text{bdy}}^{\tau\htau}$, with the $\hsig$ terms given in Eq.~\eqref{eq:C_bulk_bdy} and the $\htau$ terms defined analogously. Summing these yields

\begin{align}
\label{eq:A_N_1}
    \mathcal{A}_{\text{KW}}  = &  \sum_{i=1}^{L_x} \sum_{j=1}^{L_y} \ssi_{i,j} \left(\hssi_{i-\half,j-\half} + \hssi_{i+\half,j-\half} + \hssi_{i-\half,j+\half} + \hssi_{i+\half,j+\half}\right) + \sum_{j=1}^{L_y}  t^x_{\sigma,j} \left(\hssi_{\half,j-\half} + \hssi_{\half,j+\half}\right)   \nonumber \\
  & + \sum_{i=1}^{L_x}  t^y_{\sigma,i} \left(\hssi_{i-\half,\half} + \hssi_{i+\half,\half}\right) + t^{xy}_{\sigma} \hssi_{\half,\half}  + \sum_{i=1}^{L_x} \sum_{j=1}^{L_y} \hst_{i,j} \left(\st_{i-\half,j-\half} + \st_{i+\half,j-\half} + \st_{i-\half,j+\half}  \right. \nonumber \\
  & \left. + \st_{i+\half,j+\half}\right) + 
  \sum_{j=1}^{L_y}  \htt^x_{\tau,j} \left(\st_{\half,j-\half} + \st_{\half,j+\half}\right) 
   + \sum_{i=1}^{L_x}  \htt^y_{\tau,i} \left(\st_{i-\half,\half} + \st_{i+\half,\half}\right) + \htt^{xy}_{\tau} \st_{\half,\half},
\end{align}

and similarly, using Eq.~\eqref{eq:C_tilde_bulk_bdy}, the phase $\widetilde{\mathcal{A}}_{\text{KW}}$, analogous to $\mathcal{A}_{\text{KW}}$, is given by
\begin{align}
\label{eq:A_N_2}
    \widetilde{\mathcal{A}}_{\text{KW}}  = & \sum_{i=1}^{L_x} \sum_{j=1}^{L_y} \hssi_{i-\half,j-\half} \left( \ssi_{i-1,j-1} + \ssi_{i,j-1} + \ssi_{i-1,j} + \ssi_{i,j}\right)  + \sum_{j=1}^{L_y} \htt_{\sigma, j-\half}^x \left(\ssi_{L_x,j-1} + \ssi_{L_x,j}\right) \nonumber \\
  & + \sum_{i=1}^{L_x} \htt_{\sigma, i-\half}^y \left(\ssi_{i-1,L_y} + \ssi_{i,L_y}\right) + \htt^{xy}_{\sigma} \ssi_{L_x,L_y}   + \sum_{i=1}^{L_x} \sum_{j=1}^{L_y} \st_{i-\half,j-\half} \left( \hst_{i-1,j-1} + \hst_{i,j-1} + \hst_{i-1,j} \right. \nonumber \\
  & \left. + \hst_{i,j}\right) +  \sum_{j=1}^{L_y} t_{\tau, j-\half}^x \left(\hst_{L_x,j-1} + \hst_{L_x,j}\right)  + \sum_{i=1}^{L_x} t_{\tau, i-\half}^y \left(\hst_{i-1,L_y} + \hst_{i,L_y}\right) + t^{xy}_{\tau} \hst_{L_x,L_y}.
\end{align}

Since $\mathcal{A}_{\text{KW}} = \widetilde{\mathcal{A}}_{\text{KW}}$, the two expressions in Eq.~\eqref{eq:N_Z2Z2} are equivalent. To summarize, the generalized KW transformation maps the original degrees of freedom onto the dual one, defined on the dual lattice related via half-diagonal translation $T_{x/2,y/2}$ to the original:
\beq
\label{eq:KW-dual-variables}
\sigma_{i,j}, \tau_{i-\half, j-\half} \stackrel{\NKW}{\longrightarrow} \hsig_{i-\half,j-\half}, \htau_{i, j}
\eeq

In the previous subsection \ref{sec:SSPT_2d}, we have constructed the two-dimensional analogue of the domain-wall decorating operator $U_{\text{DW}}$ (see Eq.~\eqref{eq:U_DW_SSPT}) for models with subsystem $\mathbb{Z}_2 \times \mathbb{Z}_2$ symmetry. That operator was acting on the original spins $\{\sigma_{i,j}, \tau_{i-\half, j-\half}\}$. Anticipating that we are aiming to construct the generalized KT transformation as a composition (see Appendix~\ref{app:KT_QFT} for the field-theoretic perspective)
\beq
\label{eq:KT=NUN}
\KT = \NKW^\dagger U_{\text{DW}} \NKW
\eeq
and noting that $\NKW$ transforms to the dual degrees of freedom via Eq.~\eqref{eq:KW-dual-variables}, we must therefore define the action of $U_{\text{DW}}$ on the dual spins. Following Eq.~\eqref{eq:U_DW_SSPT}, we find 
\begin{align}
    U_{\text{DW}}  = & \prod_{i=1}^{L_x} \prod_{j=1}^{L_y} \exp \left[\frac{i \pi}{4} \left(1-\hsig^z_{i-\half,j-\half}\right) \left(1-\htau^z_{i-1,j-1}\right)\right] \exp \left[\frac{i \pi}{4} \left(1-\hsig^z_{i+\half,j-\half}\right) \left(1-\htau^z_{i-1,j-1}\right)\right] \nonumber\\
    & \times \exp \left[\frac{i \pi}{4} \left(1-\hsig^z_{i-\half,j+\half}\right) \left(1-\htau^z_{i-1,j-1}\right)\right] \exp \left[\frac{i \pi}{4} \left(1-\hsig^z_{i+\half,j+\half}\right) \left(1-\htau^z_{i-1,j-1}\right)\right] 
\end{align}   
The above expression can equivalently be written as follows: 
\begin{align}
    U_{\text{DW}} = & \prod_{i=1}^{L_x} \prod_{j=1}^{L_y} \exp \left[\frac{i \pi}{4} \left(1-\hsig^z_{i-\half,j-\half}\right) \left(1-\htau^z_{i-1,j-1}\right)\right] \exp \left[\frac{i \pi}{4} \left(1-\hsig^z_{i-\half,j-\half}\right) \left(1-\htau^z_{i,j-1}\right)\right] \nonumber\\
    & \times \exp \left[\frac{i \pi}{4} \left(1-\hsig^z_{i-\half,j-\half}\right) \left(1-\htau^z_{i-1,j}\right)\right] \exp \left[\frac{i \pi}{4} \left(1-\hsig^z_{i-\half,j-\half}\right) \left(1-\htau^z_{i,j}\right)\right].
\end{align}

The action of $U_{\text{DW}}$ on a basis state of the dual spins is given by two equivalent expressions below:
\begin{align}
\label{eq:U_DW_dual}
    U_{\text{DW}} \ket{\{\hssi_{i-\half,j-\half}, \hst_{i,j}\}} &  =  (-1)^{\mathcal{A}_{\text{DW}}(\{\hsig,\htau\})} \ket{\{\hssi_{i-\half,j-\half}, \hst_{i,j}\}} \nonumber\\
    &  = (-1)^{\widetilde{\mathcal{A}}_{\text{DW}} (\{\hsig,\htau\})} \ket{\{\hssi_{i-\half,j-\half}, \hst_{i,j}\}} 
\end{align}
where the phase $\mathcal{A}_{\text{DW}} (\{\hsig,\htau\})$ is given by
\begin{align}
\label{eq:A_U_1}
    \mathcal{A}_{\text{DW}} (\{\hsig,\htau\})  = & \sum_{i=1}^{L_x} \sum_{j=1}^{L_y} \hst_{i,j} \left(\hssi_{i-\half,j-\half} + \hssi_{i+\half,j-\half} + \hssi_{i-\half,j+\half} + \hssi_{i+\half,j+\half}\right) \nonumber\\
    & + \sum_{j=1}^{L_y}  \htt^x_{\tau,j} \left(\hssi_{\half,j-\half} + \hssi_{\half,j+\half}\right)   + \sum_{i=1}^{L_x}  \htt^y_{\tau,i} \left(\hssi_{i-\half,\half} + \hssi_{i+\half,\half}\right) + \htt^{xy}_{\tau} \hssi_{\half,\half},
\end{align}
and the phase $\widetilde{\mathcal{A}}_{\text{DW}} (\{\hsig,\htau\})$ is given by
\begin{align}
\label{eq:A_U_2}
    \widetilde{\mathcal{A}}_{\text{DW}} (\{\hsig,\htau\})  = & \sum_{i=1}^{L_x} \sum_{j=1}^{L_y} \hssi_{i-\half,j-\half} \left(\hst_{i-1,j-1} + \hst_{i,j-1} + \hst_{i-1,j} + \hst_{i,j} \right) + \sum_{j=1}^{L_y}  \htt^x_{\sigma,j-\half} \left(\hst_{L_x,j-1} + \hst_{L_x,j}\right) \nonumber\\
    &  + \sum_{i=1}^{L_x}  \htt^y_{\sigma,i} \left(\hst_{i-1,L_y} + \hst_{i,L_y}\right) + \htt^{xy}_{\sigma} \hst_{L_x,L_y}.
\end{align}

We are now in the position to finally write down the generalized KT transformation as the composition in Eq.~\eqref{eq:KT=NUN}.
Before presenting its explicit form, let us clarify how the operator acts on the basis state of the original Hilbert space. As per Eq.~\eqref{eq:KW-dual-variables}, the subsystem KW duality $\NKW$  transforms a configuration of physical spins on original lattice $\{\ssi_{i,j}, \st_{i-\half,j-\half}\}$ into a superposition of configurations of dual variables $\{\hssi_{i-\half,j-\half}, \hst_{i,j}\}$, with an appropriate sign $(-1)^{\mathcal{A}_{\text{KW}}}$ (see Eq.~\eqref{eq:N_Z2Z2}) associated with each configuration that keeps track of how the original and dual variables overlap locally. Next, we note that $U_{\text{DW}}$ in Eq.~\eqref{eq:U_DW_dual} is a diagonal unitary: it does not alter the basis, but simply assigns a ($-1$) phase to configurations where a dual $\htau$-spin coincides with a domain wall in the dual $\hsig$-spin. Thus, it modifies amplitudes without generating superpositions. Finally, conjugating this diagonal operator by $\NKW^\dagger$ maps the decorated dual wavefunction back to the spin basis on the original lattice, which is defined as $\{\ssip_{i,j}, \stp_{i-\half,j-\half}\}$, yielding the duality operator $\KT$, which acts entirely within the original degrees of freedom.

Explicitly, the operator $\KT$ acts on an arbitrary basis element of the original Hilbert space as follows (see Appendix~\ref{app:N_KT_comp} for detailed calculation):
\begin{align}
\label{eq:N_KT_def_1}
    \KT \ket{\{\ssi_{i,j}, \st_{i-\half,j-\half}\}} & = \frac{2^{L_xL_y}}{4^{L_x+L_y}}\, \mathcal{P}_{\text{KT}} \sum_{\{\ssip_{i,j},\stp_{i-\half,j-\half}\}}  (-1)^{\mathcal{A}_{\text{KT}}(\{\sigma,\sigma';\tau,\tau'\})} \ket{\{\ssip_{i,j}, \stp_{i-\half,j-\half}\}},
\end{align}
where the sum runs over all configurations of the spins defined on the original lattice $\{\ssip_{i,j}, \stp_{i-\half,j-\half}\}$ that arise when $\NKW^\dagger$ acts on the dual-spin configurations $\{\hssi_{i-\half,j-\half}, \hst_{i,j}\}$. The phase $\mathcal{A}_{\text{KT}}$ is defined as

\begin{align}
\label{eq:A_KT_def_1}
    \mathcal{A}_{\text{KT}} = & \sum_{i=1}^{L_x} \sum_{j=1}^{L_y} \left(\ssi_{i,j}+ \ssip_{i,j}\right) \left( \st_{i-\half,j-\half}+ \st_{i+\half,j-\half} + \st_{i-\half,j+\half} + \st_{i+\half,j+\half} + \stp_{i-\half,j-\half} + \stp_{i+\half,j-\half} \right. \nonumber \\
  &\left. + \stp_{i-\half,j+\half} + \stp_{i+\half,j+\half} \right) + \sum_{j=1}^{L_y} \left( t^x_{\sigma,j} + t'^x_{\sigma,j} \right) \left( \st_{\half, j-\half} + \st_{\half, j+\half} + \stp_{\half, j-\half} + \stp_{\half, j+\half} \right) \nonumber\\
  & + \sum_{i=1}^{L_x} \left( t^y_{\sigma,i} + t'^y_{\sigma,i} \right) \left( \st_{i-\half,\half} + \st_{i+\half,\half} + \stp_{i-\half,\half}+ \stp_{i+\half,\half} \right) + \left( t^{xy}_{\sigma} + t'^{xy}_{\sigma} \right) \left( \st_{\half,\half} +  \stp_{\half,\half} \right),
\end{align}
and the projector $\mathcal{P}_{\text{KT}}$ is defined as
\begin{align}
\label{eq:P_KT_def_1}
    \mathcal{P}_{\text{KT}} = & \half \prod_{j=1}^{L_y} \left[1+ (-1)^{\mt^x_{\sigma,j-\half} + \mt'^x_{\sigma,j-\half} + \hmt^x_{\tau,j-\half}} \right]\prod_{i=1}^{L_x} \left[1 + (-1)^{\mt^y_{\sigma,i-\half} + \mt'^y_{\sigma,i-\half} + \hmt^y_{\tau,i-\half}}\right].
\end{align}
We now make a few remarks to clarify and contextualize the result. First, to construct $\KT$, we use the first expression of the KW duality operator $\mN$ in Eq.~\eqref{eq:N_Z2Z2} (with the phase $\mathcal{A}_{\mN}$ in \eqref{eq:A_N_1}) and the first expression of the domain wall decoration operator $U_{\text{DW}}$ in Eq.~\eqref{eq:U_DW_dual} (with the phase $\mathcal{A}_{U}$ in \eqref{eq:A_U_1}). Second, $\hmt^x_{\sigma,j}, \hmt^y_{\sigma,i}, \hmt^x_{\tau,j-\half}, \hmt^y_{\tau,i-\half}$ the variables serve as labels for the twist sectors of the intermediate state produced after applying the first subsystem KW transformation. An important observation is that these twist variables appear only inside the projectors in Eq.~\eqref{eq:P_KT_def_1}. This means they are not independent degrees of freedom, but are instead fully determined once the twist sectors of both the initial and final states are specified \cite{Li_2023_KT}. With this dependence made explicit, Eq.~\eqref{eq:N_KT_def_1} takes on a much simpler form 
\begin{align}
\label{eq:N_KT_final_1}
    \KT \ket{\{\ssi_{i,j}, \st_{i-\half,j-\half}\}} & = 2^{L_xL_y -(L_x+L_y+1)} \sum_{\{\ssip_{i,j},\stp_{i-\half,j-\half}\}}  (-1)^{\mathcal{A}_{\text{KT}} (\{\sigma,\sigma';\tau,\tau'\})} \ket{\{\ssip_{i,j}, \stp_{i-\half,j-\half}\}},
\end{align}
where $\mathcal{A}_{\text{KT}}$ is defined in Eq.~\eqref{eq:A_KT_def_1}.

Similarly, using the second expression of $\NKW$ in Eq.~\eqref{eq:N_Z2Z2} (with the phase $\widetilde{\mathcal{A}}_{\text{KW}}$ in Eq.~\eqref{eq:A_N_2}) and the second expression of $U_{\text{DW}}$ in Eq.~\eqref{eq:U_DW_dual} (with the phase $\widetilde{\mathcal{A}}_{\text{DW}}$ in Eq.~\eqref{eq:A_U_2}), we can formulate another equivalent expression for $\KT$,
\begin{align}
\label{eq:N_KT_final_2}
    \KT \ket{\{\ssi_{i,j}, \st_{i-\half,j-\half}\}} & = 2^{L_xL_y -(L_x+L_y+1)} \sum_{\{\ssip_{i,j},\stp_{i-\half,j-\half}\}}  (-1)^{\widetilde{\mathcal{A}}_{\text{KT}} (\{\sigma,\sigma';\tau,\tau'\})} \ket{\{\ssip_{i,j}, \stp_{i-\half,j-\half}\}},
\end{align}
where the phase $\widetilde{\mathcal{A}}_{\text{KT}}$ is defined as follows:
\begin{align}
\label{eq:A_KT_def_2}
    \widetilde{\mathcal{A}}_{\text{KT}}  = & \sum_{i=1}^{L_x} \sum_{j=1}^{L_y} \left(\st_{i-\half,j-\half} + \stp_{i-\half,j-\half}\right) \left( \ssi_{i-1,j-1} + \ssi_{i,j-1} + \ssi_{i-1,j} + \ssi_{i,j} +  \ssip_{i-1,j-1} + \ssip_{i,j-1}  \right. \nonumber \\
  &\left. + \ssip_{i-1,j} + \ssip_{i,j} \right) + \sum_{j=1}^{L_y} \left( t^x_{\tau,j-\half} + t'^x_{\tau,j-\half} \right) \left(\ssi_{L_x,j-1} + \ssi_{L_x,j} + \ssip_{L_x,j-1} + \ssip_{L_x,j}\right)  + \sum_{i=1}^{L_x} \left( t^y_{\tau,i-\half} \right. \nonumber \\
  &\left. + t'^y_{\tau,i-\half} \right) \left(\ssi_{i-1,L_y} + \ssi_{i,L_y}+ \ssip_{i-1,L_y} + \ssip_{i,L_y}\right) + \left( t^{xy}_{\tau} + t'^{xy}_{\tau} \right) \left( \ssi_{L_x,L_y} +  \ssip_{L_x,L_y} \right).
\end{align}

In summary, within this subsection we have derived two equivalent formulations of the subsystem KT transformation operator $\KT$, given explicitly in Eqs.~\eqref{eq:N_KT_final_1} and \eqref{eq:N_KT_final_2}. Although these expressions are mathematically equivalent, each lends itself to different computational or conceptual advantages, and both will play an important role in the derivation of the mapping between symmetry-twist sectors in Sec.~\ref{sec:symtwist_KT_sub}.

\begin{figure}[tb]
    \centering
    \includegraphics[width=1.0\linewidth]{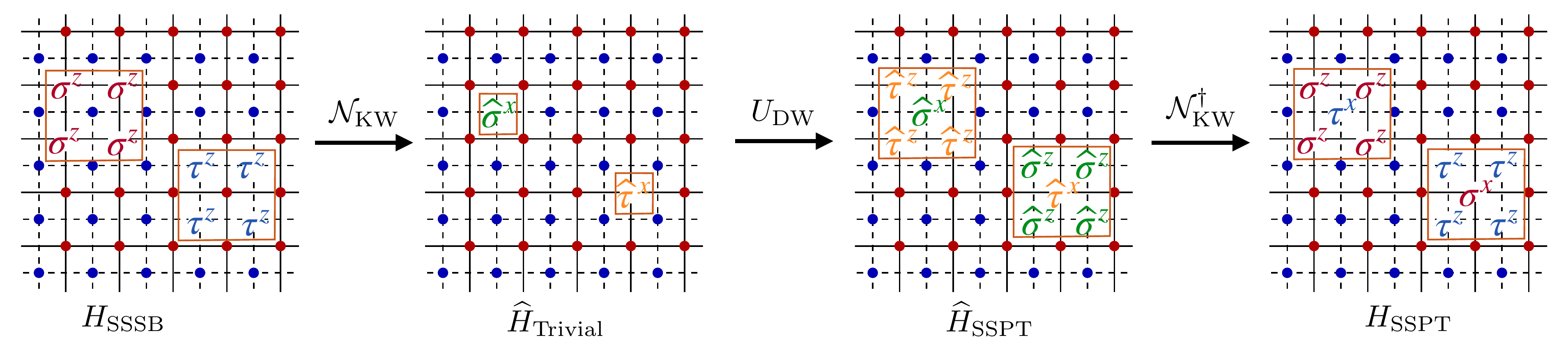}
    \caption{Sequence of transformations illustrating the construction of the subsystem KT duality $\KT = \NKW^\dagger U_{\text{DW}} \NKW$: the KW duality $\NKW$ maps the $H_\text{SSSB}$ to a trivial one in terms of dual spins; the domain-wall decoration operator $U_{\text{DW}}$ then promotes it to the $\widehat{H}_\text{SSPT}$ in the dual basis; and finally $\NKW^\dagger$ returns the SSPT Hamiltonian $H_\text{SSPT}$ in terms of the original spins.}
    \label{fig:SSSB_to_SSPT}
\end{figure}

The subsystem KT transformation \eqref{eq:N_KT_final_1} acts on the plaquette-product of the Pauli-Z operators as follows:
\begin{align}
\label{eq:Pauli_map_KT}
    & \KT \sigma^z_{i-1,j-1} \sigma^z_{i,j-1} \sigma^z_{i-1,j} \sigma^z_{i,j} \ket{\psi} = \sigma^z_{i-1,j-1} \sigma^z_{i,j-1} \sigma^z_{i-1,j} \sigma^z_{i,j} \tau^x_{i-\half,j-\half} \KT \ket{\psi}, \quad \forall \psi \in \mathcal{H}, \nonumber\\ & \KT \tau^z_{i-\half,j-\half} \tau^z_{i+\half,j-\half} \tau^z_{i-\half,j+\half} \tau^z_{i+\half,j+\half} \ket{\psi} = \tau^z_{i-\half,j-\half} \tau^z_{i+\half,j-\half} \tau^z_{i-\half,j+\half} \tau^z_{i+\half,j+\half}\sigma^x_{i,j} \KT \ket{\psi}.
\end{align}
In other words, it transforms the sum of the plaquette operators in $H_\text{SSSB}$ into the sum over the stabilizers of the 2d cluster state in $H_\text{SSPT}$.
By contrast, the KT transformation leaves the Pauli-$x$ operators invariant:
\begin{align}
\label{eq:Pauli_map_KT_2}
     \KT \sigma^x_{i,j} \ket{\psi} & = \sigma^x_{i,j} \KT \ket{\psi}, \quad \forall \psi \in \mathcal{H}, \nonumber\\  \KT \tau^x_{i-\half,j-\half} \ket{\psi} & = \tau^x_{i-\half,j-\half} \KT \ket{\psi}, \quad \forall \psi \in \mathcal{H}.
\end{align}

\subsection{Subsystem KT transformation on the two copies of Xu-Moore model}
\label{sec:closed_KT_example}

Using the mapping of operators in Eq.~\eqref{eq:Pauli_map_KT}, it follows directly that $\KT$ maps $H_{\text{SSSB}}$ in Eq.~\eqref{eq:H_SSSB} to the SSPT Hamiltonian $H_{\text{SSPT}}$ in Eq.~\eqref{eq:H_SSPT_2d}. 
We can track the composition of the operations $\KT = \NKW^\dagger U_{\text{DW}} \NKW$ step-by-step:
first, acting with the KW duality operator $\mN_{\text{KW}}$ maps $H_{\mathrm{SSSB}}$ in Eq.~\eqref{eq:H_SSSB} to a trivial Hamiltonian expressed in terms of the dual spins, 
\begin{align}
   \widehat{H}_{\text{trivial}} = -\sum_{i=1}^{L_x} \sum_{j=1}^{L_y} (\hsig^z_{i-\half,j-\half} + \htau^x_{i,j}). 
\end{align}
Subsequently, the DDW operator $U_{\text{DW}}$ transforms $\widehat{H}_{\text{trivial}}$ into the SSPT Hamiltonian, structurally analogous to $H_{\text{SSPT}}$ in Eq.~\eqref{eq:H_SSPT_2d}, but written in the dual-spin variables. A final application of $\mN_{\text{KW}}^\dagger$ brings the result back to the original spin basis, yielding the SSPT Hamiltonian $H_{\mathrm{SSPT}}$ in Eq.~\eqref{eq:H_SSPT_2d}. This sequence of mappings is schematically illustrated in Fig.~\ref{fig:SSSB_to_SSPT}.

To demonstrate that the trivial phase remains invariant under the subsystem KT transformation, we introduce transverse fields to the plaquette Ising Hamiltonian in Eq.~\eqref{eq:H_SSSB}, yielding
\begin{align}
\label{eq:H_TXM}
    H_{\text{TXM}}  = & - \sum_{i=1}^{L_x} \sum_{j=1}^{L_y} \sigma^z_{i-1,j-1} \sigma^z_{i,j-1} \sigma^z_{i-1,j} \sigma^z_{i,j} -  \sum_{i=1}^{L_x} \sum_{j=1}^{L_y} \tau^z_{i-\half,j-\half} \tau^z_{i+\half,j-\half} \tau^z_{i-\half,j+\half} \tau^z_{i+\half,j+\half} \nonumber \\
    & -h_x \sum_{i=1}^{L_x} \sum_{j=1}^{L_y} \sigma^x_{i,j} -h_x \sum_{i=1}^{L_x} \sum_{j=1}^{L_y} \tau^x_{i-\half,j-\half}.
\end{align}
This model is essentially composed of two copies of the Xu-Moore model, discussed earlier in Sec.~\ref{sec:subsym_square}. For small values of $h_x \ll 1$, the model enters a subsystem $\mathbb{Z}_2 \times \mathbb{Z}_2$ symmetry-broken phase with a ground-state degeneracy of $4^{L_x+L_y-1}$. In contrast, for large $h_x \gg 1$, the system enters a trivial paramagnetic phase. 

Importantly, by virtue of Eq.~\eqref{eq:Pauli_map_KT_2}, one finds that the transverse-field terms remain invariant under the subsystem KT transformation. Thus, applying the Pauli-operator mappings defined in Eqs.~\eqref{eq:Pauli_map_KT} and \eqref{eq:Pauli_map_KT_2}, the Hamiltonian in Eq.~\eqref{eq:H_TXM} transforms under the subsystem KT transformation into
\begin{align}
\label{eq:H_cluster}
    H_{\text{TFC}}  = & - \sum_{i=1}^{L_x} \sum_{j=1}^{L_y} \left( \sigma^z_{i-1,j-1} \sigma^z_{i,j-1} \sigma^z_{i-1,j} \sigma^z_{i,j} \tau^x_{i-\half,j-\half}  +  \tau^z_{i-\half,j-\half} \tau^z_{i+\half,j-\half} \tau^z_{i-\half,j+\half} \tau^z_{i+\half,j+\half} \sigma^x_{i,j} \right) \nonumber \\
    &  -h_x \sum_{i=1}^{L_x} \sum_{j=1}^{L_y} \sigma^x_{i,j} -h_x \sum_{i=1}^{L_x} \sum_{j=1}^{L_y} \tau^x_{i-\half,j-\half},
\end{align}

which is  the $2$d cluster model \eqref{eq:H_SSPT_2d} with an added  transverse Zeeman field \cite{Raussendorf_MBQC_2002}. 
As discussed in Ref.~\ref{sec:SSPT_2d}, for small $h_x \ll 1$, the model realizes a strong SSPT phase \cite{You_SSPT_2018,Devakul_SSPT_classification_2018}. Consequently, all the previous discussions of how our subsystem KT duality maps the SSSB phase into the SSPT phase continue to apply in the presence of small transverse fields. 
In contrast, for large $h_x \gg 1$, the model enters a trivial paramagnetic phase. In summary, we are able to organize the three phases, namely the trivial phase, the SSSB phase, and the strong SSPT phase, into a web of dualities depicted schematically in Fig.~\ref{fig:Dualities_N_KT}.

Having established the formal properties of the transformation, we now turn to the physical perspective, with a couple of remarks due:

\paragraph{Non-unitary nature of KT transformation under closed boundary conditions:}
 
In the ordered phase ($h_x \ll 1$), the two copies of the Xu-Moore model \eqref{eq:H_TXM} have $4^{L_x+L_y-1}$ degenerate ground states. Under the subsystem KT transformation, this phase is mapped to the strong SSPT phase of the transformed Hamiltonian \eqref{eq:H_cluster}, where the entire degenerate manifold collapses into a single unique ground state (under the closed boundary conditions). Such a reduction of the ground-state manifold provides direct evidence of the intrinsic non-unitarity and non-invertibility of the subsystem KT transformation. We hasten to add that while the strong SSPT phase has a unique ground state under closed boundary conditions, we shall show later that on an open square lattice it exhibits a ground-state degeneracy of $4^{L_x+L_y-1}$.

\paragraph{Enlarging the Hilbert space renders the KT transformation unitary:}
We now demonstrate that the KT transformation becomes unitary once the Hilbert space is enlarged to include twisted sectors. Similar to the discussion of the subsystem KW transformation of the Xu-Moore model in Sec.~\ref{sec:Xu-Moore_model}, in the ordered phase ($h_x \ll 1$) of $H_{\text{TXM}}$ in Eq.~\eqref{eq:H_TXM}, twisted boundary conditions cost energy because they necessarily introduce vortices. Since vortices have finite energy, the ground state in the twisted sectors lies above that in the periodic sector, so they are not degenerate. As a result, even if we formally extend the Hilbert space, the low-energy spectrum is unaffected: the ground-state degeneracy remains $4^{L_x+L_y-1}$, with origin in the spontaneous subsystem symmetry breaking.

By contrast, the KT dual of this SSSB phase is the SSPT phase of $H_{\text{TFC}}$ in Eq.~\eqref{eq:H_cluster}, where periodic and twisted boundary conditions cost the same energy in the thermodynamic limit. Thus, once we include all $4^{L_x+L_y-1}$ twisted sectors, the SSPT ground-state manifold also acquires a $4^{L_x+L_y-1}$-fold degeneracy. In this extended Hilbert space, the subsystem KT transformation maps the degenerate ground-state manifold onto itself, and hence becomes unitary.

\subsection{Mapping between symmetry-twist sectors}
\label{sec:symtwist_KT_sub}
We now turn to examining how the subsystem KT transformation \eqref{eq:N_KT_final_1} acts on the various symmetry-twist sectors. This analysis will clarify how the transformation reorganizes states between different sectors, and it will also provide our second concrete indication that $\KT$ is not a unitary operator when acting on a system with closed boundary conditions. Since each subsystem symmetry generator in Eq.~\eqref{eq:subsym_operator_sig_tau} acts by flipping all spins along a single row or column, it is convenient to introduce an explicit notation for their action.
Specifically, the operator $U^x_{\sigma,\jprime}$ flips all $\sigma$-spins along the $\jprime$-th row while leaving the rest unchanged, which we denote as
\beq
\label{eq:row-Pauli}
 U^x_{\sigma,\jprime} \ket{\{\ssi_{i,j}, \st_{i-\half,j-\half}\}} = \ket{\{\sigma_{i,j\neq j'},\bar{\sigma}_{i,j'};\st_{i-\half,j-\half}\}},
\eeq
where we denoted $\bar{\sigma} = 1-\sigma$.
The action of the symmetry generators of $\st_{i-\half,j-\half}$ on the basis states can be defined similarly.

We begin with a general state of the original spin system,
\begin{align}
    \ket{\psi} = \sum_{\{\ssi_{i,j}, \st_{i-\half,j-\half}\}} \psi_{\{\ssi_{i,j}, \st_{i-\half,j-\half}\}} \ket{\{\ssi_{i,j}, \st_{i-\half,j-\half}\}},
\end{align}  
where $\psi_{\{\ssi_{i,j}, \st_{i-\half,j-\half}\}}$ denotes the wavefunction of the spin variables. The symmetry sectors of $\ket{\psi}$ are labeled by $\{u^x_{\sigma,j}, u^y_{\sigma,i}, u^x_{\tau,j-\half}, u^y_{\tau,i-\half} \}$. For example,

\begin{align}
\label{eq:sym_sector_def_general}
    \psi_{\{\sigma_{i,j\neq j'},1-\sigma_{i,j'};\st_{i-\half,j-\half}\}} & = (-1)^{u^x_{\sigma,\jprime}} \psi_{\{\ssi_{i,j}, \st_{i-\half,j-\half}\}}, \nonumber\\ 
   \psi_{\{\ssi_{i,j}; \st_{i-\half,j\neq j'-\half}, 1-\st_{i-\half,\jprime-\half} \}}  & = (-1)^{u^x_{\tau,\jprime-\half}} \psi_{\{\ssi_{i,j}, \st_{i-\half,j-\half}\}}.
\end{align}

We now determine the symmetry-twist sector of the transformed state $\KT \ket{\psi}$ under the subsystem KT transformation in Eqs.~\eqref{eq:N_KT_final_1} and \eqref{eq:N_KT_final_2}. To this end, we evaluate
\begin{align}
    \KT \ket{\psi} = \ket{\psi'} = \sum_{\{\ssip_{i,j}, \stp_{i-\half,j-\half}\}} \psi'_{\{\ssip_{i,j}, \stp_{i-\half,j-\half}\}} \ket{\{\ssip_{i,j}, \stp_{i-\half,j-\half}\}},
\end{align}
where the transformed wavefunction is given by
\begin{align}
\label{eq:psi_prime_KT}
    \psi'_{\{\ssip_{i,j}, \stp_{i-\half,j-\half}\}} & = 
    \frac{1}{2} \sum_{\{\ssi_{i,j},\st_{i-\half,j-\half}\}} 
(-1)^{\mathcal{A}_{\text{KT}}(\{\sigma,\sigma';\tau,
    \tau'\})} 
    \psi_{\{\ssi_{i,j}, \st_{i-\half,j-\half}\}} \nonumber\\
    & = \frac{1}{2} \sum_{\{\ssi_{i,j},\st_{i-\half,j-\half}\}} 
(-1)^{\widetilde{\mathcal{A}}_{\text{KT}}(\{\sigma,\sigma';\tau,
    \tau'\})}
    \psi_{\{\ssi_{i,j}, \st_{i-\half,j-\half}\}} .
\end{align}

Here, $\mathcal{A}_{\text{KT}}$ and $\widetilde{\mathcal{A}}_{\text{KT}}$ are the configuration-dependent phases given in Eqs.~\eqref{eq:A_KT_def_1} and \eqref{eq:A_KT_def_2}, respectively, and the above two expressions are equivalent. The symmetry sectors of $\ket{\psi'}$ are labeled by $\{u'^x_{\sigma,j}, u'^y_{\sigma,i}, u'^x_{\tau,j-\half}, u'^y_{\tau,i-\half} \}$, which are defined in the same way as in Eq.~\eqref{eq:sym_sector_def_general} for $\ket{\psi}.$

\paragraph{Effect of KT transformation on the symmetry sectors:}
To determine the symmetry sectors $u'^x_{\sigma,\jprime}$ after the KT transformation, we evaluate $\psi'_{\{\ssip_{i,j}, 1-\ssip_{i,\jprime}; \stp_{i-\half,j-\half}\}}$ (the transformed wavefunction), where we consider the first expression of Eq.~\eqref{eq:psi_prime_KT}. Flipping $\ssip_{i,\jprime}$ to $1-\ssip_{i,\jprime}$ can be exactly compensated by flipping $\ssi_{i,\jprime}$ to $1-\ssi_{i,\jprime}$, since these variables always appear together in the combination $\ssip_{i,\jprime} + \ssip_{i,\jprime}$ (see Eq.~\eqref{eq:A_KT_def_1}). Therefore, we find that
\begin{align}
    u'^x_{\sigma,\jprime} = u^x_{\sigma,\jprime}, \quad u'^y_{\sigma,\iprime} = u^y_{\sigma,\iprime}.
\end{align}

The same reasoning holds for $\stp_{i-\half,\jprime-\half}$ as well. Therefore, we also find that
\begin{align}
    u'^x_{\tau,\jprime-\half} = u^x_{\tau,\jprime-\half}, \quad u'^y_{\tau,\iprime-\half} = u^y_{\tau,\iprime-\half}.
\end{align}
In other words, the subsystem KT transformation preserves all subsystem symmetry charges.

\paragraph{Effect of KT transformation on the twist sectors:} Next, to see the effect on the twist variables, let us analyze the transformed wavefunction $\psi'_{\{\ssip_{i,j\neq j'}, 1-\ssip_{i,\jprime}; \stp_{i-\half,j-\half}\}}$. When we change $\ssip_{i,\jprime}$ to $1-\ssip_{i,\jprime}$, the effect is to multiply the entire wavefunction by a phase factor $(-1)^{\phi_{\sigma}}$, with $\phi_{\sigma}$ defined as (see the expression of $\mathcal{A}_{\text{KT}}$ Eq.~\eqref{eq:A_KT_def_1})
\begin{align}
\label{eq:phi_sigma}
    \phi_{\sigma} & = \sum_{i=1}^{L_x} \left( \st_{i-\half,\jprime-\half}+ \st_{i+\half,\jprime-\half} + \st_{i-\half,\jprime+\half} + \st_{i+\half,\jprime+\half} \right. \nonumber \\
  & \phantom{=} {} \left. + \stp_{i-\half,\jprime-\half} + \stp_{i+\half,\jprime-\half}  + \stp_{i-\half,\jprime+\half} + \stp_{i+\half,\jprime+\half}  \right) \nonumber\\
    & = t^x_{\tau, \jprime-\half} + t^x_{\tau, \jprime+\half} + t'^x_{\tau, \jprime-\half} + t'^x_{\tau, \jprime+\half} \nonumber\\
    & = \mt^x_{\tau, \jprime} + \mt'^x_{\tau, \jprime}.
\end{align}
Therefore, we conclude that
\begin{align}
\label{eq:twist_KT_1}
    u'^x_{\sigma,\jprime} = \mt^x_{\tau, \jprime} + \mt'^x_{\tau, \jprime}.
\end{align}
Similarly, we can show that
\begin{align}
\label{eq:twist_KT_2}
    u'^y_{\sigma,\iprime} = \mt^y_{\tau, \iprime} + \mt'^y_{\tau, \iprime}.
\end{align}

Next, we look at how the transformed wavefunction behaves under a shift of $\stp_{i-\half,\jprime-\half}$ by one unit. Specifically, we consider $\psi' _{{\{\ssip_{i,j}; \stp_{i-\half,j \neq j'-\half}, 1-\stp_{i-\half,\jprime-\half}\}}}$ and use the second form of Eq.~\eqref{eq:psi_prime_KT} to evaluate it. Performing this shift alters the state only by introducing a phase factor, namely $(-1)^{\phi_{\tau}}$, where the quantity $\phi_{\tau}$ is defined as follows (see the expression of $\widetilde{\mathcal{A}}_{\text{KT}}$ Eq.~\eqref{eq:A_KT_def_2}):
\begin{align}
\label{eq:phi_tau}
    \phi_{\tau} & = \sum_{i=1}^{L_x} \left(  \ssi_{i-1,\jprime-1} + \ssi_{i,\jprime-1} + \ssi_{i-1,\jprime} + \ssi_{i,\jprime} +  \ssip_{i-1,\jprime-1} + \ssip_{i,\jprime-1} + \ssip_{i-1,\jprime} + \ssip_{i,\jprime} \right)\nonumber\\
    & = t^x_{\sigma, \jprime-1} + t^x_{\sigma, \jprime} + t'^x_{\sigma, \jprime-1} + t'^x_{\sigma, \jprime} \nonumber\\
    & = \mt^x_{\sigma, \jprime-\half} + \mt'^x_{\sigma, \jprime-\half}.
\end{align}
We therefore arrive at the result
\begin{align}
\label{eq:twist_KT_3}
    u'^x_{\tau,\jprime-\half} = \mt^x_{\sigma, \jprime-\half} + \mt'^x_{\sigma, \jprime-\half}.
\end{align}
In a completely analogous manner, one can verify that
\begin{align}
\label{eq:twist_KT_4}
    u'^y_{\tau,\iprime-\half} = \mt^y_{\sigma, \iprime-\half} + \mt'^y_{\sigma, \iprime-\half}.
\end{align}

Since the subsystem KT transformation preserves all subsystem symmetry sectors, the $u'$ variables appearing on the left-hand sides of Eqs.~\eqref{eq:twist_KT_1}, \eqref{eq:twist_KT_2}, \eqref{eq:twist_KT_3}, and \eqref{eq:twist_KT_4} can be replaced by their corresponding $u$’s. In summary, the KT transformation leaves the symmetry sectors unchanged:
\begin{align}
\label{eq:symmetry_mapping_KT}
    \left \{u'^x_{\sigma,j},  u'^y_{\sigma,i}, u'^x_{\tau,j-\half}, u'^y_{\tau,i-\half}\right \} = \left \{u^x_{\sigma,j}, u^y_{\sigma,i}, u^x_{\tau,j-\half}, u^y_{\tau,i-\half}\right \}
\end{align}
while the twist sectors get transformed in a way that depends on both the twist and symmetry sectors prior to the KT action:
\begin{align}
\label{eq:twist_mapping_KT}
    \left \{ \mt'^x_{\sigma,j-\half}, \mt'^y_{\sigma,i-\half}, \mt'^x_{\tau,j}, \mt'^y_{\tau,i} \right \} = \left \{\mt^x_{\sigma,j-\half} + u^x_{\tau,j-\half}, \mt^y_{\sigma,i-\half} + u^y_{\tau,i-\half}, \mt^x_{\tau,j} + u^x_{\sigma,j}, \mt^y_{\tau,i} + u^y_{\sigma,i} \right \}.
\end{align}
Hence, the subsystem KT transformation leaves all subsystem $\mathbb{Z}_2 \times \mathbb{Z}_2$ symmetry charges unchanged.
However, it shifts each $\sigma$ twist by the corresponding $\tau$ charge and each $\tau$ twist by the corresponding $\sigma$ charge. In Appendix~\ref{app:KT_QFT}, we provide an alternative derivation of how the subsystem KT transformation acts on symmetry-twist sectors, demonstrating that the mappings in Eqs.~\eqref{eq:symmetry_mapping_KT} and \eqref{eq:twist_mapping_KT} arise naturally from the partition-function formulation and a combined gauging and stacking SSPT analysis.

\paragraph{Non-invertibility of the KT transformation from the symmetry-twist mapping:}
Let us now explain how the mapping of symmetry twist sectors, following Ref.~\cite{Li_2023_KT}, makes it clear that the subsystem KT transformation is not unitary. For a meaningful physical transformation, we first fix the boundary conditions (or twist sectors). In other words, we project the final state (after the subsystem KT transformation) onto a chosen twist sector. To illustrate this, let us consider a concrete example, namely the final state in the untwisted sectors,
\begin{align}
\label{eq:untwisted_final}
    \mt'^x_{\sigma,j-\half} = \mt'^y_{\sigma,i-\half} = \mt'^x_{\tau,j} = \mt'^y_{\tau,i} = 0.
\end{align}

Using the twist-sector mapping in Eq.~\eqref{eq:twist_mapping_KT}, we find that, after the subsystem KT transformation, the projection retains only those initial states for which
\begin{align}
    \mt^x_{\sigma,j-\half} = u^x_{\tau,j-\half},\, \mt^y_{\sigma,i-\half} = u^y_{\tau,i-\half},\, \mt^x_{\tau,j} = u^x_{\sigma,j}, \, \mt^y_{\tau,i} = u^y_{\sigma,i}.
\end{align}
All other states are removed by the projection.

Now, consider an initial state in the untwisted sector, specified by
\begin{align}
\label{eq:untwisted_initial}
    \mt^x_{\sigma,j-\half} = \mt^y_{\sigma,i-\half} = \mt^x_{\tau,j} = \mt^y_{\tau,i} = 0.
\end{align}
but carrying subsystem $\mathbb{Z}_2$ odd charges, for example $u^x_{\sigma,j} = 1$. From the mapping in Eq.~\eqref{eq:twist_mapping_KT}, the transformed twist variables are given by $\mt'^x_{\tau,j} =1$, which corresponds to a twisted configuration. If we subsequently project the transformed state to the untwisted sector, this component is eliminated. Therefore, the subsystem KT transformation annihilates all untwisted states that carry nonzero subsystem charges when the projection to the untwisted sector is performed. Therefore, the probability is generally not preserved under the subsystem KT transformation, indicating that $\KT$ is nonunitary. Moreover, if a given sector is annihilated, the original state cannot be recovered from its image, which means the transformation is also non-invertible.

\subsection{Non-invertible fusion rules for the KT operators (with closed boundary conditions)}
\label{sec:fusion_KT_sub}
Our next step is to examine the fusion rules, which likewise reveal the non-unitarity and non-invertibility of $\KT$ defined in Eq.~\eqref{eq:N_KT_final_1} under the closed boundary conditions. Here, we use the terminology ``fusion'' in a loose sense, referring to how the composition of operators acts on the Hilbert space. This is distinct from the notion of the fusion of the symmetry defect operators introduced e.g. in Ref.~\cite{Seiberg_LSM_SciPostPhys2024}. Note that the KT transformation $\KT$ is not even a symmetry of the model -- it is just a duality operator and hence it does not make sense to talk about the `symmetry defects' in this context.

As a starting point, let us examine the composition of the KT duality and the generator of the subsystem symmetry $\KT \times U^x_{\sigma,j}$. From the definition given in Eq.~\eqref{eq:subsym_operator_sig_tau}, we note that
\begin{align}
   \KT \ket{\{\sigma_{i,j\neq j'},1-\sigma_{i,j'};\st_{i-\half,j-\half}\}} = \KT U^x_{\sigma,\jprime} \ket{\{\ssi_{i,j}, \st_{i-\half,j-\half}\}}.
\end{align}

From Eq.~\eqref{eq:N_KT_final_1}, the action of $\KT$ shows that shifting $\ssi_{i,\jprime}$ by $1$ produces a phase factor $(-1)^{\phi_{\sigma}}$, with $\phi_{\sigma} = \mt^x_{\tau, \jprime} + \mt'^x_{\tau, \jprime}$ as defined in Eq.~\eqref{eq:phi_sigma}. Consequently,
\begin{align}
    \KT U^x_{\sigma,\jprime} \ket{\{\ssi_{i,j}, \st_{i-\half,j-\half}\}} = (-1)^{\mt^x_{\tau, \jprime} + \mt'^x_{\tau, \jprime}} \KT \ket{\{\ssi_{i,j}, \st_{i-\half,j-\half}\}},
\end{align}
which implies the fusion rule
\begin{align}
\label{eq:fusion_NU_1}
    \KT \times U^x_{\sigma,j} = (-1)^{\mt^x_{\tau, j} + \mt'^x_{\tau, j}} \KT.
\end{align}
Using an analogous calculation, we obtain that
\begin{align}
\label{eq:fusion_NU_2}
    \KT \times U^y_{\sigma,i} & = (-1)^{\mt^y_{\tau, i} + \mt'^y_{\tau, i}} \KT, \nonumber\\
    \KT \times U^x_{\tau,j-\half} & = (-1)^{\mt^x_{\sigma, j-\half} + \mt'^x_{\sigma, j-\half}} \KT, \nonumber\\
    \KT \times U^y_{\tau,i-\half} & = (-1)^{\mt^y_{\sigma, i-\half} + \mt'^y_{\sigma, i-\half}} \KT.
\end{align}

We now highlight the utility of the fusion rules in Eqs.~\eqref{eq:fusion_NU_1} and \eqref{eq:fusion_NU_2}. In fact, all mappings between twist sectors given in Eq.~\eqref{eq:twist_mapping_KT} can be recovered directly from these fusion rules. As an illustration, consider an eigenstate of $U^x_{\sigma,j}$ satisfying
\begin{align}
    U^x_{\sigma,j} \ket{\psi} = (-1)^{u^x_{\sigma,j}} \ket{\psi}.
\end{align}
From Eq.~\eqref{eq:fusion_NU_1} it follows that
\begin{align}
    (-1)^{\mt^x_{\tau, j} + \mt'^x_{\tau, j}} \left( \KT \ket{\psi} \right) = (-1)^{u^x_{\sigma,j}} \left( \KT \ket{\psi} \right),
\end{align}
which leads to
\begin{align}
    \mt'^x_{\tau, j} = \mt^x_{\tau, j} + u^x_{\sigma,j}.
\end{align}
By a similar argument, all other twist sector mappings in Eq.~\eqref{eq:twist_mapping_KT} can be rederived from Eq.~\eqref {eq:fusion_NU_2}.

We now turn our attention to the fusion rule $\KT \times \KT$. Since the calculation is lengthy, it is presented in the Appendix~\ref{app:fusion_KTKT}. Here, we simply state the result:
\begin{align}
\label{eq:fusion_KTKT}
    \KT \times \KT = 4^{2 L_xL_y - (L_x+L_y+1)} \mathcal{P}_{\sigma} \mathcal{P}_{\tau},
\end{align}
where $\mathcal{P}_{\sigma}, \mathcal{P}_{\tau}$ are projectors onto a definite symmetry-twist sector, given by
\begin{align}
    \mathcal{P}_{\sigma} & = \half \prod_{j=1}^{L_y} \left[1+ (-1)^{\mt^x_{\tau, j} + \mt'^x_{\tau, j}} U^x_{\sigma,j}\right]\prod_{i=1}^{L_x} \left[1 + (-1)^{\mt^y_{\tau, i} + \mt'^y_{\tau, i}} U^y_{\sigma,i}\right], \nonumber\\
    \mathcal{P}_{\tau} & = \half \prod_{j=1}^{L_y} \left[1 + (-1)^{\mt^x_{\sigma, j-\half} + \mt'^x_{\sigma, j-\half}} U^x_{\tau,j-\half} \right] \prod_{i=1}^{L_x} \left[1 + (-1)^{\mt^y_{\sigma, i-\half} + \mt'^y_{\sigma, i-\half}} U^y_{\tau,i-\half} \right].
\end{align}

We now address the question of whether the subsystem KT transformation defined in Eq.~\eqref{eq:N_KT_final_1} is unitary. By definition, unitarity would require that $\KT^\dagger \KT$ be exactly equal to the identity operator. Examining Eq.~\eqref{eq:fusion_KTKT}, however, we see that its right-hand side is not identical to the identity. This deviation means that $\KT$ fails to satisfy the unitarity condition, and is therefore non-unitary. In particular, under periodic boundary conditions (untwisted) as specified in \eqref{eq:untwisted_final} and \eqref{eq:untwisted_initial}, $\KT$ annihilates any state that is odd with respect to any of the subsystem $\mathbb{Z}_2$ symmetry operators defined in Eq.~\eqref{eq:subsym_operator_sig_tau}. In other words, the kernel of the operator $\KT$ is not empty, thus establishing its non-invertibility.

\section{Invertability of the Subsystem Kennedy-Tasaki transformation with open boundary conditions} 
\label{sec:KT_open}

Let us now examine the subsystem KT transformation for a spin-$1/2$ system on an open square lattice. Much like the subsystem KW transformation, we will see that while the $\KT$ operator is non-unitary and non-invertible in general, it becomes unitary when considered under free open boundary conditions. We consider an open square lattice of dimensions $L_x \times L_y$, where just like in Sec.~\ref{sec:KT_SSPT_strong}
the $\sigma_{i,j}$ spin degrees of freedom live on the vertices of the lattice, while $\tau_{i-\half,j-\half}$ spins reside on the centers of plaquettes. We begin by imposing free boundary conditions on Eqs.~\eqref{eq:N_KT_final_1} and \eqref{eq:A_KT_def_1}, where contributions involving sites outside the square lattice are omitted. Retaining in the exponent only those terms that are entirely supported on the open square lattice, we obtain that the spin eigenstates transform as follows under the KT transformation:
\begin{align}
\label{eq:N_KT_open}
    \KT^{\text{open}} \ket{\{\ssi_{i,j}, \st_{i-\half,j-\half}\}} & = \frac{1}{2^{L_x+L_y}} \sum_{\{\ssip_{i,j},\stp_{i-\half,j-\half}\}} (-1)^{\mathcal{A}^{\text{open}}_{\text{KT}} (\{\sigma,\sigma';\tau,
    \tau'\})} \ket{\{\ssip_{i,j}, \stp_{i-\half,j-\half}\}},
\end{align}
where the phase $\mathcal{A}^{\text{open}}_{\text{KT}} (\{\sigma,\sigma';\tau,
    \tau'\})$ is defined as
\begin{align}
\label{eq:A_KT_open}
    \mathcal{A}^{\text{open}}_{\text{KT}} (\{\sigma,\sigma';\tau,
    \tau'\}) = & \sum_{i=1}^{L_x} \sum_{j=1}^{L_y} \left(\ssi_{i,j}+ \ssip_{i,j}\right) \left( \st_{i-\half,j-\half} + \stp_{i-\half,j-\half}\right) + \sum_{i=1}^{L_x-1} \sum_{j=1}^{L_y} \left(\ssi_{i,j}+ \ssip_{i,j}\right) \nonumber\\
    & \times \left( \st_{i+\half,j-\half} + \stp_{i+\half,j-\half}\right) + \sum_{i=1}^{L_x} \sum_{j=1}^{L_y-1} \left(\ssi_{i,j}+ \ssip_{i,j}\right) \left( \st_{i-\half,j+\half} + \stp_{i-\half,j-\half}\right) \nonumber\\
    & + \sum_{i=1}^{L_x-1} \sum_{j=1}^{L_y-1} \left(\ssi_{i,j}+ \ssip_{i,j}\right) \left( \st_{i+\half,j+\half}+ \stp_{i+\half,j+\half}\right). 
\end{align}
Note that the terms in Eq.~\eqref{eq:A_KT_def_1} containing twisted-sector labels $\left(t^x_{\sigma,j} + t'^x_{\sigma,j}\right)$, $\left(t^y_{\sigma,i} + t'^y_{\sigma,i}\right)$, and $\left(t^{xy}_{\sigma} + t'^{xy}_{\sigma}\right)$ have been omitted, as these correspond to twisted boundary conditions that are meaningful only on closed lattices.

By carrying out a calculation analogous to that in Sec.~\ref{sec:KW_operator_open}, where we established the unitarity of the subsystem KW transformation on an open square lattice, we can similarly demonstrate that
\begin{align}
    \bra{\{\ssipp_{i,j}, \stpp_{i-\half,j-\half}\}} \KT^{\text{open} \, \dagger} \KT^{\text{open}} \ket{\{\ssi_{i,j}, \st_{i-\half,j-\half}\}}  = \prod_{i=1}^{L_x} \prod_{j=1}^{L_y} \delta_{\ssi_{i,j},\ssipp_{i,j}} \delta_{\st_{i-\half,j-\half},\stpp_{i-\half,j-\half}}.
\end{align}
Hence, $ \KT^{\text{open}}$ is unitary and invertible.

We now proceed to discuss the mapping between Pauli operators under $\KT^{\text{open}} = \NKW^{\text{open}\, \dagger} \allowbreak U^{\text{open}}_{\text{DW}} \NKW^{\text{open}} $ on an open square lattice. The subsystem KW operator $\NKW^{\text{open}}$ on an open square lattice is given in Eq.~\eqref{eq:N_sub_open}, which we have already discussed in Sec.~\ref{sec:KW_operator_open}. For the case of an open square lattice, the domain-wall decoration operator $U^{\text{open}}_{\text{DW}}$ is obtained by starting from Eq.~\eqref{eq:U_DW_dual} and retaining in the exponent only those contributions from Eqs.~\eqref{eq:A_U_1} and \eqref{eq:A_U_2} that are completely supported within the open lattice.

Under $\NKW^{\text{open}}$, both $\sigma^x_{i,j}$ and $\tau^x_{i-\half,j-\half}$ are mapped to products of $\hsig^z_{i-\half,j-\half}$ and $\htau^z_{i,j}$ operators, respectively (see Eq.~\eqref{eq:Pauli_map_KW_open}), which are diagonal in the $(\hsig^z, \htau^z)$ basis. Because such diagonal operators commute with the domain-wall operator $U_{\mathrm{DW}}$ appearing in $\KT^{\text{open}}$, they are mapped back to the original $\sigma^x_{i,j}$ and $\tau^x_{i-\half,j-\half}$ operators by $\NKW^{\text{open}\, \dagger}$. As a result, the $x$-components remain unchanged:
\begin{align}
\label{eq:Pauli_map_KT_open_1}
    \KT^{\text{open}} \sigma^x_{i,j} \KT^{\text{open}\, \dagger} & = \sigma^x_{i,j}, \nonumber\\
    \KT^{\text{open}} \tau^x_{i-\half,j-\half} \KT^{\text{open}\, \dagger} & = \tau^x_{i-\half,j-\half}.
\end{align}

From Eq.\eqref{eq:Pauli_map_KT_open_1}, it follows that the subsystem $\mathbb{Z}_2 \times \mathbb{Z}_2$ symmetry generators \eqref{eq:subsym_operator_sig_tau} remain invariant under $\KT^{\text{open}}$. This marks an important distinction between the subsystem KT and subsystem KW transformations on an open square lattice: while the subsystem KT transformation preserves the symmetry generators, the subsystem KW transformation maps them to local boundary operators (see Eq.\eqref{eq:U_map_KW_open}).

Let us now discuss how $\NKW^{\text{open}}$ acts on $\sigma^z_{i,j}$ and $\tau^z_{i-\half,j-\half}$. From Eq.~\eqref{eq:Pauli_map_KW_open}, we find that
\begin{align}
\label{eq:Pauli_map_KW_open_2}
    \NKW^{\text{open}} \sigma^z_{i,j} \NKW^{\text{open}\, \dagger} & = \prod_{\iprime=1}^{i}  \prod_{\jprime=1}^{j} \hsig^x_{\iprime-\half,\jprime-\half}, \nonumber\\
    \NKW^{\text{open}} \tau^z_{i-\half,j-\half} \NKW^{\text{open}\, \dagger} & = \prod_{\iprime=i}^{L_x}  \prod_{\jprime=j}^{L_y} \htau^x_{\iprime,\jprime}.
\end{align}

The action of $U^{\text{open}}_{\text{DW}}$ on the spins in the dual Hilbert space is given by
\begin{align}
\label{eq:Pauli_map_U_open_sigx}
    &  U^{\text{open}}_{\text{DW}} \hsig^x_{i-\half,j-\half} U^{\text{open}\, \dagger}_{\text{DW}} = \begin{cases} \htau^z_{i-1,j-1} \htau^z_{i,j-1} \htau^z_{i-1,j} \htau^z_{i,j} \hsig^x_{i-\half,j-\half}, & i=2,\dots,L_x,\, j = 2,\dots,L_y\\\htau^z_{1,j-1} \htau^z_{1,j} \hsig^x_{\half,j-\half}, &i=1,\, j = 2,\dots,L_y \\\htau^z_{i-1,1} \htau^z_{i,1} \hsig^x_{i-\half,\half}, &i=2,\dots,L_x,\, j = 1 \\ \htau^z_{1,1} \hsig^x_{1,1}, & i=1,\, j=1\end{cases},
\end{align}
and 

\begin{align}
\label{eq:Pauli_map_U_open_taux}
    &  U^{\text{open}}_{\text{DW}} \htau^x_{i,j} U^{\text{open}\, \dagger}_{\text{DW}} = \begin{cases} \hsig^z_{i-\half,j-\half} \hsig^z_{i+\half,j-\half} \hsig^z_{i-\half,j+\half} \hsig^z_{i+\half,j+\half} \htau^x_{i,j}, & i=1,\dots,L_x-1,\, j = 1,\dots,L_y-1\\\hsig^z_{L_x-\half,j-\half} \hsig^z_{L_x-\half,j+\half} \htau^x_{L_x,j}, &i=L_x,\, j = 1,\dots,L_y-1 \\\hsig^z_{i-\half,L_y-\half} \hsig^z_{i+\half,L_y-\half} \htau^x_{i,L_y}, &i=1,\dots,L_x-1,\, j = L_y \\ \hsig^z_{L_x-\half,L_y-\half} \htau^x_{L_x,L_y}, & i=L_x,\, j=L_y\end{cases}.
\end{align}

Combining Eqs.~\eqref{eq:Pauli_map_KW_open_2}, \eqref{eq:Pauli_map_U_open_sigx} and \eqref{eq:Pauli_map_U_open_taux}, we obtain
\begin{align}
\label{eq:Pauli_map_KT_open_2}
    \KT^{\text{open}} \sigma^z_{i,j} \KT^{\text{open}\, \dagger} & =\left( \prod_{\iprime=1}^{i}  \prod_{\jprime=1}^{j} \tau^x_{\iprime-\half,\jprime-\half} \right) \sigma^z_{i,j} \nonumber\\
    \KT^{\text{open}} \tau^z_{i-\half,j-\half} \KT^{\text{open}\, \dagger} & = \tau^z_{i-\half,j-\half} \left( \prod_{\iprime=i}^{L_x}  \prod_{\jprime=j}^{L_y} \sigma^x_{\iprime,\jprime} \right).
\end{align}
We note that a similar mapping of the $\sigma^z$ and $\tau^z$ operators under the subsystem KT transformation also appears in the closed-lattice case and has been discussed previously in Refs.~\cite{Doherty_PRL_2009, Aswin_KT_sub_PRB2024}.The relevance of these maps becomes central in two places: first, in establishing the duality between the SSSB phase and the strong SSPT phase on an open square lattice (Sec.~\ref{sec:gapped_SSPT_open}); and second, in deriving the operator-level expressions for the transformed bulk and edge invariants that diagnose the strong SSPT phase (Sec.~\ref{sec:bulk-edge}).

Similarly, by applying Eqs.~\eqref{eq:Pauli_map_KT_open_1} and \eqref{eq:Pauli_map_KT_open_2}, we find that the two copies of the Xu-Moore model in Eq.~\eqref{eq:H_TXM} are mapped, under the open-boundary subsystem KT transformation $\KT^{\text{open}}$ defined in Eq.~\eqref{eq:N_KT_open}, to the two-dimensional cluster model in a transverse Zeeman field in Eq.~\eqref{eq:H_cluster} on an open square lattice.

We now examine how the invertibility or unitarity of the KT transformation manifests when comparing the models in Eqs.~\eqref{eq:H_TXM} and \eqref{eq:H_cluster} on such open geometries. In the trivial paramagnetic phase, both models clearly share the same nondegenerate ground state. In contrast, for small transverse fields, where the system enters the ordered (SSSB) regime, the open-boundary ground-state degeneracy becomes $4^{L_x + L_y - 1}$. Remarkably, this same degeneracy appears in the KT-dual SSPT phase as well, where it originates from protected edge states. The edge structure of the strong SSPT phase in open geometries is discussed in detail in Sec.~\ref{sec:SSPT_edge}. While this observation does not constitute a rigorous proof, it strongly suggests that the subsystem KT transformation acts unitarily in this context, as it preserves the ground-state degeneracy structure across the two models.

\section{Correspondence between SSSB ground states and SSPT edge degeneracy}
\label{sec:gapped_SSPT_open}

The subsystem KT transformation not only connects the bulk Hamiltonians of the SSSB and SSPT phases but also maps their boundary degrees of freedom. In this section, we show that under the open transformation $\KT^{\text{open}}$, the extensive ground-state degeneracy of the SSSB phase with open boundary conditions is converted into the boundary-protected degeneracy of the strong $\mathbb{Z}_2 \times \mathbb{Z}_2$ SSPT.   We derive the resulting boundary symmetry algebra, which exhibits a \textit{chain-like anticommutation pattern} forming a non-factorizable projective representation of the subsystem symmetry. This provides a microscopic signature that distinguishes the strong SSPT from its weak, stackable counterpart.

To set the stage, we write the Hamiltonian of the two decoupled plaquette Ising models, Eq.~\eqref{eq:H_SSSB}, now defined on an open square lattice that realizes the gapped $\mathbb{Z}_2 \times \mathbb{Z}_2$ SSSB phase:
\begin{align}
\label{eq:H_SSSB_open}
    H_{\text{SSSB}}^{\text{open}}  = - \sum_{i=2}^{L_x} \sum_{j=2}^{L_y} \sigma^z_{i-1,j-1} \sigma^z_{i,j-1} \sigma^z_{i-1,j} \sigma^z_{i,j} -  \sum_{i=1}^{L_x-1} \sum_{j=1}^{L_y-1} \tau^z_{i-\half,j-\half} \tau^z_{i+\half,j-\half} \tau^z_{i-\half,j+\half} \tau^z_{i+\half,j+\half}.
\end{align}

Each plaquette term enforces a local ferromagnetic constraint, minimizing the energy when the product of the four spins around every plaquette equals $+1$. Consequently, the $\sigma$- and $\tau$-sectors each possess 
$2^{L_x+L_y-1}$ degenerate ground states, giving an overall ground-state degeneracy $4^{L_x + L_y - 1}$. The generators of the subsystem $\mathbb{Z}_2 \times \mathbb{Z}_2$ symmetry, defined in Eq.~\eqref{eq:subsym_operator_sig_tau}, act by flipping all spins along a fixed row or column.  
Because $H_{\text{SSSB}}^{\text{open}}$ commutes with the symmetry generators, the entire ground-state manifold can be organized into simultaneous eigenstates of these subsystem symmetries.  

Explicit symmetry-eigenstate ground states can be constructed by projecting a reference configuration $\ket{\mathrm{ref}}$ (e.g., all spins up) into the desired $\sigma$-symmetry sector labeled by $u^x_{\sigma,j}$ and $u^y_{\sigma,i}$ as follows:
\begin{equation}
    \ket{\sigma; \{u^x_{\sigma,j},u^y_{\sigma,i}\}}
    =
    \frac{1}{2^{L_x+L_y-1}}
    \left[\prod_{j=1}^{L_y} \left(1 + (-1)^{u^x_{\sigma,j}} U^x_{\sigma,j}\right) \right]
    \left[\prod_{i=1}^{L_x}  \left(1 + (-1)^{u^y_{\sigma,i}} U^y_{\sigma,i}\right) \right]
    \ket{\mathrm{ref}}_{\sigma},
\end{equation}
and similarly for the $\tau$-sector.  
The full ground states are tensor products of the eigenfunctions in the $\sigma$ and $\tau$ sectors,
\begin{equation}
\label{eq:sigma-tau-tensor-product}
    \ket{\Psi_{\{u^x_{\sigma},u^y_{\sigma};u^x_{\tau},u^y_{\tau}\}}}
    =
    \ket{\sigma; \{u^x_{\sigma,j},u^y_{\sigma,i}\}}
    \otimes
    \ket{\tau; \{u^x_{\tau,j-\frac12},u^y_{\tau,i-\frac12}\}}.
\end{equation}

Under the open subsystem KT transformation $\KT^{\text{open}}$ defined in the previous section, this subsystem symmetry-broken manifold is mapped to the ground-state subspace of the two-dimensional cluster model, Eq.~\eqref{eq:H_SSPT_2d}, now defined on an open square lattice. The resulting model, obtained using the Pauli-operator mapping defined in Eqs.~\eqref{eq:Pauli_map_KT_open_1} and \eqref{eq:Pauli_map_KT_open_2}, realizes the gapped $\mathbb{Z}_2 \times \mathbb{Z}_2$ strong SSPT and takes the explicit form as given in Eq.~\eqref{eq:H_SSPT_open}.

\section{Bulk and Edge Invariants and Their Invariance under the KT Transformation}
\label{sec:bulk-edge}

Having introduced the bulk and edge invariants characterizing weak and strong SSPT phases in Sec.~\ref{sec:invariants}, we now study their behavior under the subsystem KT transformation. Since the subsystem KT duality generally maps local operators to highly nonlocal ones, it is not immediately obvious whether these SSPT invariants remain unchanged. In this section, we show that both the bulk invariant $\beta(g)$ in Eq.~\eqref{eq:beta_def} and the edge invariant $\phi_{\mathrm{top}}$ in Eq.~\eqref{eq:phi_top_def} are preserved under the subsystem KT transformation, demonstrating that the underlying algebraic structure characterizing SSPT order remains invariant under the duality.


\subsection{Invariance of bulk invariants under the subsystem KT transformation}
Although the subsystem KT transformation maps the strong SSPT phase to the SSSB phase, the transformed quantity $\beta_{ij}'^{R}(g)$ obtained from the image of the local operators under $\KT^{\text{open}}$ does not correspond to any intrinsic bulk invariant of the SSPT itself.  
Indeed, the KT transformation renders the local repair operator $V_{ij}(g)$ highly nonlocal, as seen below, even though the map acts locally at the operator level.  
Nevertheless, it is instructive to examine the explicit transformed form of these operators and to verify that the commutation structure defining $\beta_{ij}^{R}(g)$ is preserved.

Let us focus on the nontrivial case $\beta_{ij}^{R}(g_\sigma g_\tau) = -1$ in Eq.~\eqref{eq:beta_SSPT} in the 2d cluster state. Under the open-boundary KT transformation, the ground state $\ket{\psi}_{\text{\smaller{SSPT}}}$ is mapped to $\ket{\psi}_{\text{\smaller{SSSB}}}$, while the half-space symmetry operator in Eq.~\eqref{eq:half_space_sym} and local repair operators in Eq.~\eqref{eq:repair_operator} transform as
\begin{align}
\mathcal{S}_i^R(g_\sigma g_\tau) & \stackrel{\smaller \KT^{\text{open}}}{\longrightarrow} \mathcal{S}_i'^R(g_\sigma g_\tau) = \prod_{i'=i}^{L_x} \prod_{j=1}^{L_y}
\sigma^x_{i',j}\tau^x_{i'-\half,j-\half}, \nonumber\\
V_{ij}(g_\sigma g_\tau) & \stackrel{\smaller \KT^{\text{open}}}{\longrightarrow} V'_{ij}(g_\sigma g_\tau) = \left[\tau^z_{i-\half,j-\half} \left( \prod_{\iprime=i}^{L_x}  \prod_{\jprime=j}^{L_y} \sigma^x_{\iprime,\jprime} \right)\right] \left[\left( \prod_{\iprime=1}^{i-1}  \prod_{\jprime=1}^{j-1} \tau^x_{\iprime-\half,\jprime-\half} \right) \sigma^z_{i-1,j-1}\right].
\end{align}
Although $V'_{ij}(g_\sigma g_\tau)$ is manifestly nonlocal, its leading local component $\tau^z_{i-\half,j-\half}$ anticommutes with the $\tau^x_{i-\half,j-\half}$ contained in $\mathcal{S}_i'^R(g_\sigma g_\tau)$, while all other factors commute.
As a result, the commutation relation between $\mathcal{S}_i'^R(g_\sigma g_\tau)$ and $V'_{ij}(g_\sigma g_\tau)$ remains identical to that of the original operators, giving
\begin{align}
\beta_{ij}'^{R}(g_\sigma g_\tau) = -1.
\end{align}
Similarly,
\begin{align}
\beta_{ij}'^{R}(1) = \beta_{ij}'^{R}(g_\sigma) = \beta_{ij}'^{R}(g_\tau) = 1.
\end{align}

These results demonstrate that while one should no longer think of  $\beta_{ij}'^{R}(g)$ as being the intrinsic bulk invariants of the SSPT -- indeed, the KT-dual model in Eq.~\eqref{eq:H_SSSB_open} is not at all an SSPT -- their algebraic values nevertheless coincide with those of $\beta_{ij}^{R}(g)$ in Eq.~\eqref{eq:beta_SSPT}, reflecting the preservation of the underlying commutation structure under the subsystem KT transformation.


\subsection{Invariance of edge projective representation under the KT transformation}
Having shown that the bulk invariant $\beta(g)$ remains unchanged under the subsystem KT transformation, we now verify that its boundary counterpart, the edge commutation phase $\phi_{\mathrm{top}}$ as defined in Sec.~\ref{ssec:edge_invariants}, is likewise invariant. Applying the Pauli operator mappings in Eqs.~\eqref{eq:Pauli_map_KT_open_1} and \eqref{eq:Pauli_map_KT_open_2}, the edge operators in Eq.~\eqref{eq:edge_SSPT} transform as $U_{\sigma, \tau} \stackrel{\smaller \KT^{\text{open}}}{\longrightarrow} U'_{\sigma, \tau}$, where the transformed operators $U'_{\sigma, \tau}$ are explicitly given by

\begin{align}
\label{eq:edge_SSSB}
    \left(U'^x_{\sigma,j}\right)_L & = \tau^z_{\half, j-\half} \tau^z_{\half, j+\half} \left(\prod_{i=1}^{L_x} \sigma^x_{i,j} \right), \, \left(U'^x_{\sigma,j}\right)_R = \tau^z_{L_x-\half, j-\half} \tau^z_{L_x-\half, j+\half}, \, (j=1,\dots,L_y-1), \nonumber\\
    \left(U'^y_{\sigma,i}\right)_B & = \tau^z_{i-\half,\half} \tau^z_{i+\half,\half} \left(\prod_{j=1}^{L_y} \sigma^x_{i,j} \right),  \, \left(U'^y_{\sigma,i}\right)_T = \tau^z_{i-\half,L_y-\half} \tau^z_{i+\half,L_y-\half}, \, (i=1,\dots,L_x-1),\nonumber\\
    \left(U'^x_{\tau,j-\half}\right)_L & = \sigma^z_{1,j-1} \sigma^z_{1,j},  \, \left(U'^x_{\tau,j-\half}\right)_R = \left(\prod_{i=1}^{L_x} \tau^x_{i-\half,j-\half} \right) \sigma^z_{L_x,j-1} \sigma^z_{L_x,j}, \, (j=2,\dots,L_y), \nonumber\\
    \left(U'^y_{\tau,i-\half}\right)_B & = \sigma^z_{i-1,1} \sigma^z_{i,1},  \, \left(U'^y_{\tau,i-\half}\right)_T = \left(\prod_{j=1}^{L_y} \tau^x_{i-\half,j-\half} \right) \sigma^z_{i-1,L_y} \sigma^z_{i,L_y}\, (i=1,\dots,L_x).
\end{align}
On the top boundary, $U^{\mathrm{top}}_i(g_\sigma)$ becomes a pair of $\tau^z$ operators on the adjacent plaquettes, while $U^{\mathrm{top}}_{i-\half}(g_\tau)$ acquires an extended $\tau^x$ string along the column together with $\sigma^z$ endcaps.  
Despite these nonlocal attachments, the local commutation structure at the edge remains identical:  
the overlap of $\tau^z$ from $U'^{\mathrm{top}}_i(g_\sigma)$ with $\tau^x$ from $U'^{\mathrm{top}}_{i\pm\half}(g_\tau)$ on the same plaquette still produces an anticommuting phase,
\begin{align}
    \phi'_{\mathrm{top}} \left(h^{g_\sigma}_i,h^{g_\tau}_{i\pm\half}\right)=-1,
    \qquad
    \phi'_{\mathrm{top}}=+1\ \text{otherwise}.
\end{align}
All other pairs commute as before, so the edge link diagram--the pattern of $\phi_{\mathrm{top}}=-1$ connections--is unaltered by $\KT^{\mathrm{open}}$.  
Since $\beta(g)$ depends only on the parity of these nontrivial links across any cut, the invariance of $\phi_{\mathrm{top}}$ guarantees the invariance of $\beta(g)$ at the boundary level as well.  
Thus, the subsystem KT transformation leaves both the bulk and edge topological data unchanged, while transforming the strong SSPT into an SSSB phase.

\section{Conclusions}
\label{sec:conclusion}

In this work, we have extended the Kennedy--Tasaki duality from $(1+1)$D to $(2+1)$D lattice systems endowed with one-dimensional subsystem symmetries, and developed a complementary formulation of the generalized KW duality for the same class of models. Focusing on the paradigmatic case of $\mathbb{Z}_2\times\mathbb{Z}_2$ subsystem symmetry on the square lattice, our construction establishes a precise duality web connecting three distinct gapped phases: the trivial paramagnet, SSPT phases (both weak and strong), and SSSB phases, as illustrated schematically in Fig.~\ref{fig:Dualities_N_KT}.

First, in Sec.~\ref{sec:subsym_KW} we have formulated a generalized (Kramers--Wannier-like) duality  for models with row/column subsystem symmetries and applied it to the Xu--Moore model (a.k.a. Ising plaquette model in transverse field). At the self-dual point of the model, the duality can be regarded as a symmetry, but its algebraic character depends sensitively on boundary conditions: on open lattices, the duality is implemented by a unitary operator, whereas on closed (periodic or twisted) manifolds the corresponding operator is non-invertible and satisfies nontrivial fusion relations. We have derived these fusion relations for the subsystem defect operators, providing an explicit lattice realization of a non-invertible symmetry in $(2+1)$D.

Building on this subsystem duality construction, in Secs.~\ref{sec:KT_closed}--\ref{sec:KT_SSPT_strong} we have introduced a generalized Kennedy--Tasaki map for $\mathbb{Z}_2\times\mathbb{Z}_2$ subsystem-symmetric models and showed that this transformation furnishes a one-to-one correspondence between subsystem-SPT phases and subsystem symmetry-broken (SSSB) phases. Concretely, the weak SSPT maps to a stack of decoupled Ising chains, while the strong SSPT (realized by the square-lattice cluster model) maps onto two copies of the Ising plaquette model. 

A central technical result is the dichotomy between closed and open geometries: on closed square lattices, the subsystem KT transformation is intrinsically non-unitary and non-invertible when acting within the physical Hilbert space. We have demonstrated this non-invertibility through three complementary perspectives. 
First, by implementing the subsystem KT map on two copies of the Xu-Moore model, we showed that it sends a $\mathbb{Z}_2\times \mathbb{Z}_2$ SSSB phase to the square-lattice cluster Hamiltonian realizing the strong SSPT phase, while leaving the trivial paramagnet invariant--revealing a ground-state obstruction to invertibility. 
Second, an explicit analysis of how symmetry-twist sectors transform shows that closed (periodic or twisted) boundary conditions are mixed in a way incompatible with any unitary action. 
Third, the fusion algebra between the subsystem KT operator and the subsystem symmetry generators directly encodes the lack of an inverse. Together, these arguments establish that the subsystem KT map acts as a genuinely non-invertible duality on closed manifolds.
Importantly, we have further shown that by enlarging the Hilbert space to include all twisted sectors, the subsystem KT transformation becomes fully unitary and invertible, in agreement with recent generalized Wigner-theorem formulations that incorporate non-invertible symmetries~\cite{Wigner-theorem-Ortiz2025}. 
To complement these lattice considerations, Appendix~\ref{app:KT_QFT} provides a field-theoretic perspective: by combining the gauging procedure of subsystem symmetries with the stacking of SSPT phases, we have re-derived the subsystem KT correspondence and thereby supplied an independent, continuum-level check of the lattice constructions we have developed.

In contrast, on open lattices, the subsystem KT transformation becomes manifestly unitary and invertible (Sec.~\ref{sec:KT_open}). In Sec.~\ref{sec:gapped_SSPT_open}, we have shown that in this open geometry, the subsystem KT map provides a one-to-one correspondence not only between the bulk Hamiltonians but also between the degeneracy structures: the spontaneous ground-state degeneracy of the SSSB phase is mapped to the boundary-mode degeneracy of the SSPT phase.

We have also constructed bulk and edge diagnostics that detect strong SSPT order (Sec.~\ref{sec:invariants}) and examined their behavior under the KT map (Sec.~\ref{sec:bulk-edge}). The bulk invariant remains unchanged under the subsystem KT mapping. Similarly, the edge invariant, encoded in the projective algebra of the boundary symmetry generators, preserves its essential algebraic structure. Although strictly local repair operators in the SSPT description map to highly extended, nonlocal objects in the dual SSSB picture, the commutation structure that distinguishes the strong SSPT from its weak cousin persists. This observation yields an explicit bulk-edge correspondence in which the bulk invariant equals the commutation phase between short edge segments across a cut -- a relation that is manifest in both the SSPT and SSSB descriptions, mapped onto one another by the KT duality.

\section*{Acknowledgments}
A.M. and V.T. acknowledge the support of the Department of Atomic Energy, Government of India, under Project Identification No. RTI 4002. A.H.N. was supported by the U.S. Department of Energy (DOE) Office of Basic
Energy Sciences (BES), Division of Materials Sciences and Engineering under contract No. DE-SC0012704 with Brookhaven National Laboratory.

\appendix
\section{The subsystem Kramers-Wannier transformation as a gauging procedure}
\label{app:gauging_KW}

In this appendix, we explain how the subsystem KW transformation can be understood as the gauging of a non-anomalous subsystem $\mathbb{Z}_2$ global symmetry, as outlined in Sec.~\ref{sec:KW_operator_closed}. We remark that an alternative, lattice-resolved formulation of the gauging procedure, complete with the derivation of the associated duality operator, can be found in the recent work by one of the present authors~\cite{XM_Giridhar2025}.
Before proceeding, we first review how to formulate the partition function in an arbitrary symmetry-twisted sector and how to couple the system to a background gauge field. Throughout this discussion, we follow the notation of Refs.~\cite{Cao_sub_duality_2022,Cao_sub_KW_2023,Cao_subSymTFT_2024JHEP}.

\subsection{Implementing background gauge fields on the lattice}

Let $\mathcal{X}$ be a $(2+1)$D QFT with an anomaly-free subsystem $\mathbb{Z}_2$ global symmetry, defined on a closed three-dimensional spacetime $X_3$. For simplicity, we take $X_3$ to be a cubic spacetime lattice $T^3$ with $L_x L_y L_z$ sites, where the $z$-direction is identified as the time direction. 
We denote the partition function, within the fixed twist sector specified by $\mt := \{\mt^x_{j-\frac{1}{2}}, \mt^y_{i-\frac{1}{2}}\}$ and symmetry sector specified by $u := \{u^x_j, u^y_i\}$, as follows \cite{Cao_sub_duality_2022,Cao_sub_KW_2023,Cao_subSymTFT_2024JHEP}:
\begin{align}
\label{eq:Z_ut}
    Z_{\mathcal{X}} [u^x_j, u^y_i, \mt^x_{j-\frac{1}{2}}, \mt^y_{i-\frac{1}{2}}]  = \text{Tr}_{\mathcal{H}_{\mt}} \left(\prod_{j=1}^{L_y} \frac{1+(-1)^{u^x_j}U^x_j}{2} \right) \left(\prod_{i=1}^{L_x} \frac{1+(-1)^{u^y_i}U^y_i}{2} \right) e^{-\beta H},
\end{align}
where $\mathcal{H}_{\mt}$ denotes the Hilbert space corresponding to the twist sector labeled by $(\mt^x_{j-\frac{1}{2}}, \mt^y_{i-\frac{1}{2}})$. The symmetry-twist labels satisfy the constraints given in Eq.~\eqref{eq:symtwist_constraint}. 

To couple the theory to the subsystem $\mathbb{Z}_2$ symmetry, we introduce a background gauge field $A = (A^z,A^{xy}) \in \{0,1\} $. Here, the time-like component $A^z_{i,j,k-\half}$ is defined on the $z$-links, and the spatial component $A^{xy}_{i-\half,j-\half,k}$ is defined on the $xy$-plaquette. 

We next clarify the correspondence between the symmetry-twist sectors $(u,\mt)$ and the background gauge fields $(A^z,A^{xy})$. Since the theory is defined on the three-torus $T^3$, the background gauge fields can be equivalently described in terms of their $\mathbb{Z}_2$-valued Wilson loops.
The Wilson loop of $A^z$ can be expressed, following Ref.~\cite{Cao_sub_duality_2022} as follows:
\begin{align}
    W_{z;i,j} = \sum_{k=1}^{L_z} A^z_{i,j,k-\half}.
\end{align}
At first sight, one might expect $L_x L_y$ distinct variables $W_{z;i,j}$. However, their number of independent degrees of freedom is actually reduced to $L_x + L_y - 1$. The reason is that $W_{z;i,j}$, being the Wilson loop in the time direction, effectively measures whether the subsystem $\mathbb{Z}_2$ symmetry operators defined in Eq.~\eqref{eq:subsym_operator} are inserted along the spatial directions. Concretely, acting with a symmetry operator of the form $
\prod_{j=1}^{L_y} \left(U^x_j\right)^{\alpha_j} \prod_{i=1}^{L_x} \left(U^y_i\right)^{\beta_i}
$ on the ground state corresponds to turning on background fields such that $
W_{z;i,j} = \alpha_j + \beta_i 
$ \cite{Cao_sub_duality_2022}. This immediately implies that not all $L_x L_y$ Wilson loops are independent; only $L_x + L_y - 1$ of them are, matching the number of independent subsystem symmetry operators. Equivalently, one can decompose the time-like Wilson loop as
\begin{align}
    W_{z;i,j} = W_{z,x;j} + W_{z,y;i},
\end{align}
where $W_{z,x;j}$ and $W_{z,y;i}$ are the $\mathbb{Z}_2$-valued Wilson loops associated with the row- and column-type subsystem symmetries at time slice $z$. Concretely, $W_{z,x;j}$ ($W_{z,y;i}$) detects the insertion of the operator $(U^x_j)^{W_{z,x;j}}$ ($ (U^y_i)^{W_{z,y;i}})$, respectively), corresponding to subsystem symmetries acting along the $j$-th row ($i$-th column).

On the other hand, the gauge invariant Wilson loops of $A^{xy}$ along $x$ and $y$ directions are
\begin{align}
    W_{x;j-\half} = \sum_{i=1}^{L_x} A^{xy}_{i-\half,j-\half,k} = \mt^x_{j-\half}, \quad W_{y;i-\half} = \sum_{j=1}^{L_y} A^{xy}_{i-\half,j-\half,k} = \mt^y_{i-\half}.
\end{align}
They capture the insertion of symmetry defects along the $z$-direction and are identified with the twist variables $\mt^x_{j-\frac{1}{2}}, \mt^y_{i-\frac{1}{2}}$. The constraints on the symmetry-twist labels given in Eq.~\eqref{eq:symtwist_constraint} translate into corresponding conditions on the Wilson loops, which can be written as 
\begin{align}
\label{eq:constraint_W}
    \left(W_{z,x;j}, W_{z,y;i}\right) \sim  \left(W_{z,x;j}+1, W_{z,y;i}+1\right), \quad \prod_{j = 1}^{L_y} (-1)^{W_{x;j-\half}} \prod_{i = 1}^{L_x} (-1)^{W_{y;i-\half}}  = 1.
\end{align}

We denote the partition function in terms of the Wilson loop of the gauge fields as \cite{Cao_sub_duality_2022,Cao_sub_KW_2023,Cao_subSymTFT_2024JHEP}
\begin{align}
\label{eq:Z_W}
    Z_{\mathcal{X}}[W_{z,x;j}, W_{z,y;i}, W_{x;j-\half}, W_{y;i-\half}] = \text{Tr}_{\mathcal{H}_{\mt}}  \left(\prod_{j=1}^{L_y} \left(U^x_j\right)^{W_{z,x;j}}\right) \left(\prod_{i=1}^{L_x} \left(U^y_i\right)^{W_{z,y;i}}\right) e^{-\beta H}.
\end{align}
The partition function in different symmetry-twist sectors \eqref{eq:Z_ut} is related to the partition function in terms of the Wilson loops \eqref{eq:Z_W} by a discrete Fourier transformation
\begin{align}
\label{eq:FT_ut_W}
    Z_{\mathcal{X}} [u^x_j, u^y_i, \mt^x_{j-\frac{1}{2}}, \mt^y_{i-\frac{1}{2}}]  = & \frac{1}{2^{L_x+L_y-1}} \sum_{W_{z,x;j}, W_{z,y;i}} (-1)^{\sum_{j=1}^{L_y} u^x_j W_{z,x;j} + \sum_{i=1}^{L_x} u^y_i W_{z,y;i}} \nonumber\\
     & {} \times Z_{\mathcal{X}}[W_{z,x;j}, W_{z,y;i}, \mt^x_{j-\half}, \mt^y_{i-\half}],
\end{align}
where the sum over $(W_{z,x;j}, W_{z,y;i})$ should obey the constraint \eqref{eq:constraint_W} and the converse relation is
\begin{align}
\label{eq:FT_W_ut}
    Z_{\mathcal{X}}[W_{z,x;j}, W_{z,y;i}, W_{x;j-\half}, W_{y;i-\half}] = & \sum_{u^x_j, u^y_i} (-1)^{\sum_{j=1}^{L_y} u^x_j W_{z,x;j} + \sum_{i=1}^{L_x} u^y_i W_{z,y;i}} \nonumber\\
    & {} \times Z_{\mathcal{X}} [u^x_j, u^y_i, W_{x;j-\half}, W_{y;i-\half}],
\end{align}
where the sum over $(u^x_j, u^y_i)$ should obey the gauge redundancy \eqref{eq:symtwist_constraint}.

We now turn to the dual theory (after the subsystem KW transformation) $\widehat{\mathcal{X}}$, where the twist sectors are labeled by $\hmt := \{\hmt^x_{j}, \hmt^y_{i}\}$ and symmetry sectors by $\hu := \{\hu^x_{j-\half}, \hu^y_{i-\half}\}$. Similarly, to couple the dual theory to the dual subsystem $\mathbb{Z}_2$ symmetry, we introduce a dual background gauge field $\hA = (\hA^z,\hA^{xy}) \in \{0,1\} $. The Wilson loop variables of the dual gauge fields are $(\hW_{z,x;j-\half}, \hW_{z,y;i-\half}, \hW_{x;j}, \hW_{y;i})$ and satisfy the gauge redundancy and constraints
\begin{align}
\label{eq:constraint_W_dual}
    \left(\hW_{z,x;j-\half}, \hW_{z,y;i-\half}\right) \sim  \left(\hW_{z,x;j-\half}+1, \hW_{z,y;i-\half}+1\right), \quad \prod_{j = 1}^{L_y} (-1)^{\hW_{x;j}} \prod_{i = 1}^{L_x} (-1)^{\hW_{y;i}}  = 1.
\end{align}
Analogous to Eqs.~\eqref{eq:FT_ut_W} and \eqref{eq:FT_W_ut}, one can define a Fourier transform relation between the dual partition function in a given dual symmetry-twist sector and the corresponding partition function in terms of the dual Wilson loops. 

\subsection{Gauging subsystem $\mathbb{Z}_2$ symmetry}
\label{sec:sector_mapping_from_gauging}

\subsubsection*{Review: gauging global $\mathbb{Z}_2$ symmetry}
Next, we relate the partition function of the original theory $\mathcal{X}$ with gauge field $A$ to that of the dual theory $ \widehat{\mathcal{X}}$ with $\widehat{A}$ via gauging. In general, for a theory with an ordinary $\mathbb{Z}_2$ global symmetry, the gauging procedure (also known in field theory as the $S$ transformation) yields a dual theory $ S \mathcal{X} =\mathcal{X}/\mathbb{Z}_2 \equiv \widehat{\mathcal{X}}$ defined on the dual lattice and endowed with a new  $\mathbb{Z}_2$ symmetry. Its partition function takes the form \cite{Gaiotto_Generalized_2015JHEP, Bhardwaj_gauging_2018JHEP, Roumpedakis_Gauging_2023CMaPh,Kaidi_SymTFT_2023CMaPh,Li_2023_KT}
\begin{align}
\label{eq:gauging_Xd+1}
    Z_{\widehat{\mathcal{X}}} [X_{d+1},\hA] = \frac{1}{|H^0(X_{d+1},\mathbb{Z}_2)|} \sum_{a\in H^1(X_{d+1},\mathbb{Z}_2)} Z_{\mathcal{X}} [X_{d+1},a] (-1)^{\int_{X_{d+1}} a \hA}.
\end{align}

The role of the two cohomology groups in this expression is as follows. The factor $H^1(X_{d+1},\allowbreak \mathbb{Z}_2)$ classifies all $\mathbb{Z}_2$ gauge fields on the manifold $X_{d+1}$, so that the sum over $a\in H^1(X_{d+1},\mathbb{Z}_2)$ implements the promotion of the original background field ($A$) to a dynamical one ($a$) by summing over all possible periodic/twisted boundary conditions. Meanwhile, $H^0(X_{d+1},\mathbb{Z}_2)$ encodes the constant $\mathbb{Z}_2$ gauge transformations, one for each connected component of $X_{d+1}$, which act trivially on the theory; dividing by $|H^0|$ removes this redundancy and gives the correct normalization of the gauged partition function.

The additional sign $(-1)^{\int a\widehat A}$ is determined by the cup product pairing $a\cup \widehat{A}$ (denoted as $a\widehat{A}$ for brevity)  between the dynamical gauge field ($a$) and the new background field ($\widehat A$), and can be interpreted as counting their intersections modulo two. In this way, the relation above provides the modern perspective on the KW transformation, viewing it as nothing but the gauging of the underlying global $\mathbb{Z}_2$ symmetry.    
\subsubsection*{Gauging subsystem $\mathbb{Z}_2$ symmetry}

We now generalize the above gauging prescription to the case of \textit{subsystem} global symmetries in two spatial dimensions, following Ref.~\cite{Cao_subSymTFT_2024JHEP}. In this setting, the usual cohomology groups are replaced by their subsystem analogues, and the dual partition function is then
\begin{align}
\label{eq:gauging_X3}
    Z_{\widehat{\mathcal{X}}} [X_3,\hA] = \frac{1}{|H^0_{\text{sub}}(X_3,\mathbb{Z}_2)|} \sum_{a\in H_{\text{sub}}^1(X_3,\mathbb{Z}_2)} Z_{\mathcal{X}} [X_3,a] (-1)^{\int_{X_3} a \hA}.
\end{align}
Here $H^1_{\text{sub}}(X_3,\mathbb{Z}_2)$ is the group of subsystem $\mathbb{Z}_2$ gauge fields, while $H^0_{\text{sub}}(X_3,\mathbb{Z}_2)$ consists of the constant subsystem gauge transformations that act trivially and must be divided out to normalize the partition function. As in the global symmetry case, the additional phase factor encodes the subsystem analogue of the cup-product pairing, given explicitly by $a \widehat A \equiv a\cup \widehat{A} = a^{xy}\widehat A^z \allowbreak + a^z \widehat A^{xy}$.

To demonstrate the utility of this general construction, we now rederive the mapping between the symmetry-twist sectors under the subsystem KW transformation given in \eqref{eq:mapping_symtwist_KW}. To define the symmetry and twist sectors, we place the theory on the three-torus $T^3$, so that Eq.~\eqref{eq:gauging_X3} can be expressed in terms of the Wilson loops of the gauge fields \cite{Cao_subSymTFT_2024JHEP}:
\begin{align}
\label{eq:gauging_T3}
   & Z_{\widehat{\mathcal{X}}} [\hW_{z,x;j-\half}, \hW_{z,y;i-\half}, \hW_{x;j}, \hW_{y;i}] \nonumber\\
    & \quad = \frac{1}{2^{L_x+L_y-1}}\sum_{\substack{w_{z,x;j}, w_{z,y;i},\\
   w_{x;j-\half}, w_{y;i-\half}}}  Z_{\mathcal{X}}[w_{z,x;j}, w_{z,y;i}, w_{x;j-\half}, w_{y;i-\half}] \nonumber\\
    & \hspace{1cm} \times (-1)^{\sum_{j=1}^{L_y} (\hW_{z,x;j-\half} w_{x;j-\half} + \hW_{x;j} w_{z,x;j}) + \sum_{i=1}^{L_x} (\hW_{z,y;i-\half} w_{y;i-\half} + \hW_{y;i}  w_{z,y;i})}.
\end{align}

Combining \eqref{eq:FT_ut_W}, \eqref{eq:FT_W_ut} and \eqref{eq:gauging_T3}, we obtain
\begin{align}
    Z_{\widehat{\mathcal{X}}} [\hu^x_{j-\half},\hu^y_{i-\half},\hmt^x_{j}, \hmt^y_{i}]   = {} & \frac{1}{2^{L_x+L_y-1}} \sum_{\hW_{z,x;j-\half}, \hW_{z,y;i-\half}} Z_{\widehat{\mathcal{X}}}[\hW_{z,x;j-\half}, \hW_{z,y;i-\half}, \hmt^x_{j}, \hmt^y_{i}] \nonumber\\
    {} &  \times (-1)^{\sum_{j=1}^{L_y} \hu^x_{j-\half} \hW_{z,x;j-\half} + \sum_{i=1}^{L_x} \hu^y_{i-\half} \hW_{z,y;i-\half}} \nonumber\\ 
    = {} & \frac{1}{4^{L_x+L_y-1}} \sum_{\hW_{z,x;j-\half}, \hW_{z,y;i-\half}} \sum_{\substack{w_{z,x;j}, w_{z,y;i},\\
   w_{x;j-\half}, w_{y;i-\half}}}  Z_{\mathcal{X}}[w_{z,x;j}, w_{z,y;i}, w_{x;j-\half}, w_{y;i-\half}] \nonumber\\
    {} & \times (-1)^{\substack{\sum_{j=1}^{L_y} \hu^x_{j-\half} \hW_{z,x;j-\half} + \sum_{i=1}^{L_x} \hu^y_{i-\half} \hW_{z,y;i-\half} + \sum_{j=1}^{L_y} (\hW_{z,x;j-\half} w_{x;j-\half} + \hmt^x_j w_{z,x;j})\\
     + \sum_{i=1}^{L_x} (\hW_{z,y;i-\half} w_{y;i-\half} + \hmt^y_i  w_{z,y;i})} } \nonumber\\ 
    = {} & \frac{1}{4^{L_x+L_y-1}} \sum_{\hW_{z,x;j-\half}, \hW_{z,y;i-\half}} \sum_{\substack{w_{z,x;j}, w_{z,y;i},\\
   w_{x;j-\half}, w_{y;i-\half}}} \sum_{u^x_j,u^y_i}  Z_{\mathcal{X}}[u^x_j,u^y_i, w_{x;j-\half}, w_{y;i-\half}] \nonumber\\
     {} & \times (-1)^{\substack{\sum_{j=1}^{L_y} \hu^x_{j-\half} \hW_{z,x;j-\half} + \sum_{i=1}^{L_x} \hu^y_{i-\half} \hW_{z,y;i-\half} + \sum_{j=1}^{L_y} (\hW_{z,x;j-\half} w_{x;j-\half} + \hmt^x_j w_{z,x;j})\\
     + \sum_{i=1}^{L_x} (\hW_{z,y;i-\half} w_{y;i-\half} + \hmt^y_i  w_{z,y;i}) + \sum_{j=1}^{L_y} u^x_j w_{z,x;j} + \sum_{i=1}^{L_x} u^y_i w_{z,y;i}}}.
\end{align}

After carrying out the sum over $(\hW_{z,x;j-\half}, \hW_{z,y;i-\half})$ and $(w_{z,x;j}, w_{z,y;i})$, we obtain
\begin{align}
    Z_{\widehat{\mathcal{X}}} [\hu^x_{j-\half},\hu^y_{i-\half},\hmt^x_{j}, \hmt^y_{i}] = & \sum_{\substack{
   w_{x;j-\half}, w_{y;i-\half}}} \sum_{u^x_j,u^y_i}  Z_{\mathcal{X}}[u^x_j,u^y_i, w_{x;j-\half}, w_{y;i-\half}] \nonumber\\
   & \times \delta_{\hu^x_{j-\half}, w_{x;j-\half}} \delta_{\hu^y_{i-\half}, w_{y;i-\half}} \delta_{\hmt^x_j, u^x_j} \delta_{\hmt^y_i, u^y_i}\nonumber\\
    = {} &  Z_{\mathcal{X}} [\hmt^x_{j}, \hmt^y_{i},\hu^x_{j-\half},\hu^y_{i-\half}].
\end{align}
Thus, we conclude that the symmetry-twist sectors before and after gauging are related by 
\begin{align}
    \left(\hu^x_{j-\half},\hu^y_{i-\half},\hmt^x_{j}, \hmt^y_{i}\right) = \left(\mt^x_{j-\frac{1}{2}}, \mt^y_{i-\frac{1}{2}},u^x_j,u^y_i\right),
\end{align}
which exactly reproduces the mapping \eqref{eq:mapping_symtwist_KW} obtained in the main text from lattice considerations. 

\section{Subsystem Kennedy-Tasaki transformation in a field-theory framework}
\label{app:KT_QFT}
In this Appendix, we present a field-theoretic formulation of the subsystem KT transformation, following the approach of Ref.~\cite{Li_2023_KT}. For a generic QFT $\mathcal{X}$ with a subsystem $\mathbb{Z}_2^\sigma \times \mathbb{Z}_2^\tau$ symmetry, we write its partition function as $Z_{\mathcal{X}} [A_\sigma,A_\tau]$, where both $\mathbb{Z}_2$ symmetries are taken to be non-anomalous. Following the notation in Appendix~\ref{app:gauging_KW}, we denote the background gauge field for the first subsystem symmetry $\mathbb{Z}_2^\sigma$ as $A_\sigma = (A^z_\sigma,A^{xy}_\sigma) \in \{0,1\} $ and for the second subsystem symmetry $\mathbb{Z}_2^\tau$ as $A_\tau = (A^z_\tau,A^{xy}_\tau) \in \{0,1\} $.

When $\mathcal{X}$ is in the trivial phase, the corresponding fixed-point partition function can be set to unity \cite{Jia_SSPTSymTFT_JHEP2025}
\begin{align}
    Z_{\text{Tri}} [A_\sigma,A_\tau] = 1.
\end{align}

For the SSSB phase, the fixed-point partition function is given by \cite{Jia_SSPTSymTFT_JHEP2025}
\begin{align}
     Z_{\text{SSSB}} [A_\sigma,A_\tau] = \delta(A_\sigma) \delta(A_\tau).
\end{align}

The fixed point partition function for the global $\mathbb{Z}_2 \times \mathbb{Z}_2$ SSB phase was constructed in Refs.~\cite{Wang_SPT_2015PRL,Wang_nonabelian_2015PRB}. Extending this framework to the subsystem setting, we obtain for the strong $\mathbb{Z}_2^\sigma \times \mathbb{Z}_2^\tau$ SSPT phase
\begin{align}
    Z_{\text{SSPT}} [A_\sigma,A_\tau] = (-1)^{\int_{X_3} A_\sigma A_\tau}.
\end{align}

To clarify how the different fixed-point theories are connected, it is useful to introduce certain transformations that act directly on the partition function:
\begin{align}
   & S : Z_{S\mathcal{X}} [\hA_\sigma,\hA_\tau] = \frac{1}{|H^0_{\text{sub}}(X_3,\mathbb{Z}_2)|^2} \sum_{\substack{a_\sigma \in H_{\text{sub}}^1(X_3,\mathbb{Z}_2^\sigma),\\
    a_\tau \in H_{\text{sub}}^1(X_3,\mathbb{Z}_2^\tau)}} Z_{\mathcal{X}} [a_\sigma,a_\tau] (-1)^{\int_{X_3} a_\sigma \hA_\tau + a_\tau \hA_\sigma}, \nonumber\\
    & T : Z_{T\mathcal{X}} [A_\sigma,A_\tau] = Z_{\mathcal{X}} [A_\sigma,A_\tau] (-1)^{\int_{X_3} A_\sigma A_\tau}. 
\end{align}
Here, $S$ corresponds to gauging the pair of subsystem $\mathbb{Z}_2$ symmetries, and $T$ corresponds to ``stacking'' with a nontrivial $\mathbb{Z}_2^\sigma \times \mathbb{Z}_2^\tau$ SSPT phase~\cite{Li_2023_KT}. The unique combination of topological operations that maps the strong SSPT phase to the SSSB phase (and vice versa), while leaving the trivial phase invariant, is given by the composition
\begin{align}
    STS = TST.
\end{align}

\begin{figure}[tb]
    \centering
    \includegraphics[width=0.50\linewidth]{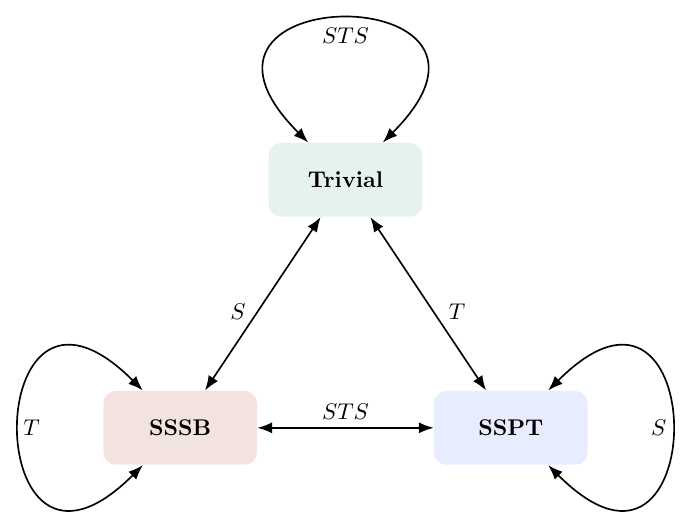}
    \caption{Relation between the three gapped phases with subsystem $\mathbb{Z}_2^\sigma \times \mathbb{Z}_2^\tau$ symmetry through duality transformations.}
    \label{fig:Dualities_STS}
\end{figure}

The connections among the trivial, SSSB, and SSPT phases through the duality transformations are explicitly summarized in Fig.~\ref{fig:Dualities_STS}. In this work, we employ the $STS$ sequence to implement the subsystem KT transformation, as it provides a considerably simpler and more transparent construction \cite{Li_2023_KT}. The action of the $STS$ transformation on the partition function can be expressed as follows
\begin{align}
\label{eq:Z_STS_X3}
    Z_{STS\mathcal{X}} [A'_\sigma,A'_\tau] & = \frac{1}{|H^0_{\text{sub}}(X_3,\mathbb{Z}_2)|^4} \sum_{\substack{a_\sigma, a_\tau, \ha_\sigma, \ha_\tau}} Z_{\mathcal{X}} [a_\sigma,a_\tau] (-1)^{\int_{X_3} a_\sigma \ha_\tau + a_\tau \ha_\sigma + \ha_\sigma \ha_\tau + \ha_\sigma A'_\tau + \ha_\tau A'_\sigma} \nonumber\\
    & = \frac{|H^1_{\text{sub}}(X_3,\mathbb{Z}_2)|}{|H^0_{\text{sub}}(X_3,\mathbb{Z}_2)|^4} \sum_{\substack{a_\sigma, a_\tau}} Z_{\mathcal{X}} [a_\sigma,a_\tau] (-1)^{\int_{X_3} a_\sigma a_\tau + a_\sigma A'_\tau + a_\tau A'_\sigma + A'_\sigma A'_\tau},
\end{align}
where, $a_\sigma$ and $a_\tau$ ($\ha_\sigma$ and $\ha_\tau$) represent the dynamical $\mathbb{Z}_2^\sigma \times \mathbb{Z}_2^\tau$ gauge fields that are introduced during the first $S$ (second) operation.

We now derive the mapping between the symmetry-twist sectors under the STS transformation starting from the partition function of a generic QFT $\mathcal{X}$ with a subsystem $\mathbb{Z}_2^\sigma \times \mathbb{Z}_2^\tau$ symmetry. To define the symmetry and twist sectors in a precise manner, we consider the theory compactified on the three-torus $T^3$. On this geometry, Eq.~\eqref{eq:Z_STS_X3} naturally takes the form of an expression written in terms of the Wilson loops of the background gauge fields
\begin{align}
     Z_{STS\mathcal{X}} & [W'^{\sigma}_{z,x;j}, W'^{\sigma}_{z,y;i}, W'^{\tau}_{z,x;j-\half}, W'^{\tau}_{z,y;i-\half},  W'^{\sigma}_{x;j-\half}, W'^{\sigma}_{y;i-\half}, W'^{\tau}_{x;j}, W'^{\tau}_{y;i}] \nonumber\\
    &  = \frac{1}{4^{L_x+L_y-1}} \sum_{\substack{w^{\sigma}_{z,x;j}, w^{\sigma}_{z,y;i}, w^{\tau}_{z,x;j-\half}, w^{\tau}_{z,y;i-\half},\\
    w^{\sigma}_{x;j-\half}, w^{\sigma}_{y;i-\half}, w^{\tau}_{x;j}, w^{\tau}_{y;i}}} \nonumber\\
    & \phantom{=} \times (-1)^{\mathcal{Q}} Z_{\mathcal{X}} [w^{\sigma}_{z,x;j}, w^{\sigma}_{z,y;i}, w^{\tau}_{z,x;j-\half}, w^{\tau}_{z,y;i-\half},  w^{\sigma}_{x;j-\half}, w^{\sigma}_{y;i-\half}, w^{\tau}_{x;j}, w^{\tau}_{y;i}],
\end{align}
where the phase $\mathcal{Q}$ is given by
\begin{align}
   \mathcal{Q} = &  \sum_{j=1}^{L_y} (w^{\sigma}_{x;j-\half} w^{\tau}_{z,x;j-\half}  + w^{\sigma}_{z,x;j} w^{\tau}_{x;j} ) + \sum_{i=1}^{L_x} ( w^{\sigma}_{y;i-\half} w^{\tau}_{z,y;i-\half} + w^{\sigma}_{z,y;i} w^{\tau}_{y;i}) \nonumber\\
   & + \sum_{j=1}^{L_y} (w^{\sigma}_{x;j-\half} W'^{\tau}_{z,x;j-\half}  + w^{\sigma}_{z,x;j} W'^{\tau}_{x;j} ) + \sum_{i=1}^{L_x} ( w^{\sigma}_{y;i-\half} W'^{\tau}_{z,y;i-\half} + w^{\sigma}_{z,y;i} W'^{\tau}_{y;i}) \nonumber\\
   & + \sum_{j=1}^{L_y} (w^{\tau}_{z,x;j-\half} W'^{\sigma}_{x;j-\half} + w^{\tau}_{x;j} W'^{\sigma}_{z,x;j}) + \sum_{i=1}^{L_x} (w^{\tau}_{z,y;i-\half} W'^{\sigma}_{y;i-\half} + w^{\tau}_{y;i}  W'^{\sigma}_{z,y;i}) \nonumber\\
   & + \sum_{j=1}^{L_y} (W'^{\sigma}_{x;j-\half} W'^{\tau}_{z,x;j-\half}  + W'^{\sigma}_{z,x;j} W'^{\tau}_{x;j} ) + \sum_{i=1}^{L_x} ( W'^{\sigma}_{y;i-\half} W'^{\tau}_{z,y;i-\half} + W'^{\sigma}_{z,y;i} W'^{\tau}_{y;i}).
\end{align}

The relation between the partition function of the original theory $\mathcal{X}$ in different symmetry-twist sectors and its expression in terms of gauge-field Wilson loops is established via a discrete Fourier transformation
\begin{align}
   & Z_{\mathcal{X}} [u^x_{\sigma,j}, u^y_{\sigma,i}, u^x_{\tau,j-\half}, u^y_{\tau,i-\half}, \mt^x_{\sigma,j-\half}, \mt^y_{\sigma,i-\half}, \mt^x_{\tau,j}, \mt^y_{\tau,i}]  \nonumber\\
   & \quad = \frac{1}{4^{L_x+L_y-1}} \sum_{\substack{w^{\sigma}_{z,x;j}, w^{\sigma}_{z,y;i},\\
   w^{\tau}_{z,x;j-\half}, w^{\tau}_{z,y;i-\half}}} Z_{\mathcal{X}}[w^{\sigma}_{z,x;j}, w^{\sigma}_{z,y;i}, w^{\tau}_{z,x;j-\half}, w^{\tau}_{z,y;i-\half}, \mt^x_{\sigma,j-\half}, \mt^y_{\sigma,i-\half}, \mt^x_{\tau,j}, \mt^y_{\tau,i}] \nonumber\\
   & \qquad \times (-1)^{\sum_{j=1}^{L_y} u^x_{\sigma,j} w^{\sigma}_{z,x;j} + \sum_{i=1}^{L_x} u^y_{\sigma,i} w^{\sigma}_{z,y;i} + \sum_{j=1}^{L_y} u^x_{\tau,j-\half} w^{\tau}_{z,x;j-\half} + \sum_{i=1}^{L_x} u^y_{\tau,i-\half} w^{\tau}_{z,y;i-\half}},
\end{align}
and the inverse relation is
\begin{align}
   & Z_{\mathcal{X}} [w^{\sigma}_{z,x;j}, w^{\sigma}_{z,y;i}, w^{\tau}_{z,x;j-\half}, w^{\tau}_{z,y;i-\half},  w^{\sigma}_{x;j-\half}, w^{\sigma}_{y;i-\half}, w^{\tau}_{x;j}, w^{\tau}_{y;i}] \nonumber\\
    & \quad = \sum_{\substack{u^x_{\sigma,j}, u^y_{\sigma,i}, \\
    u^x_{\tau,j-\half}, u^y_{\tau,i-\half}}} Z_{\mathcal{X}} [u^x_{\sigma,j}, u^y_{\sigma,i}, u^x_{\tau,j-\half}, u^y_{\tau,i-\half},  w^{\sigma}_{x;j-\half}, w^{\sigma}_{y;i-\half}, w^{\tau}_{x;j}, w^{\tau}_{y;i}]\nonumber\\
   & \qquad \times (-1)^{\sum_{j=1}^{L_y} u^x_{\sigma,j} w^{\sigma}_{z,x;j} + \sum_{i=1}^{L_x} u^y_{\sigma,i} w^{\sigma}_{z,y;i} + \sum_{j=1}^{L_y} u^x_{\tau,j-\half} w^{\tau}_{z,x;j-\half} + \sum_{i=1}^{L_x} u^y_{\tau,i-\half} w^{\tau}_{z,y;i-\half}}.
\end{align}
From the relations established above, we can now combine them to deduce the desired mapping
\begin{align}
    & Z_{STS\mathcal{X}} [u'^x_{\sigma,j},  u'^y_{\sigma,i}, u'^x_{\tau,j-\half}, u'^y_{\tau,i-\half},\mt'^x_{\sigma,j-\half}, \mt'^y_{\sigma,i-\half}, \mt'^x_{\tau,j}, \mt'^y_{\tau,i}] \nonumber \\ & \quad = Z_{\mathcal{X}} [u'^x_{\sigma,j}, u'^y_{\sigma,i}, u'^x_{\tau,j-\half}, u'^y_{\tau,i-\half}, \mt'^x_{\sigma,j-\half} + u'^x_{\tau,j-\half}, \mt'^y_{\sigma,i-\half} + u'^y_{\tau,i-\half}, \mt'^x_{\tau,j} + u'^x_{\sigma,j}, \mt'^y_{\tau,i} + u'^y_{\sigma,i}] \nonumber\\
    & \quad := Z_{\mathcal{X}} [u^x_{\sigma,j}, u^y_{\sigma,i}, u^x_{\tau,j-\half}, u^y_{\tau,i-\half}, \mt^x_{\sigma,j-\half}, \mt^y_{\sigma,i-\half}, \mt^x_{\tau,j}, \mt^y_{\tau,i}],
\end{align}
which reproduces the symmetry-twist mapping obtained earlier in Eqs.~\eqref{eq:symmetry_mapping_KT} and \eqref{eq:twist_mapping_KT}. Namely, the KT transformation leaves the symmetry sectors unchanged, while the twist sectors get transformed in a way that depends on both the twist and symmetry sectors of the model.

\section{Explicit computation of non-invertible KT transformation}
In this Appendix, we present the detailed derivation of the expression for $\KT$ given in Eq.~\eqref{eq:N_KT_def_1} of the main text, and explicitly work out the steps leading to the fusion rule $ \KT \times \KT$ stated in Eq.~\eqref{eq:fusion_KTKT}.

\subsection{Computation of $\KT$}
\label{app:N_KT_comp}
With the explicit forms of the subsystem KW duality operator $\mN$ in Eq.~\eqref{eq:N_Z2Z2} and the domain wall decoration operator $U_{\text{DW}}$ in Eq.~\eqref{eq:U_DW_dual} at hand, we can now evaluate $\KT$, defined by $\KT = \mN^\dagger U_{\text{DW}} \mN$. Applying this definition to an arbitrary basis state in the original Hilbert space, we obtain the explicit form of the KT transformation operator,
\begin{align}
\label{eq:N_KT_1}
    \KT \ket{\{\ssi_{i,j}, \st_{i-\half,j-\half}\}} & = \frac{1}{4^{L_x+L_y}}  \sum_{\substack {\{\hssi_{i-\half,j-\half}, \hst_{i,j}\},\\
    \{\ssip_{i,j},\stp_{i-\half,j-\half}\}}} (-1)^{\mathcal{A}_{\text{KT}}(\{\sigma,\sigma';\tau,\tau'\}) } \ket{\{\ssip_{i,j}, \stp_{i-\half,j-\half}\}},
\end{align}
where the phase $\mathcal{A}_{\text{KT}}(\{\sigma,\sigma';\tau,\tau'\})$ is given by
\begin{align}
\label{eq:A_KT_1}
    \mathcal{A}_{\text{KT}}   = & \sum_{i=1}^{L_x} \sum_{j=1}^{L_y} \ssi_{i,j} \left(\hssi_{i-\half,j-\half} + \hssi_{i+\half,j-\half} + \hssi_{i-\half,j+\half} + \hssi_{i+\half,j+\half}\right)  + \sum_{j=1}^{L_y}  t^x_{\sigma,j} \left(\hssi_{\half,j-\half}   + \hssi_{\half,j+\half}\right) \nonumber \\
  & + \sum_{i=1}^{L_x}  t^y_{\sigma,i} \left(\hssi_{i-\half,\half} + \hssi_{i+\half,\half}\right) + t^{xy}_{\sigma} \hssi_{\half,\half} + \sum_{i=1}^{L_x} \sum_{j=1}^{L_y} \hst_{i,j} \left(\st_{i-\half,j-\half} + \st_{i+\half,j-\half} + \st_{i-\half,j+\half} \right. \nonumber \\
  &\left. + \st_{i+\half,j+\half}\right)   + \sum_{j=1}^{L_y}  \htt^x_{\tau,j} \left(\st_{\half,j-\half} + \st_{\half,j+\half}\right) + \sum_{i=1}^{L_x}  \htt^y_{\tau,i} \left(\st_{i-\half,\half} + \st_{i+\half,\half}\right) + \htt^{xy}_{\tau} \st_{\half,\half} \nonumber \\
  & + \sum_{i=1}^{L_x} \sum_{j=1}^{L_y} \hst_{i,j} \left(\hssi_{i-\half,j-\half}  + \hssi_{i+\half,j-\half} + \hssi_{i-\half,j+\half} + \hssi_{i+\half,j+\half}\right) + \sum_{j=1}^{L_y}  \htt^x_{\tau,j} \left(\hssi_{\half,j-\half} + \hssi_{\half,j+\half}\right) \nonumber \\
  &  + \sum_{i=1}^{L_x}  \htt^y_{\tau,i} \left(\hssi_{i-\half,\half} + \hssi_{i+\half,\half}\right) + \htt^{xy}_{\tau} \hssi_{\half,\half}  + \sum_{i=1}^{L_x} \sum_{j=1}^{L_y} \ssip_{i,j} \left(\hssi_{i-\half,j-\half} + \hssi_{i+\half,j-\half} + \hssi_{i-\half,j+\half} \right. \nonumber \\
  &\left. + \hssi_{i+\half,j+\half}\right) + \sum_{j=1}^{L_y}  t'^x_{\sigma,j} \left(\hssi_{\half,j-\half} + \hssi_{\half,j+\half} \right) + \sum_{i=1}^{L_x}  t'^y_{\sigma,i} \left(\hssi_{i-\half,\half} + \hssi_{i+\half,\half}\right) + t'^{xy}_{\sigma} \hssi_{\half,\half} \nonumber \\
  &  + \sum_{i=1}^{L_x} \sum_{j=1}^{L_y} \hst_{i,j} \left(\stp_{i-\half,j-\half} + \stp_{i+\half,j-\half} + \stp_{i-\half,j+\half} + \stp_{i+\half,j+\half}\right)  + \sum_{j=1}^{L_y}  \htt^x_{\tau,j} \left(\stp_{\half,j-\half} + \stp_{\half,j+\half}\right) \nonumber \\
  & + \sum_{i=1}^{L_x}  \htt^y_{\tau,i} \left(\stp_{i-\half,\half} + \stp_{i+\half,\half}\right) + \htt^{xy}_{\tau} \stp_{\half,\half}.
\end{align}
We then carry out the summation over the $\hst_{i,j}$ degrees of freedom in the dual Hilbert space, which yields a product of delta-function constraints
\begin{align}
   & \left(\st_{i-\half,j-\half} + \st_{i+\half,j-\half} + \st_{i-\half,j+\half} + \st_{i+\half,j+\half} + \hssi_{i-\half,j-\half} + \hssi_{i+\half,j-\half} + \hssi_{i-\half,j+\half} + \hssi_{i+\half,j+\half} \right. \nonumber \\
  &\left. \quad + \stp_{i-\half,j-\half} + \stp_{i+\half,j-\half} + \stp_{i-\half,j+\half} + \stp_{i+\half,j+\half}\right) = 0, \quad \forall i,j.
\end{align}
The general solutions of the constraints are
 \begin{align}
 \label{eq:sol_N_KT}
     \hssi_{i-\half,j-\half}  = \st_{i-\half,j-\half} + \stp_{i-\half,j-\half} + m^x_{j-\half} + m^y_{i-\half}, \quad m^x_{j-\half}, m^y_{i-\half} \in \{0,1\}, \quad \forall i,j,
 \end{align}
The general solutions of the constraints further imply that
\begin{align}
    \hmt_{\sigma,j}^x = \mt^x_{\tau,j} +  \mt'^x_{\tau,j}, \quad \hmt_{\sigma,i}^y = \mt^y_{\tau,i} +  \mt'^y_{\tau,i}.
\end{align}

We subsequently sum over all distinct solutions
\begin{align}
\label{eq:M_N_KT}
    M = \bra{\left( m^x_{j-\half}, m^y_{i-\half} \right)} \ket{m^x_{j-\half}, m^y_{i-\half} \in \{0,1\}, \left( m^x_{j-\half}, m^y_{i-\half} \right) \simeq \left( m^x_{j-\half}+1, m^y_{i-\half}+1 \right)}, \quad \forall i,j.
\end{align}
By substituting the solutions \eqref{eq:sol_N_KT} into Eqs.~\eqref{eq:N_KT_1} and \eqref{eq:A_KT_1}, we obtain
\begin{align}
\label{eq:N_KT_2}
    \KT \ket{\{\ssi_{i,j}, \st_{i-\half,j-\half}\}} & = \frac{2^{L_xL_y}}{4^{L_x+L_y}} \sum_{\{\ssip_{i,j},\stp_{i-\half,j-\half}\}} \sum_{( m^x_{j-\half}, m^y_{i-\half}) \in M} (-1)^{\mathcal{A}_{\text{KT}}(\{\sigma,\sigma';\tau,\tau'\})} \ket{\{\ssip_{i,j}, \stp_{i-\half,j-\half}\}},
\end{align}
where the prefactor $2^{L_xL_y}$ arises from the summation over $\hst_{i,j}$ and the phase $\mathcal{A}_{\text{KT}}(\{\sigma,\sigma';\tau,\tau'\})$ in Eq.~\eqref{eq:A_KT_1} becomes
\begin{align}
    \mathcal{A}_{\text{KT}} = & \sum_{i=1}^{L_x} \sum_{j=1}^{L_y} \left(\ssi_{i,j}+ \ssip_{i,j}\right) \left( \st_{i-\half,j-\half}+ \st_{i+\half,j-\half} + \st_{i-\half,j+\half} + \st_{i+\half,j+\half} + \stp_{i-\half,j-\half} + \stp_{i+\half,j-\half} \right. \nonumber \\
  &\left. + \stp_{i-\half,j+\half} + \stp_{i+\half,j+\half} \right) + \sum_{j=1}^{L_y} \left( t^x_{\sigma,j} + t'^x_{\sigma,j} \right) \left( \st_{\half, j-\half} + \st_{\half, j+\half} + \stp_{\half, j-\half} + \stp_{\half, j+\half} \right) \nonumber\\
  & + \sum_{i=1}^{L_x} \left( t^y_{\sigma,i} + t'^y_{\sigma,i} \right) \left( \st_{i-\half,\half} + \st_{i+\half,\half} + \stp_{i-\half,\half}+ \stp_{i+\half,\half} \right) + \left( t^{xy}_{\sigma} + t'^{xy}_{\sigma} \right) \left( \st_{\half,\half} +  \stp_{\half,\half} \right) \nonumber\\
  & + \sum_{j=1}^{L_y} m^x_{j-\half} \left( \mt^x_{\sigma,j-\half} + \mt'^x_{\sigma,j-\half} + \hmt^x_{\tau,j-\half} \right) + \sum_{i=1}^{L_x} m^y_{i-\half} \left( \mt^y_{\sigma,i-\half} + \mt'^y_{\sigma,i-\half} + \hmt^y_{\tau,i-\half} \right).
\end{align}
By performing the sum over $( m^x_{j-\half}, m^y_{i-\half}) \in M$ subject to the constraint specified in \eqref{eq:M_N_KT}, we can derive the compact expression for $\KT$ presented in \eqref{eq:N_KT_def_1} of the main text.

\subsection{Computation of the fusion rule: $\KT \times \KT$}
\label{app:fusion_KTKT}
Here, we outline the derivation of the fusion rule $\KT \times \KT$ given in Eq.~\eqref{eq:fusion_KTKT}. By the definition of $\KT$ in Eq.~\eqref{eq:N_KT_final_1}, we have
\begin{align}
\label{eq:fusion_KTKT_1}
    \KT \times \KT \ket{\{\ssi_{i,j}, \st_{i-\half,j-\half}\}} = 4^{L_xL_y - (L_x+L_y+1)} \sum_{\substack{\{\ssip_{i,j}, \stp_{i-\half,j-\half}\},\\
    \{\ssipp_{i,j}, \stpp_{i-\half,j-\half}\}}} (-1)^{\mathcal{B}} \ket{\{\ssipp_{i,j}, \stpp_{i-\half,j-\half}\}},
\end{align}
where the phase $\mathcal{B}(\{\sigma,\tau;\sigma',\tau'; \sigma'',\tau''\})$ is given by
\begin{align}
\label{eq:B_KTKT_1}
    \mathcal{B} = & \sum_{i=1}^{L_x} \sum_{j=1}^{L_y} \left(\ssi_{i,j}+ \ssip_{i,j}\right) \left( \st_{i-\half,j-\half}+ \st_{i+\half,j-\half} + \st_{i-\half,j+\half} + \st_{i+\half,j+\half} + \stp_{i-\half,j-\half} + \stp_{i+\half,j-\half} \right. \nonumber \\
  &\left. + \stp_{i-\half,j+\half} + \stp_{i+\half,j+\half} \right) + \sum_{j=1}^{L_y} \left( t^x_{\sigma,j} + t'^x_{\sigma,j} \right) \left( \st_{\half, j-\half} + \st_{\half, j+\half} + \stp_{\half, j-\half} + \stp_{\half, j+\half} \right) \nonumber\\
  & + \sum_{i=1}^{L_x} \left( t^y_{\sigma,i} + t'^y_{\sigma,i} \right) \left( \st_{i-\half,\half} + \st_{i+\half,\half} + \stp_{i-\half,\half}+ \stp_{i+\half,\half} \right) + \left( t^{xy}_{\sigma} + t'^{xy}_{\sigma} \right) \left( \st_{\half,\half} +  \stp_{\half,\half} \right) \nonumber\\
  & + \sum_{i=1}^{L_x} \sum_{j=1}^{L_y} \left(\ssip_{i,j}+ \ssipp_{i,j}\right) \left( \stp_{i-\half,j-\half}+ \stp_{i+\half,j-\half} + \stp_{i-\half,j+\half} + \stp_{i+\half,j+\half} + \stpp_{i-\half,j-\half} + \stpp_{i+\half,j-\half} \right. \nonumber \\
  &\left. + \stpp_{i-\half,j+\half} + \stpp_{i+\half,j+\half} \right) + \sum_{j=1}^{L_y} \left( t'^x_{\sigma,j} + t''^x_{\sigma,j} \right) \left( \stp_{\half, j-\half} + \stp_{\half, j+\half} + \stpp_{\half, j-\half} + \stpp_{\half, j+\half} \right) \nonumber\\
  & + \sum_{i=1}^{L_x} \left( t'^y_{\sigma,i} + t''^y_{\sigma,i} \right) \left( \stp_{i-\half,\half} + \stp_{i+\half,\half} + \stpp_{i-\half,\half}+ \stpp_{i+\half,\half} \right) + \left( t'^{xy}_{\sigma} + t''^{xy}_{\sigma} \right) \left( \stp_{\half,\half} +  \stpp_{\half,\half} \right).
\end{align}

We next perform the summation over $\ssip_{i,j}$, which produces a factor of $2^{L_xL_y}$ along with a set of delta-function constraints
\begin{align}
     \Big{(}\st_{i-\half,j-\half} & + \st_{i+\half,j-\half}  + \st_{i-\half,j+\half} + \st_{i+\half,j+\half}  \nonumber \\
  & + \stpp_{i-\half,j-\half} + \stpp_{i+\half,j-\half} + \stpp_{i-\half,j+\half} + \stpp_{i+\half,j+\half} \Big{)} = 0, \quad \forall i,j.
\end{align}

The constraints admit the following general solutions
\begin{align}
\label{eq:sol_KTKT_tau}
     \stpp_{i-\half,j-\half}  = \st_{i-\half,j-\half} + m^x_{\tau,j-\half} + m^y_{\tau,i-\half}, \quad m^x_{\tau,j-\half}, m^y_{\tau,i-\half} \in \{0,1\}, \quad \forall i,j,
\end{align}
In this notation, setting $m^x_{\tau,j-\half} = 1$ corresponds to inserting $U^x_{\tau,j-\half}$, and $m^x_{\tau,i-\half} = 1$ corresponds to inserting $U^y_{\tau,i-\half}$ \cite{Li_2023_KT}. The next step is to sum over all the unique solutions,

\begin{align}
\label{eq:M_KTKT_tau}
    M_{\tau} =  \bigg{\langle} & \left(  m^x_{\tau,j-\half}, m^y_{\tau,i-\half} \right) \bigg{|} m^x_{\tau,j-\half}, m^y_{\tau,i-\half} \in \{0,1\}, \nonumber\\
    & {} \left( m^x_{\tau,j-\half}, m^y_{\tau,i-\half} \right) \simeq \left( m^x_{\tau,j-\half}+1, m^y_{\tau,i-\half}+1 \right) \bigg{\rangle}, \quad \forall i,j.
\end{align}

From these solutions in \eqref{eq:sol_KTKT_tau}, it naturally follows that
\begin{align}
    \mt''^x_{\tau,j} = \mt^x_{\tau,j}, \quad \mt''^y_{\tau,i} = \mt^y_{\tau,i}.
\end{align}
Upon substitution of \eqref{eq:sol_KTKT_tau} into Eqs.~\eqref{eq:fusion_KTKT_1} and \eqref{eq:B_KTKT_1} followed by simplification, the expression reduces to
\begin{align}
\label{eq:fusion_KTKT_2}
    \KT \times \KT \ket{\{\ssi_{i,j}, \st_{i-\half,j-\half}\}} = & 4^{L_xL_y - (L_x+L_y+1)} 2^{L_xL_y}  \sum_{\substack{\{\stp_{i-\half,j-\half}\},
    \{\ssipp_{i,j}\}}} \sum_{( m^x_{\tau,j-\half}, m^y_{\tau,i-\half}) \in M_{\tau} } \nonumber\\
    & \times  (-1)^{\mathcal{B}} \ket{\{\ssipp_{i,j}, \st_{i-\half,j-\half} + m^x_{\tau,j-\half} + m^y_{\tau,i-\half}\}},
\end{align}
where $\mathcal{B}$ in \eqref{eq:B_KTKT_1} reduces to
\begin{align}
\label{eq:B_KTKT_2}
    \mathcal{B} = & \sum_{i=1}^{L_x} \sum_{j=1}^{L_y} \left(\st_{i-\half,j-\half} + \stp_{i-\half,j-\half}\right) \left( \ssi_{i-1,j-1} + \ssi_{i,j-1} + \ssi_{i-1,j} + \ssi_{i,j} +  \ssipp_{i-1,j-1} + \ssipp_{i,j-1} \right. \nonumber \\
  &\left.  + \ssipp_{i-1,j} + \ssipp_{i,j} \right)  + \sum_{j=1}^{L_y} \left( t^x_{\tau,j-\half} + t'^x_{\tau,j-\half} \right) \left(\ssi_{L_x,j-1} + \ssi_{L_x,j} + \ssipp_{L_x,j-1} + \ssipp_{L_x,j}\right) \nonumber\\
  & + \sum_{i=1}^{L_x} \left( t^y_{\tau,i-\half} + t'^y_{\tau,i-\half} \right) \left(\ssi_{i-1,L_y} + \ssi_{i,L_y}+ \ssipp_{i-1,L_y} + \ssipp_{i,L_y}\right) + \left( t^{xy}_{\tau} + t'^{xy}_{\tau} \right) \left( \ssi_{L_x,L_y} +  \ssipp_{L_x,L_y} \right) \nonumber\\
  & + \sum_{j=1}^{L_y} \left( t'^x_{\sigma,j} + t''^x_{\sigma,j} \right) \left( m^x_{\tau,j-\half} + m^x_{\tau,j+\half}\right) + \sum_{i=1}^{L_x} \left( t'^y_{\sigma,i} + t''^y_{\sigma,i} \right) \left( m^y_{\tau,i-\half} + m^y_{\tau,i+\half} \right) \nonumber\\
  &   + \left( t'^{xy}_{\sigma} + t''^{xy}_{\sigma} \right) \left( m^x_{\tau,\half} +  m^y_{\tau,\half} \right).
\end{align}

We further sum over $\stp_{i-\half,j-\half}$, which produces another factor of $2^{L_xL_y}$ along with a set of delta-function constraints
\begin{align}
    \ssi_{i-1,j-1} + \ssi_{i,j-1} + \ssi_{i-1,j} + \ssi_{i,j} +  \ssipp_{i-1,j-1} + \ssipp_{i,j-1} + \ssipp_{i-1,j} + \ssipp_{i,j} = 0, \quad \forall i,j.
\end{align}
Solving the constraints yields the following general solutions,
\begin{align}
\label{eq:sol_KTKT_sig}
    \ssipp_{i,j} = \ssi_{i,j} + m^x_{\sigma,j} + m^y_{\sigma,i}, \quad m^x_{\sigma,j}, m^y_{\sigma,i} \in \{0,1\}, \quad \forall i,j.
\end{align}

Here, $m^x_{\sigma,j} = 1$ indicates an insertion of $U^x_{\sigma,j}$, while $m^y_{\sigma,i} = 1$ indicates an insertion of $U^y_{\sigma,i}$. We then perform the sum over all distinct configurations that satisfy the constraints \eqref{eq:sol_KTKT_sig},
\begin{align}
\label{eq:M_KTKT_sig}
    M_{\sigma} = \bra{\left( m^x_{\sigma,j}, m^y_{\sigma,i} \right)} \ket{m^x_{\sigma,j}, m^y_{\sigma,i} \in \{0,1\}, \left( m^x_{\sigma,j}, m^y_{\sigma,i} \right) \simeq \left(m^x_{\sigma,j}+1, m^y_{\sigma,i}+1 \right)}, \quad \forall i,j.
\end{align}

The general solutions of the constraints in \eqref{eq:sol_KTKT_sig}, in turn, imply that
\begin{align}
    \mt''^x_{\sigma,j-\half} = \mt^x_{\sigma,j-\half}, \quad \mt''^y_{\sigma,i-\half} = \mt^y_{\sigma,i-\half}.
\end{align}
Inserting the solutions \eqref{eq:sol_KTKT_sig} into Eqs.~\eqref{eq:fusion_KTKT_2} and \eqref{eq:B_KTKT_2}, and after a few lines of calculation, the expression simplifies to
\begin{align}
\label{eq:fusion_KTKT_3}
    \KT \times \KT \ket{\{\ssi_{i,j}, \st_{i-\half,j-\half}\}} = & 4^{L_xL_y - (L_x+L_y+1)} 4^{L_xL_y} \sum_{\left( m^x_{\sigma,j}, m^y_{\sigma,i} \right) \in M_{\sigma} }  \sum_{( m^x_{\tau,j-\half}, m^y_{\tau,i-\half}) \in M_{\tau} } \nonumber\\
    & \times  (-1)^{\mathcal{B}} \ket{\{\ssi_{i,j} + m^x_{\sigma,j} + m^y_{\sigma,i}, \st_{i-\half,j-\half} + m^x_{\tau,j-\half} + m^y_{\tau,i-\half}\}},
\end{align}
where the exponent $\mathcal{B}$ in \eqref{eq:B_KTKT_2} reduces to
\begin{align}
\label{eq:B_KTKT_3}
    \mathcal{B} = & \sum_{j=1}^{L_y} m^x_{\sigma,j} \left(\mt^x_{\tau, j} + \mt'^x_{\tau, j}\right) + \sum_{i=1}^{L_x} m^y_{\sigma,i} \left(\mt^y_{\tau, i} + \mt'^y_{\tau, i}\right) + \sum_{j=1}^{L_y} m^x_{\tau,j-\half} \left(\mt^x_{\sigma, j-\half} + \mt'^x_{\sigma, j-\half}\right) \nonumber\\
    & + \sum_{i=1}^{L_x} m^y_{\tau,i-\half} \left(\mt^y_{\sigma, i-\half} + \mt'^y_{\sigma, i-\half}\right).
\end{align}

We may express Eq.~\eqref{eq:fusion_KTKT_3} in a more formal form as
\begin{align}
\label{eq:fusion_KTKT_4}
    \KT \times \KT \ket{\{\ssi_{i,j}, \st_{i-\half,j-\half}\}} = {} &  4^{2 L_xL_y - (L_x+L_y+1)} \sum_{\left( m^x_{\sigma,j}, m^y_{\sigma,i} \right) \in M_{\sigma} }  \sum_{( m^x_{\tau,j-\half}, m^y_{\tau,i-\half}) \in M_{\tau} } \nonumber\\
    & \times \mathcal{P}_{\mathcal{B}} \ket{\{\ssi_{i,j}, \st_{i-\half,j-\half}\}},
\end{align}
where $\mathcal{P}_{\mathcal{B}}$  is a projector onto a definite symmetry-twist sector given by
\begin{align}
    \mathcal{P}_{\mathcal{B}} = &  \left[(-1)^{\sum_{j=1}^{L_y} m^x_{\sigma,j} \left(\mt^x_{\tau, j} + \mt'^x_{\tau, j}\right)} \prod_{j=1}^{L_y} \left(U^x_{\sigma,j}\right)^{m^x_{\sigma,j}} \right] \left[(-1)^{\sum_{i=1}^{L_x} m^y_{\sigma,i} \left(\mt^y_{\tau, i} + \mt'^y_{\tau, i}\right)} \prod_{i=1}^{L_x} \left(U^y_{\sigma,i}\right)^{m^y_{\sigma,i}} \right] \nonumber\\
    & \times \left[(-1)^{\sum_{j=1}^{L_y} m^x_{\tau,j-\half} \left(\mt^x_{\sigma, j-\half} + \mt'^x_{\sigma, j-\half}\right)} \prod_{j=1}^{L_y} \left(U^x_{\tau,j-\half}\right)^{m^x_{\tau,j-\half}}\right] \nonumber\\
    & \times \left[(-1)^{\sum_{i=1}^{L_x} m^y_{\tau,i-\half} \left(\mt^y_{\sigma, i-\half} + \mt'^y_{\sigma, i-\half}\right)} \prod_{i=1}^{L_x} \left(U^y_{\tau,i-\half}\right)^{m^y_{\tau,i-\half}}\right].
\end{align}

By summing over $\left( m^x_{\sigma,j}, m^y_{\sigma,i} \right) \in M_{\sigma}$ and $( m^x_{\tau,j-\half}, m^y_{\tau,i-\half}) \in M_{\tau} $ subject to the constraint specified in \eqref{eq:M_KTKT_sig} and \eqref{eq:M_KTKT_tau}, respectively, we can obtain the final expression for $ \KT \times \KT$ presented in Eq.~\eqref{eq:fusion_KTKT} of the main text.

\bibliography{bibliography}

\end{document}